\documentclass[11pt,letterpaper, table, xcdraw]{article}

\newcommand*{\addFileDependency}[1]{
\typeout{(#1)}
\@addtofilelist{#1}
\IfFileExists{#1}{}{\typeout{No file #1.}}
}\makeatother

\usepackage{xr}
\usepackage{etoc}
\usepackage{latexsym}
\usepackage{amssymb,amsmath, bm}
\usepackage{dsfont}
\usepackage{mathtools}
\usepackage{graphicx}
\graphicspath{{./figs/}}
\usepackage{marvosym}
\usepackage{multirow}
\usepackage{caption}
\usepackage{subcaption}
\usepackage{comment}

\DeclareUnicodeCharacter{0301}{{\'o}}

\usepackage{tikz}
\usepackage{inputenc}
\usetikzlibrary{shapes,arrows,trees,fit,positioning,shapes.misc}
\usetikzlibrary{bayesnet}

\usepackage[backend=biber,articlein=false,style=ext-authoryear,autocite=inline,maxbibnames=99,doi=false,isbn=false,url=false,uniquename=false,natbib]{biblatex}
\AtEveryBibitem{%
  \clearlist{language}%
}
\usepackage{dcolumn}
\newcolumntype{.}{D{.}{.}{-1}}
\newcolumntype{d}[1]{D{.}{.}{#1}}

\usepackage{theorem}
\theoremstyle{plain}
\theoremheaderfont{\scshape}
\newcommand{\qed}{\hfill \ensuremath{\Box}}

\usepackage{kantlipsum}
\allowdisplaybreaks

\usepackage{rotating}

\usepackage{arydshln}
\usepackage[compact]{titlesec}
\usepackage{soul}
\newcommand{\blind}{0}

\usepackage{times}

\usepackage[bookmarksopen=true, bookmarksnumbered=true,
pdfstartview=FitH, breaklinks=true, urlbordercolor={0 1 0}, citebordercolor={0 0 1}]{hyperref}

\usepackage{longtable}
\usepackage{multirow}
\usepackage{array}
\usepackage{wrapfig}
\usepackage{float}
\usepackage{colortbl}
\usepackage{pdflscape}
\usepackage{tabu}
\usepackage{threeparttable} 
\usepackage{threeparttablex} 
\usepackage[normalem]{ulem} 
\usepackage{makecell}
\usepackage{xcolor}
\usepackage{booktabs}
\usepackage{siunitx}
 \definecolor{demblue}{rgb}{0.2, 0.2, 0.6}
 \definecolor{repred}{rgb}{0.7, 0.11, 0.11}

\begin{document}

\newcommand\const{\mathrm{const.}}
\newcommand*\diff{\mathop{}\!\mathrm{d}}
\newcommand*\expec{\mathop{}\mathbb{E}}
\newcommand\numberthis{\addtocounter{equation}{1}\tag{\theequation}}
\newcommand{\appropto}{\mathrel{\vcenter{
  \offinterlineskip\halign{\hfil$##$\cr
    \propto\cr\noalign{\kern2pt}\sim\cr\noalign{\kern-2pt}}}}}

\newcommand\dynMMSBM{\textsf{dynMMSBM}}
\newcommand\dist{\buildrel\rm d\over\sim}
\newcommand\ind{\stackrel{\rm indep.}{\sim}}
\newcommand\iid{\stackrel{\rm i.i.d.}{\sim}}
\newcommand\logit{{\rm logit}}
\renewcommand\r{\right}
\renewcommand\l{\left}
\newcommand\E{\mathbb{E}}
\newcommand\PP{\mathbb{P}}
\newcommand{\argmax}{\operatornamewithlimits{argmax}}

\newcommand\spacingset[1]{\renewcommand{\baselinestretch}%
  {#1}\small\normalsize}

\spacingset{1.25}

\newcommand{\tit}{A Statistical Model of Bipartite Networks:\\
  Application to Cosponsorship in the United States Senate}


\if0\blind

{\title{\tit\thanks{The methods described in this paper can be
     implemented via the open-source statistical software, {\sf
       NetMix}, available at
     \url{https://CRAN.R-project.org/package=NetMix}. The authors are grateful for comments from Alison Craig, Skyler Cranmer, Sarah Shugars and the participants of the Harvard IQSS Applied Statistics seminar.
     }}

 \author{Adeline Lo\thanks{Assistant Professor of Political Science, UW-Madison. Email: \href{mailto:aylo@wisc.edu}{\texttt{aylo@wisc.edu}}, URL:\href{https://www.loadeline.com/}{\tt https://www.loadeline.com/}
     }  \quad \quad Santiago Olivella\thanks{Associate Professor of Political Science, UNC-Chapel Hill. Email: \href{mailto:olivella@unc.edu}{\texttt{olivella@unc.edu}}} \quad \quad Kosuke
   Imai\thanks{Professor, Department of Government and Department of Statistics, Harvard
     University.  1737 Cambridge Street, Institute for Quantitative
     Social Science, Cambridge 02138. Email:
     \href{mailto:imai@harvard.edu}{\texttt{imai@harvard.edu}}, URL:\href{https://imai.fas.harvard.edu/}{\tt https://imai.fas.harvard.edu/}
     }
     }

  \date{}

\maketitle
}\fi

\if1\blind \title{\bf \tit} \date{October 2024} \author{Word Count: 9310}
\maketitle
\fi

\pdfbookmark[1]{Title Page}{Title Page}

\thispagestyle{empty}
\setcounter{page}{0}


\begin{abstract}
Many networks in political and social research are bipartite, with edges connecting exclusively across two distinct types of nodes. A common example includes cosponsorship networks, in which legislators are connected indirectly through the bills they support. Yet most existing network models are designed for unipartite networks, where edges can arise between any pair of nodes. However, using a unipartite network model to analyze bipartite networks, as often done in practice, can result in aggregation bias and artificially high-clustering --- a particularly insidious problem when studying the role groups play in network formation. To address these methodological problems, we develop a statistical model of bipartite networks theorized to be generated through group interactions by extending the popular mixed-membership stochastic blockmodel.  Our model allows researchers to identify the groups of nodes, within each node type in the bipartite structure, that share common patterns of edge formation.  The model also incorporates both node and dyad-level covariates as the predictors of group membership and of observed dyadic relations.  We develop an efficient computational algorithm for fitting the model, and apply it to cosponsorship data from the United States Senate. We show that legislators in a Senate that was perfectly split along party lines were able to remain productive and pass major legislation by forming non-partisan, power-brokering coalitions that found common ground through their collaboration on low-stakes bills. We also find evidence for norms of reciprocity, and uncover the substantial role played by policy expertise in the formation of cosponsorships between senators and legislation. We make an open-source software package available that makes it possible for other researchers to uncover similar insights from bipartite networks.
 
\end{abstract}
  \bigskip 
  
  \noindent {\bf Keywords:} bipartite network, mixed-membership model, social network, stochastic blockmodel, variational inference, Congress, cosponsorship

\newpage
\spacingset{1.83}

\section{Introduction}
\label{sec:intro}

Bipartite networks, where ties connect two distinct actor types without intra-type connections, are common in political and social research. Examples include ethnic group memberships \citep{larson2017}, U.S. state policy adoptions \citep{desmarais_etal_2015}, and product-level trade \citep{kim:liau:imai:20}. These affiliation networks also appear in customer-product relationships \citep{huang_etal_2005}, actor-movie ties \citep{peixoto2014}, and even document-word
occurrences of the kind typically used in text-as-data analyses
\citep[e.g.][]{lancichinetti_etal_2015} can be represented as
bipartite networks.

\begin{figure}[h]
\centering\spacingset{1}
\includegraphics[scale=0.75]{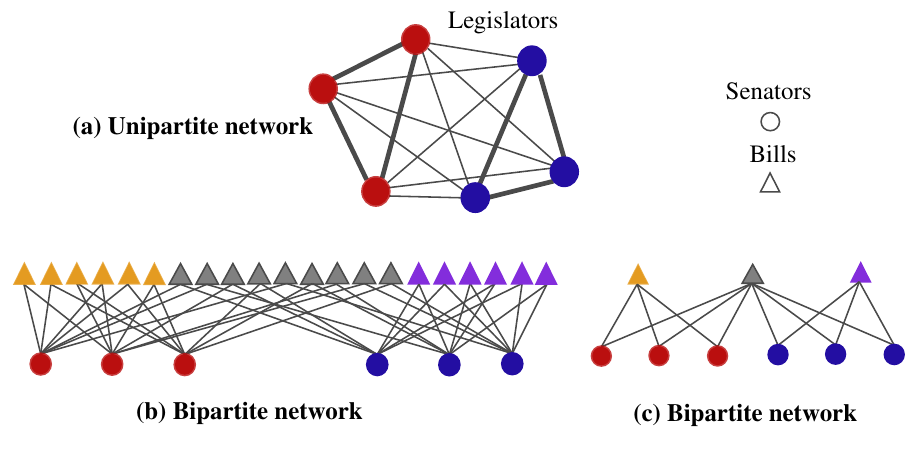}
\caption{{\bf Example networks for bill cosponsorship in bipartite and unipartite forms.} Panels (b) and (c) show different bipartite networks that project to the same unipartite network in panel (a). This projection loses information about bill types (triangle colors) and cosponsorship details (e.g., number of cosponsors, number of bills).}
\label{fig:toy1}
\end{figure}

Despite their ubiquity, bipartite networks are often analyzed by aggregating them into a \emph{unipartite} network, focusing on relationships among one node type.  Consider a stylized example of bipartite network depicted in
panel (b) of Figure~\ref{fig:toy1}, in which legislators (circles) and
bills (triangles) represent two separate types of nodes with
cosponsorship ties occurring only between the two types rather than
within each type. Researchers commonly project this onto a unipartite network of legislators.  Panel~(a) of
Figure~\ref{fig:toy1} shows the resulting projected unipartite
network, where edges between legislators indicate the number of cosponsored bills \citep[e.g.,][]{tam_cho_legislative_2010,muraoka2020}.

Such projections are a common practice whenever researchers analyze bipartite networks. After examining recent publications in
top political science journals, we find that 26 (28\%) of 93 articles model relational data as bipartite, with two exclusive actor types and cross-type connections.  All but one study projects onto a unipartite network (see Table~\ref{tab:netapp} in Appendix~\ref{app:projection} for details on actor types, such as country-product trade networks).

The popularity of this projection strategy is unsurprising, as commonly used statistical models for static network formation in social sciences, such as the latent-space model \citep{hoff2002} and exponential random graph model (ERGM) \citep{frank_strauss1986, wasserman_pattison_1996}, were initially developed for unipartite networks. This holds true for newer approaches like the generalized ERGM \citep{desmarais2012}, additive and multiplicative effects models \citep{minhas2019}, stochastic blockmodel (SBM) \citep{karrer2011}, and dynamic mixed membership stochastic blockmodel (dynMMSBM) \citep{oliv:prat:imai:22}.

Unfortunately, the projection of a bipartite network onto a unipartite
network leads to substantial loss of information, possibly resulting
in misleading estimates of the determinants of network ties
\citep[e.g.][]{marrs2020inferring}.   In Figure~\ref{fig:toy1}, panels (b) and (c) depict entirely different bipartite networks that yield the same unipartite network in panel (a). Furthermore, such projections tend to inflate clustering coefficients \citep[e.g.,][]{newman:etal:2001, guillaume:latapy:2004}, which can misestimate network community structure and polarization. This issue is critical for theories focusing on group dynamics, community formation, and coalition behaviors \citep[e.g.,][]{gonzalez:wang:2016, larson:etal:2019, sunstein:2009, sunstein:2018}.

To minimize the risks brought about by these issues and avoid the need
for projections altogether, we extend the popular mixed membership
stochastic blockmodel (\texttt{MMSBM}; \citeauthor{airoldi2008}
\citeyear{airoldi2008}, and \texttt{dynMMSBM};
\citeauthor{oliv:prat:imai:22} \citeyear{oliv:prat:imai:22}) to
bipartite networks, in which groups are theorized to play an
influential role. The proposed model, which we call \texttt{biMMSBM},
allows researchers to discover the groups of nodes, within each node
type or family, that share common probabilistic patterns of edge
formation (so called \emph{stochastic equivalence classes}). In the example of cosponsorship, \texttt{biMMSBM} categorizes legislators and bills into meaningful groups based on their cosponsorship behaviors, avoiding artificial polarization and hyper-partisanship in congress.

The \texttt{biMMSBM} is based on a mixed-membership (or admixture)
structure, allowing nodes of one type to belong to multiple latent groups depending on interactions with nodes of the other type.  This flexibility allows us to capture nuanced social
interactions, in which actors adopt different roles when interacting
with others.  It also sets our model appart from most of the existing
bipartite community detection models that assume every node (or every
edge) belongs to a single group
\citep[e.g.][]{govaert_clustering_2003,larremore_efficiently_2014,
  zhou_analysis_2019, kim2020}.\footnote{Allowing for
  mixed-memberships also
  sidesteps the serious estimation issues common in ERGM-style
  modeling approaches \citep{schweinberger2011, chatterjee2013}.
  Issues like inferential degeneracy and ill-behaved likelihood
  surfaces, which plague unipartite ERGMs, are also present in their
  bipartite extensions --- even after resorting to common
  regularization strategies \citep[such as geometric weighting; see,
  for instance,][]{skvoretz1999, wang2009}.}

In cosponsoring networks, for example, a legislator may cooperate with
a different group of colleagues when considering the cosponsorship of
bills in different policy areas.  Our model allows for this
possibility by letting each senator belong to a different latent
group, depending on the types of a bill that is under consideration
for co-sponsorship.  In contrast, most of the existing bipartite
network models assume that each senator belongs to the same latent
group, regardless of bills to be co-sponsored.  Similarly, under the
proposed model, each bill can be part of different latent groups,
depending on which senator is considering its cosponsorship.  Thus,
\texttt{biMMSBM} can capture the complexity of social networks through
a wide range of edge formation patterns.

In addition, our model supports the use of covariates to explain edge formation between nodes of different types \citep{white_murphy_2016, razaee_matched_2019}. It incorporates two types of covariates: node-level covariates describe learned group memberships, such as legislators' ideology and partisanship, and bills' policy content or author characteristics in the cosponsorship example. Additionally, dyadic covariates predict edge formation directly, relaxing the reliance on latent group structures alone in network generation. This capability accommodates theoretically relevant variables defined for pairs of nodes of different types, like whether a legislator belongs to committee(s) a bill was referred to.

In contrast, many existing modeling approaches force researchers to
adopt a two-step analytic strategy, conducting standard regression
analyses of network model outputs \citep[e.g.][]{maoz_etal_2006,
  handcock2007,zhang_community_2008, cao2009,
  tam_cho_legislative_2010,battaglini_etal_2020}. The few approaches
that could accommodate bipartite networks and both nodal and dyadic
covariates either require changing the inferential goal \citep[e.g., 
identifying the ``backbone'' of a network; see][]{neal2014backbone},  lack available software implementations (e.g., the
otherwise excellent AME model of \citeauthor{hoff2021additive},
\citeyear{hoff2021additive}), or suffer from issues inherited from
their unipartite counterparts (e.g., the bipartite ERGM model
\citeauthor{agneessens2004}, \citeyear{agneessens2004}, implemented in
the R package \texttt{ergm}, can experience severe practical issues of
degeneracy and misspecification). Our model offers a single-step,
comprehensive approach to network analysis with well-behaved posterior
distributions, facilitating applied research testing group roles in
network formation or predicting new actor behaviors.

One disadvantage of \texttt{MMSBM}-type network models is that a fully
Bayesian inference strategy relying on Markov chain Monte Carlo
simulation is computationally prohibitive for networks of medium or
large size.  To overcome this, we follow the computational strategy
used in the existing methodological literature and develop a
computationally efficient variational Bayes approximation to our
model's collapsed posterior using stochastic optimization
\citep{teh_collapsed_2007,airoldi2008,hoffman_etal2013,gopalan_blei2013,oliv:prat:imai:22}. \if0\blind{
  We implement this fitting algorithm in the open-source software
  package {\tt NetMix} \citep{olivella_etal_2021} so that researchers
  can use the proposed model in their own research.  }\fi \if1\blind
{We implement this fitting algorithm in an open-source software
  package for R so that researchers can use the proposed model in
  their own research.}  \fi

To demonstrate \texttt{biMMSBM}'s applicability in studying political
interrelations, we apply the model to a network of cosponsorship decisions in the U.S. Congress. As coalitions are at the
heart of legislative politics \citep{riker:1962}, a model adept at identifying and explaining latent groups is ideal for understanding the politics of legislative support that underlie these cosponsorship decisions. More specifically, we study the patterns of
legislative cosponsorship during the penultimate instance of a
perfectly split Senate in U.S. history --- the 107th Congress. We
model the bipartite network connecting Senators to legislation (or
``bills'') through the discovery of latent groups, while examining the
roles of Senator characteristics (party, ideology, seniority, and
gender, bill characteristics (time of introduction, the party,
ideology, seniority, and gender of the bill's sponsor; and the bill's
substantive topic), and Senator-bill dyadic features (reciprocity and
shared legislative committees).

Contrary to the results of a unipartite network analysis, our modeling strategy uncovers cross-party collaboration among senators in cosponsoring low-stakes legislation, which later facilitates consideration of more contentious bills. Junior senators from both parties notably drive this cooperation. Additionally, shared committee memberships and bill-specific reciprocity norms in cosponsorship are crucial findings often missed in projected network measures.

In what follows, we discuss this motivating application --- the
politics of cosponsorship in the U.S. --- and explain the risk of
aggregation bias and artificial inflation of network clustering when
projecting bipartite networks to unipartite ones
(Section~\ref{sec:data}). We then detail our modeling
approach in Section~\ref{sec:model}, and present empirical findings from the cosponsorship network during the 107th Congress in
Section~\ref{sec:empirical}.  Finally, in Section~\ref{sec:conclude},
we conclude with implications for other domains and future research.

\section{The Cosponsorship Network Among Senators}
\label{sec:data}

In this section, we introduce the cosponsorship network data among
legislators in the U.S. Senate, which serves as our motivating
example. We point out the bipartite nature of the data and explain why
projecting this bipartite network onto a unipartite one results in
loss of information, an incorrect (and inflated) notion of how polarized relations are,
and possibly aggregation bias.

\subsection{Background}

Cosponsorship among U.S. senators reveals their legislative interests and goals, as it signifies public endorsement of specific legislation \citep[see e.g.,][]{koger:2003, tam_cho_legislative_2010, arnold_friendship_2000, kirkland_relational_2011}. In the Senate, sponsorship constraints make cosponsorship crucial for indicating broader support, increasing media attention, and serving as a credible commitment device \citep{krutz_issues_2005, bernhard:sulkin:2013}. The Senate's 60-vote threshold for overcoming filibusters amplifies the significance of cosponsorships, particularly bipartisan ones \citep{rippere2016polarization}.

Furthermore, previous work has noted the importance of collaboration
among senators in legislative productivity and influence
\citep{fowler_legislative_2006,holman2022let} and the costs of
reneging on cosponsorships \citep{bernhard2013commitment}. Although
there are limits to what cosponsorships can help us understand (for
instance, \citet{wilson1997cosponsorship} and \citet{anderson2003keys}
suggest limits to how much cosponsorship can predict ultimate bill
passage), evidence suggests they significantly impact legislators' effectiveness \citep[e.g.,][]{harbridge:etal:2023} and represent clear issue positions \citep{desposato:etal:2011, lawless_nice_2018}.

Scholars have sought to identify the factors influencing cosponsorship \citep[see, e.g.,][]{campbell1982cosponsoring, krutz_issues_2005, grossmann2013lobbying, fong2020expertise}. This study examines cosponsorship to explore partisan gridlock and collaborative dynamics in the Senate, particularly how cosponsorship patterns reflect bipartisan cooperation amid potential stalemates from partisan divides. What insights can these networks provide into overcoming hyper-partisanship? Can the Senate maintain productivity during perfect partisan splits? Addressing these questions requires assessing collaboration breadth, including variations in the types of legislation senators choose to cosponsor.

\begin{figure}[p]
\centering\spacingset{1}
\begin{subfigure}[b]{0.48\textwidth}
  \centering
  \includegraphics[width=0.975\textwidth]{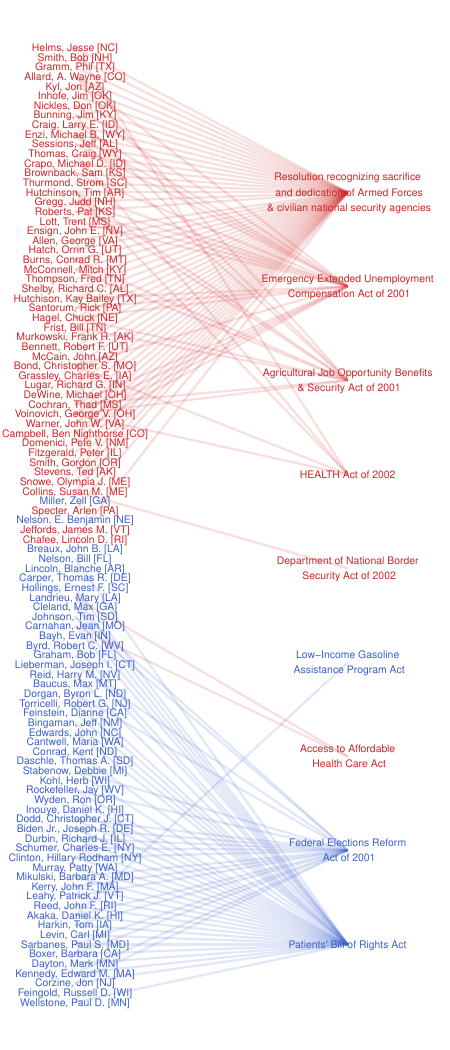}
  \caption{Sample of Partisan Bills}
\end{subfigure}
\hfill
\begin{subfigure}[b]{0.48\textwidth}
  \centering
  \includegraphics[width=0.975\textwidth]{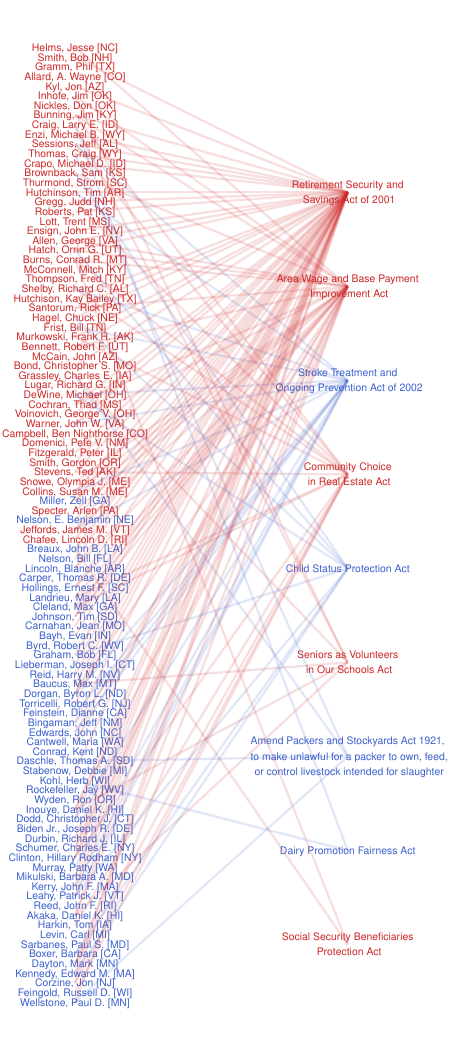}
    \caption{Sample of Bipartisan Bills}
\end{subfigure}
    \caption{ {\bf Cosponsorship Networks among Senators in the
        107th Congress.}  The figure shows two bipartite networks sampled from the 107th Congress, with all 100 senators sorted by ideology (most conservative senators at the top) and a sample of bills sorted by node degree. The left panel depicts a network where bills are predominantly cosponsored either by Republicans or Democrats, while the right panel shows bills with highly bipartisan cosponsorship compositions. These networks highlight significant heterogeneity in composition and degree across bill nodes in our dataset.}
\label{fig:107webUB}
\end{figure}

This last point is important, because such heterogeneity is common in
cosponsorship networks. Consider, for example, the two sample
bipartite networks in Figure~\ref{fig:107webUB}. These are drawn from
the full network of consponsorships during 107th Congress, which was
in session between 2001 and 2003, as the presidency transitioned from
the Clinton to the Bush administration. The networks in both panels
contain all 100 senators (left-side nodes) and two samples of bills
(right-side nodes). On the left panel, we look at senator-bill
bipartite patterns for bills that are heavily partisan in
cosponsorship --- either primarily cosponsored by Republicans or by
Democrats only). On the right, Figure~\ref{fig:107webUB} presents
bipartite patterns for bills that are heavily \emph{bipartisan} in
cosponsorship. Over these two subsets of bills, we observe substantial
heterogeneity in cosponsorship behaviors: while a few bills attract
many cosponsors (represented by multiple drawn edges to the bill),
many more have relatively few.\footnote{More formally, the degree
  distribution of bills is best captured using a power law, with many
  bills having few cosponsors no bills, and a few bills being
  cosponsored by many senators. In contrast, the degree distribution
  of senators is far less heavy tailed, suggesting less heterogeneity
  in behavior. See Section~\ref{app:degrees} of the Online Supplementary Information for
  details. } Some bills are highly As demonstrated next, aggregating
over bills by projecting bipartite cosponsorships onto a network among
senators, as commonly practiced, would erase this heterogeneity. If
cross-party collaboration is systematically associated with whatever
brings about this heterogeneity, we risk painting an incorrect picture
of how partisanship drives collaboration among legislators.

\subsection{Projection onto a unipartite network can be misleading}
\label{sec:agg}

Despite this risk, the common practice in studying bipartite networks
involves projecting them onto unipartite ones. In this case, one might
transform the senator-bill bipartite network into a senator-only
unipartite graph, where edges between senators indicate cosponsorship
of bills together \citep[or weighted by the number of bills; see
e.g.,][]{tam_cho_legislative_2010}. However, this approach risks
omitting important bill-specific and senator-bill pair-specific
information crucial for understanding cosponsorship decisions,
including policy content, timing of bill introduction and endorsement,
and collaboration extent through popular legislation
\citep{neal_backbone_2014, neal:2020, kirkland:gross:2014}.

Thus, aggregating over potentially relevant heterogeneity in bill- and
senator-bill level data can lead to incorrect substantive
takeaways. To see how this may be the case, revisit the stylized
scenario presented in Figure~\ref{fig:toy1} of
Section~\ref{sec:intro}. Panel~(a) shows a weighted unipartite
representation of a cosponsorship network where circle nodes denote
senators and triangle nodes represent bills, with edge weights
indicating cosponsorships. In this graph, the circle nodes
representing senators have weighted edges connecting them to
cosponsored triangle bills, with weights given by the number of such
cosponsorships. Node colors denote party and policy area. A cursory
view of this graph would suggest that while there is some cross-party
collaboration, the bulk of it is expected to occur within parties ---
a pattern that could be taken as evidence of polarization in the
collaboration network.

Yet, \emph{the same} unipartite representation can be derived from two
disparate bipartite network structures. Indeed, projecting the
networks in panels~(b) and (c) at the bottom of Figure~\ref{fig:toy1}
using sums over shared cosponsorships results in exactly the same
weighted network depicted in panel~(a), despite representing two
\emph{fundamentally different} legislative collaboration environments.

In panel~(b), legislators are highly productive in terms of
legislative output (given by the large number of bills represented by
triangular nodes in the graph), and cross-party work is common (as
indicated by the relatively large number of gray triangular nodes in
the graph). Indeed, the bipartisan vs. within-party cosponsored
legislation ratio is 3 to 4, and the average probability that two bill
cosponsors selected at random belong to the same party is
0.43.  The graph in
panel~(c), in contrast, tells a different story altogether. Senators
in this network are far less productive in terms of individual bills,
more unified in terms of within party cosponsorships, and collaborate
only once across the aisle through a single, widely cosponsored
bill.\footnote{In formal graph-theoretic terms, the vertex
  connectivity of network~(c) is much smaller than that of
  network~(b), with a cutset of only a single bill cosponsored by all
  legislators: it would only take removing the only shared bill (gray
  triangle) to separate the network into two disconnected components.}
Likewise, the bipartisan versus within-party cosponsored legislation
ratio is much lower --- namely, 1 to 2 --- and the average probability
that two cosponsors of a bill chosen at random are, in turn,
copartisans, is about double --- namely 0.84.\footnote{For the network in this panel, the distribution of these probabilities across bills is also bimodal, but now left-skewed, with masses at 0.5 (for the gray bill in the middle) and 1 (for the remaining two bills).}

This substantive differentiation of bipartisan versus polarized behavior is completely
obscured in the network in the top of Figure~\ref{fig:toy1}, which
depicts the result of projecting either of these bipartite graphs using weights that track the number of bills through which connections happen.\footnote{Some recent work develops bipartite (weighted) projection approaches that at least permit some inferences on network degree heterogeneity, but the main issue of non-injectiveness remains \citep[see, among others,][]{marrs2020inferring,baltakys2023inference}.} In this
new network among senators, the strength of within-party ties is much
higher than that of ties across party lines, and the average
probability that a connection is made with a copartisan is thus fairly
high --- despite how different the probabilities of randomly picking
cosponsors of the same party are in the two networks. Overall,
then, relevant information about the nature of collaboration can be
lost in aggregating over node types.

\begin{figure}[h]
  \centering \spacingset{1} 
  \includegraphics[scale=0.7]{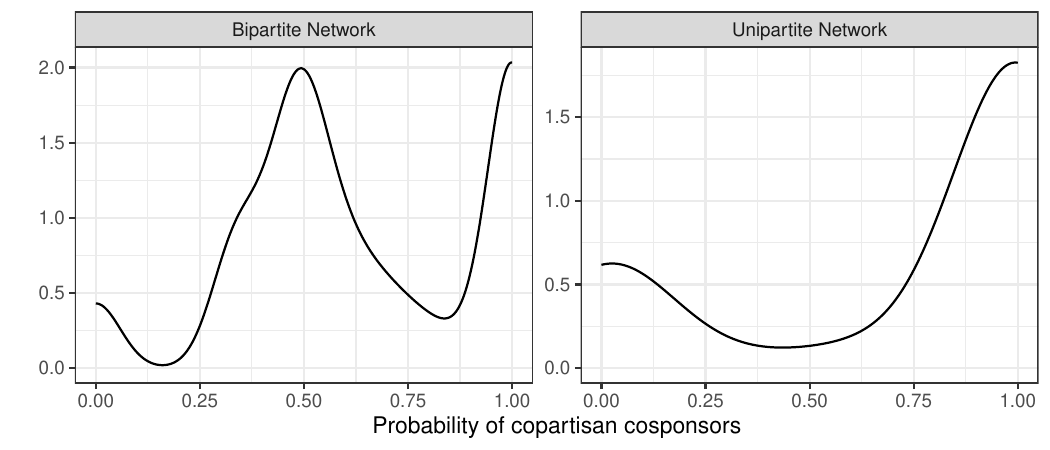}
  \caption{\textbf{Probability of copartisan cosponsors during the
      107th Senate}: The figure displays distributions of probabilities: left panel shows probabilities that any two distinct cosponsors of a bill are from the same party, and right panel shows probabilities that a senator's randomly chosen pair of cosponsors are copartisan. The bipartite network reveals substantial bipartisan cosponsorship, while the weighted unipartite network among senators indicates less cooperation.}
  \label{fig:CopartCospon}
\end{figure}

Such aggregation bias is palpable when considering the 107th
Senate. Figure~\ref{fig:CopartCospon} shows the distribution of
probabilities that a randomly selected pair of cosponsors belong to
the same party, computed for each bill in the original bipartite
cosponsorship network (left panel) and for each senator in the
unipartite senator-only, weighted projection of the bipartite
network. When considering the bipartite network, the average
probability that a randomly selected pair of each bill's cosponsors
belong to the same party is about 0.63. The distribution over these
probabilities of copartisan cosponsors is distinctly multimodal,
revealing that we are just as likely to find perfectly bipartisan
bills (i.e. bills for which the probability of drawing two cosponsors
who are also copartisans is about 1/2) as we are to find a perfectly
partisan bill (i.e., a bill for which all cosponsors belong to the
same party).\footnote{To ensure the measure is not distorted by
  comparing a cosponsor to herself, we compute the bill-specific
  probability of two copartisan cosponsors being drawn without
  replacement.}

The distribution associated with the projected unipartite network,
shown on the right panel of Figure~\ref{fig:CopartCospon}, paints a
completely different picture. Here, the average probability that each
senator's randomly selected cosponsor belongs to the same party is
about 0.75, suggesting a typical legislator will cosponsor with a set
of others who are highly likely to be copartisans.\footnote{Since we
  compute a weighted projection, we sample distinct neighbors of a
  senator in proportion to the strength of their connections to the
  later.}  Furthermore, the distribution is markedly skewed left, with
a vast majority of senators collaborating mostly with copartisans.

Unfortunately, this kind of strong, artificial clustering present in
both the real projected network of the 107th Senate and the simple
example in panel (a) of Figure~\ref{fig:toy1} is a manifestation of a
more general and systematic phenomenon
\citep{newman:etal:2001,latapy2008basic}, and can lead to incorrect
conclusions about the extent and nature of polarization in Congress,
as we show in Section~\ref{sec:empirical} below. To address these
concerns, we introduce a new modeling strategy in the following
section. 

\section{The Proposed Methodology}
\label{sec:model}

In this section, we begin by describing the core intuition behind the
model, which we refer to as the bipartite mixed-membership stochastic
blockmodel (\texttt{biMMSBM}). We then formally present the full
modeling approach, and discuss estimation strategies that enable the
analysis of large networks.

\subsection{Modeling strategy}

We represent an observed network as a bipartite graph, for which there
exist two disjoint sets or families of nodes.  For a bipartite graph,
a set of undirected edges are formed between pairs of nodes belonging
to these two different families.  In other words, no edge exists
between any two nodes of the same family.  In our application,
senators and legislative bills correspond to these two families. The
edges represent cosponsorship relationships, which only occur between
senators and bills and do not exist directly among legislators or
bills.

The bipartite mixed-membership stochastic blockmodel
(\texttt{biMMSBM}) assumes that a node belongs to one of the several
latent groups when interacting with each node of the other family.
For any dyadic relationship between two nodes of different families,
the latent group memberships of the nodes determine the likelihood of
forming an edge. Thus, a senator may belong to different latent
communities when deciding whether to co-sponsor different
legislations.  Similarly, bills can be sorted into separate latent
groups across senator-bill dyads.  For instance, John McCain (R-AZ)
might have behaved similarly to other Republicans when deciding
whether to cosponsor bills related to national security, but might
have acted differently when considering bills related to campaign
finance reform --- a pattern that could help us understand his
reputation as a party maverick.  This contrasts with many existing
bipartite network models, under which senators would have to belong to
a single latent group regardless of types of bills to be co-sponsored.

\begin{figure}[t]
  \centering \spacingset{1}
  \includegraphics[scale=0.65]{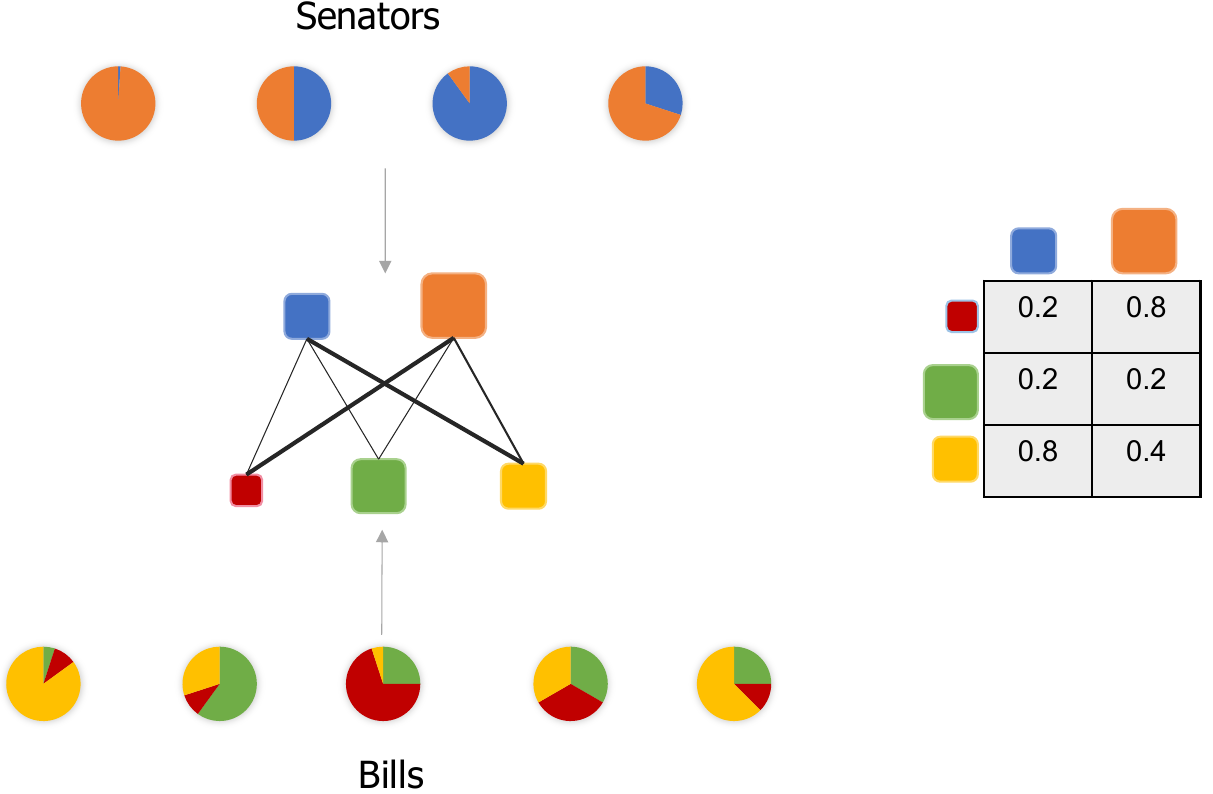}
  \caption{{\bf Mixed-Membership Stochastic Blockmodel for Bipartite Networks.} The schematic depicts a 2-by-3 latent community model, where four senators exhibit mixed memberships across two communities (blue and orange), and five bills exhibit mixed memberships across three communities (yellow, red, green). Community affinities between senators and bills are encoded in the block model matrix on the right, illustrated by edge thickness in the left graph.
}
    \label{fig:intuition}
\end{figure}

To capture this, we define a probabilistic model to account for diverse latent community memberships.Figure~\ref{fig:intuition} presents a schematic representation of our
model, where different colors in piecharts represent the relative
likelihoods of four senators belonging to two different communities
(blue and orange).  The same idea applies to bills, which might
attract different sets of communities of senators as cosponsors. Here,
each of five bills belongs to a mixture of three different communities
(represented by red, green, and yellow).  The fact that each node
belongs to their corresponding communities corresponds to the
``mixed-membership'' portion of the model.  Mixed-memberships can be
explained using node-level covariates --- characteristics of senators
and bills, such as ideology for senators and policy area for bills.

The $2 \times 3$ matrix on the right of Figure~\ref{fig:intuition},
shows the \emph{blockmodel}, indicating probabilities of cosponsorship between senator and bill communities. Certain community pairs (e.g., orange senators and red bills) show higher cosponsorship probabilities than others (e.g., blue senators and green bills), reflecting diverse coalitional strategies among senators towards legislation.

Cosponsorship networks often exhibit this stochastic equivalence, where \textit{coalitions} of senators support similar \textit{classes} of legislation \citep[e.g.,][]{bratton_networks_2011}. Similar group-based dynamics are found in other networks, like economic trade between countries and co-occurrence of words in documents. We now proceed with a formal presentation of our full model.

\subsection{The Bipartite Mixed-Membership Stochastic Blockmodel}

Formally, let $G(V_1,V_2,E)$ represent a bipartite graph where $V_1$
and $V_2$ denote the two disjoint families of nodes
($V_1 \cap V_2 = \emptyset$), and $E$ represents the set of undirected
edges, or pairs of nodes of different families.  Suppose that family~1
has a total of $N_1 = |V_1|$ nodes, and that the number of nodes in
family~2 is $N_2=|V_2|$.  For each dyad, we let
$z_{pq}\in \{1,\ldots,K_1\}$ denote the latent group, to which node
$p \in V_1$ of family~1 belongs when interacting with node $q \in V_2$
of family~2, whose latent group membership is denoted by
$u_{pq}\in \{1, \ldots, K_2\}$. In general, we allow $K_1\neq
K_2$. Further, we use $y_{pq} = 1$ to denote the existence of an edge
between these two nodes $\{p,q\}\in E$ while $y_{pq} = 0$ indicates
its absence.

We assume that the probability of edge formation is a function of
dyadic predictors $\bm{d}_{pq}$ and a blockmodel $\bm{B}$, which is a
$K_1\times K_2$ matrix representing the log odds of forming an edge
between any two latent group members, as illustrated on the right of
Figure~\ref{fig:intuition},
\begin{equation}
y_{pq} \mid z_{pq},u_{pq}, \bm{B}, \boldsymbol{\gamma} \ \ind \
\text{Bernoulli}\left(\text{logit}^{-1}(B_{z_{pq},u_{pq}} + \bm{d}_{pq}^\top \boldsymbol{\gamma})\right),
\end{equation}
where  $\boldsymbol{\gamma}$
denotes a vector of dyad-level regression coefficients. By including a
set of dyadic predictors, the model allows edge
formation probabilities to be different even for pairs of nodes that
have instantiated the same latent groups, thus relaxing the common
assumption of stochastic equivalence made by stochastic blockmodels
(or SBMs). Substantively, this allows for
scenarios where senator-bill dyads whose respective nodes sort into
the same pairs of latent communities to be further differentiated in
cosponsorship likelihood by characteristics pertinent to their
particular dyad --- such as a senator's collaborative history with the
author of a bill.

As is common in mixed-membership SBMs, we define a categorical
sampling model for the dyad-specific group memberships, $z_{pq}$ and
$u_{pq}$, so that
\begin{equation}
       z_{pq} \mid \boldsymbol{\pi}_p \sim \text{Categorical}(\boldsymbol{\pi}_p), \quad u_{pq} \mid \boldsymbol{\psi}_q \sim \text{Categorical}(\boldsymbol{\psi}_q), 
\end{equation}
where the probability that node $p$ of family~1 (node $q$ of family~2)
belongs to a latent group on any possible interaction is given by
$\boldsymbol{\pi}_p$ ($\boldsymbol{\psi}_q$) --- a $K_1$-dimensional
($K_2$-dimensional) probability vector usually known as the
\emph{mixed-membership} vector (represented as the pie charts for each
node in Figure~\ref{fig:intuition}).

Furthermore, our model incorporates node-level information. This is a
critical feature of the model because it incorporates (monadic)
senator and bill level predictor information directly into the
definition of the mixed-membership probabilities of latent groups.
These covariates themselves predict the cosponsorship likelihood
through the resulting instantiated element of the blockmodel.
Specifically, we assume that the mixed-membership probability vectors
are generated by a Dirichlet distribution with concentration
parameters that are a function of node covariates,\footnote{Refer to Section S.1 
 of the Online SI for a plate diagram illustrating the full model, including monadic and dyadic components.}
\begin{equation}
       \boldsymbol{\pi}_{p} \mid \boldsymbol{\beta}_1 \sim \text{Dirichlet}\l(\l\{\exp(\bm{x}_{p}^\top\boldsymbol{\beta}_{1g})\r\}_{g=1}^{K_1}\r), \quad \boldsymbol{\psi}_{q} \mid \boldsymbol{\beta}_2 \sim \text{Dirichlet}\l(\l\{\exp(\bm{w}_{q}^\top\boldsymbol{\beta}_{2h})\r\}_{h=1}^{K_2}\r), 
\end{equation}
where hyper-parameter vectors $\boldsymbol{\beta}_{1g}$ and
$\boldsymbol{\beta}_{2h}$ contain regression coefficients associated
with the $g$th and $h$th groups of vertex families~1~and~2,
respectively.

Putting it all together, the full joint distribution of data and
latent variables is given by,
\begin{align}
f(\bm{Y},\bm{Z},\bm{U},\boldsymbol{\Pi},\boldsymbol{\Psi}, \mid \bm{B}, \boldsymbol{\beta}, \boldsymbol{\gamma}) 
& = \prod_{p,q \in V_1\times V_2} f(y_{pq}\mid z_{pq},
u_{qp}, \bm{B}, \boldsymbol{\gamma}) f(z_{pq}\mid \boldsymbol{\pi}_{p}) f(u_{pq}
\mid \boldsymbol{\psi}_{q}) \notag\\
&\quad\times 
\prod_{p\in V_{1}} f(\boldsymbol{\pi}_{p}\mid \boldsymbol{\beta}_1)
\prod_{q\in V_{2}} f(\boldsymbol{\psi}_{q}\mid \boldsymbol{\beta}_2).
\label{eq:joint}
\end{align}

This specification allows us to more formally describe the potential
issues raised by aggregation illustrated informally on
Section~\ref{sec:agg}.  A typical aggregation strategy simply sums the
number of connections to a member of family $V_2$ shared by two
members of family $V_1$, which can be obtained by defining an
aggregated sociomatrix $\widetilde{\bm{Y}}=\bm{Y}\bm{Y}^\top$. Under
this aggregation strategy, and in the absence of dyadic covariates,
the model in Equation~\eqref{eq:joint} implies
\begin{equation}
  \expec[\widetilde{\bm{Y}}] = \expec[\bm{Y}\bm{Y}^\top] >
  \boldsymbol{\Pi}\left[\bm{B}\boldsymbol{\Psi}^\top\boldsymbol{\Psi}\bm{B}^\top
  \right]\boldsymbol{\Pi}^\top
  \label{eq:agg_expec}
\end{equation}
where $\boldsymbol{\Pi}$ is a $N_1\times K_1$ matrix that stacks mixed
memberships $\boldsymbol{\pi}_p$ for all $p\in V_1$, and similarly for
$\boldsymbol{\Psi}$. The issue arises because the bracketed terms in Equation~\eqref{eq:agg_expec} (i.e., the blockmodel and the mixed-memberships of Family $V_2$) cannot be separately identified from the aggregated sociomatrix $\widetilde{\bm{Y}}$, leading to observational equivalence as shown in Figure~\ref{fig:toy1}. This can lead to misconstrued relationships among members of Family 1 when relying on aggregated data. Our model avoids this issue by directly modeling the original bipartite network without aggregation or projection.

\subsection{Estimation}

With the thousands of vertices and millions of potential edges
involved in an application such as bill cosponsorships, sampling
directly from the posterior distribution given in
Equation~\eqref{eq:joint} is computationally prohibitive. To obtain
estimates of quantities of interest in a reasonable amount of time, we
follow the computational strategy of \citet{oliv:prat:imai:22} by
first marginalizing latent variables and then defining a stochastic
variational approximation to the full posterior.  We briefly summarize
these computational strategies here.

\paragraph{Marginalization.}

At first, and given their high dimensionality, marginalizing out
dyad-specific group memberships (i.e. $\bm{Z}$ and $\bm{U}$)
may seem attractive. Doing so, however, would result in variational
updates that cannot be computed in closed-form, as the
Dirichlet-distributed mixed-membership vectors are not conjugate with
respect to the Bernoulli likelihood we have adopted. Instead, we
collapse the full posterior over the mixed-membership vectors
(i.e. $\boldsymbol{\Pi}$ and $\boldsymbol{\Psi}$):
\begin{align}
& f(\bm{Y},\bm{Z},\bm{U}, \mid \bm{B}, \boldsymbol{\beta}, \boldsymbol{\gamma}) \notag\\
 = \ & \int\int f(\bm{Y},\bm{Z},\bm{U},\boldsymbol{\Pi},\boldsymbol{\Psi},\mid \bm{B}, \boldsymbol{\beta}, \boldsymbol{\gamma})\; d\boldsymbol{\Pi} d\boldsymbol{\Psi}\notag\\
=\ & \prod_{p,q\in  V_{1}\times V_{2}} \Biggl[\theta_{pq,z_{pq},u_{pq}}^{y_{pq}}(1-\theta_{pq,z_{pq},u_{pq}})^{1-y_{pq}}\notag\\
& \qquad \times \left(\frac{\Gamma(\xi_{p})}{\Gamma(\xi_{p}+ N_{2})} \prod_{g=1}^{K_1} \frac{\Gamma(\alpha_{pg}+ C_{pg})}{\Gamma(\alpha_{pg})}\right) \left(\frac{\Gamma(\xi_{q})}{\Gamma(\xi_{q}+ N_{1})} \prod_{h=1}^{K_2} \frac{\Gamma(\alpha_{qh}+ C_{qh})}{\Gamma(\alpha_{qh})}\right)\Biggr] 
\label{eq:coll} 
\end{align}
where $\Gamma(\cdot)$ is the Gamma function,
$\alpha_{pg} = \exp(\bm{x}_{p}^\top\boldsymbol{\beta}_{1g})$,
$\xi_{p} = \sum_{g=1}^{K_1} \alpha_{pg}$ (and similarly for
$\alpha_{qh}$ and $\xi_{q}$), and
$C_{pg} = \sum_{q \in V_{2}}\mathds{1}(z_{pq}=g)$ is a count statistic
representing the number of times node $p$ instantiates group $g$
across its interactions with nodes in family~2 (and similarly for
$C_{qh}$). Lastly, 
$\theta_{pq, z_{pq}, u_{qp}} = \text{logit}^{-1}(B_{z_{pq}, u_{qp}} +
\bm{d}_{pq}^\top \boldsymbol{\gamma})$ is the probability of a tie
between the vertices in dyad $p,q$.

\paragraph{Stochastic Variational Inference.}
To enhance scalability, we employ two strategies. First, we rely on a mean-field variational
approximation to the collapsed posterior in Equation~\eqref{eq:coll}
\citep{blei:etal:2017}, which first defines a lower bound
$\mathcal{L}(\boldsymbol{\Phi})$ for this collapsed posterior, and
then tightens the bound by updating the parameters $\boldsymbol{\Phi}$
of the approximating distributions by following a strategy similar to
that of the EM algorithm. Previous studies indicate that similar marginalization approaches enhance variational approximation quality
\citep{teh_collapsed_2007}.

Second, we rely on stochastic optimization to find the maximum of the
lower bound \citep{hoffman_etal2013, foulds_etal2013,
  dulac_etal2020}. To do so, our algorithm follows, with decreasing
step sizes, a noisy estimate of the gradient of
$\mathcal{L}(\boldsymbol{\Phi})$ formed by subsampling dyads in the
original network. Provided the schedule of step sizes satisfies the
Robbins-Monro conditions, and the gradient estimate is unbiased, the
procedure is guaranteed to find a local optimum of the variational
target \citep{hoffman_etal2013}. Critically, it does so while using a
fraction of the available data at each iteration, thus dramatically
improving estimation time. Details of our exact estimation procedures
--- including a description of how we compute measures of uncertainty,
initialize all relevant parameters and latent variables, and sample
dyads to form the sub-network on which gradient estimates are based
--- are available in Section~\ref{app:model} of the Online
Supplementary Information.\footnote{To support claims about the
  ability of our model to recover meaningful quantities of interest,
  our Online SI (Section S.2 
) also contains results from simulations. These assess our estimation's accuracy in determining mixed-membership, its ability to capture network structure, the properties of our uncertainty approximation, and scalability across varied sample sizes.}

\subsection{Methodological contributions}

While the \texttt{biMMSBM} model is an extension of the unipartite
\texttt{MMSBM} \citep{airoldi2008} and its structural variant
\citep{oliv:prat:imai:22}, we believe it makes three methodological
contributions.  First, while the \texttt{MMSBM} is a popular modeling
framework for network data across disciplines, there is no version of
it that can be applied to bipartite network data, which are common in
political science. The model we propose can take full advantage of
information about both kinds of vertices involved in bipartite
networks, without the need to aggregate and ignore either. Second, by
avoiding projections that are common in practice, our model allows
researchers to avoid biased results (such artificially higher
clustering) related to either of the types of vertices under
study. Finally, and particularly by incorporating node-level
predictors, our model allows researchers to make predictions about
specific pairs of actors (e.g., which senators will support which
specific bills). Such granular predictions are not possible when
working with the projected network, which most prior models forced
researchers to do.

While bipartite versions of the popular \texttt{ERGM} model theoretically offer similar capabilities, practical deployment even in medium-sized data contexts can be challenging. To explore these
issues, we compare the fit of \texttt{biMMSBM} to that of a bipartite
\texttt{ERGM} fit using the \texttt{ergm} package in R on a subset of
our data (see Section S.3.2 
 of the Online SI). This is the only alternative implementation of a model designed to predict edges on bipartite networks that we are aware of. Our approach outperforms in classification accuracy and calibration of predicted probabilities, with simpler user input requirements. Unlike \texttt{biMMSBM}, which only requires the number of latent communities for both families as input, bipartite \texttt{ERGM} necessitates choosing from numerous sufficient statistic terms.

\section{Empirical Analysis of the 107th U.S. Senate}
\label{sec:empirical}

Before 2021, the Senate had only been perfectly split three other times --- with
the 107th being the most recent instance of this rare event in the
Senate's history. Despite this, the 107th Senate was not unusual in
terms of its productivity, passing about 17\% of the 3,242 pieces of
legislation introduced between 2000 and 2002 --- close to the average
of 22\% passage rate during the modern Senate --- and adopting major
legislation, including the Patriot Act and the so-called No Child Left
Behind bill.

In fact, such sustained productivity during times of narrow or
non-existent partisan majorities is not uncommon, with many major
bipartisan pieces of legislation in U.S. history passing under similar
circumstances --- including the legislation that made the interstate
highway system possible, the National Housing Act of 1954, and the
Civil Rights Act of 1957.  In all, there were a total of 20,660
instances of cosponsorships where a senator attached his/her name to a
piece of legislation during the 107th Senate, with about half of them
having at least one cosponsor from each party.

To explore the drivers of collaboration in cosponsorship, we use the
proposed \texttt{biMMSBM} model to better understand why this session
of the Senate remained legislatively active, avoiding the gridlock
that many associate with partisan divisions. Our model identifies the
important role played by junior, bipartisan power
brokers. Specifically, they were able to establish the common ground
by supporting low-stakes, non-biding resolutions and popular social
programs.  In addition, the model detects the important influences of
\emph{quid pro quo} cosponsorship behaviors and senate committee
experience on cosponsorship likelihood, confirming prior research on
the relevance of these features.

Moreover, as we show at the end of this section, fitting a unipartite
version of our model would make it impossible to identify these
pathways to collaboration. As we would expect given the descriptive
analysis in Section~\ref{sec:agg}, the unipartite network model
reveals little other than partisanship as the main driver of
coalitional politics, making it hard to understand how a perfectly
divided legislature was able to remain productive.

\subsection{Model specification and fit}

Our goal, then, is to understand the structural and contextual
features that made collaboration possible during the 107th Senate. A
rich literature on collaboration in Congress suggests that legislators
make cosponsorship decisions based on partisanship, seniority, and
personal political history
\citep{bratton_networks_2011,rippere_polarization_2016}. Accordingly,
we include senators' \textit{party affiliation}, \textit{ideology},
\textit{seniority} and \textit{gender} in our model as monadic
predictors for how senators sort into latent communities.

\citet{harward_calculus_2010} articulates that senators are more
likely to cosponsor bills when they share closer preferences with the
\emph{sponsor} of the bill, and when they are more connected to their
colleagues. To capture this, we model legislation groups as a function
of their corresponding sponsors' \textit{party}, \textit{ideology},
\textit{seniority} and \textit{gender} (we remove dyads where senators
are the sponsors of bills in question).

Lastly, senators tend to cosponsor bills that cover specific policy
domains \citep{harward_calculus_2010} and may choose to opt into
bipartisan cosponsorships based on topics covered in the legislative
bill \citep{harbridge_is_2015}. This inclination cannot be modeled in
a senator-only unipartite network, but can be directly accounted for
when modeling the bipartite structure.  We address this possibility by
including the \textit{substantive topic} as a bill-level
covariate.\footnote{We also consider two notable alternative specifications:
  first, we consider a specification in which no additional predictors are included (i.e., the classic mixed-membership stochastic blockmodel). Second, we fit a model that also includes a dyadic indicator of shared state between a senator cosponsor and the sponsor of a bill. We present the results of both exercises in Section S.3.6 
   of the Online SI.}

To capture the described shifts in the \textit{temporal context} in
which bills are introduced, we also include a bill-level (i.e.,
monadic) covariate indicating whether a bill was presented in the
first phase of the Congress (prior to Jeffords leaving the Republican
party), in the second phase (post Jeffords leaving and prior
to 9/11) or in the third phase (after 9/11). The temporal context under which a
piece of legislation is introduced is an additional example of
information that would inevitably be discarded in a projected
unipartite analysis of the bipartite network \citep{kirkland:gross:2014}.

As we indicated earlier, we also pay close attention to two additional
forces that can be expected to affect the likelihood of
cosponsorship. First, we aim to capture \textit{reciprocity}
behaviors, or favor-trading on the Senate floor
\citep{brandenberger_trading_2018, harbridge:etal:2023}.  The model
therefore includes a dyadic predictor measuring, for each senator-bill
dyad, the (log) proportion of times the bill's sponsor in turn
cosponsored proposals introduced by the corresponding senator in the
previous Congress. As this proportion of reciprocity is heavily skewed
and contains a number of zeros, we use the log transformation of
non-zero values and an indicator variable for the cases of
zeros.

Second, our dyadic model includes the \textit{number of
  committees} shared by a senator and a piece of legislation. A greater number of
shared committees indicates a higher chance that the senator has
overseen the development of a bill and holds relevant substantive
expertise.  While the roles of committees have been studied previously
\citep{porter_network_2005,cirone_cabinets_2018}, our analysis
directly examines how overlap in committees between legislator and
legislation relates to cosponsorship.  Relatedly,
\citet{gross_kirkland2019} find evidence of strong predictive power of
shared committee \emph{leadership} among the subset of ranking
legislators when exploring cosponsorship decisions.

With predictors at the monadic and dyadic levels in place, we
determine the number of latent groups for senators and bills. To do
this, we first randomly select 25\% of data as a test set, and
compare models with a range of possible latent group-size pairings
through the area under the ROC curve (AUROC) values for the
out-of-sample edges. We then
select the group sizes offering the best fit according to this
criterion, resulting in 3 groups each for legislators and bills, i.e.,
$K_1=K_2=3$.

\begin{figure}[h]
  \centering \spacingset{1}
  \includegraphics[width=\linewidth]{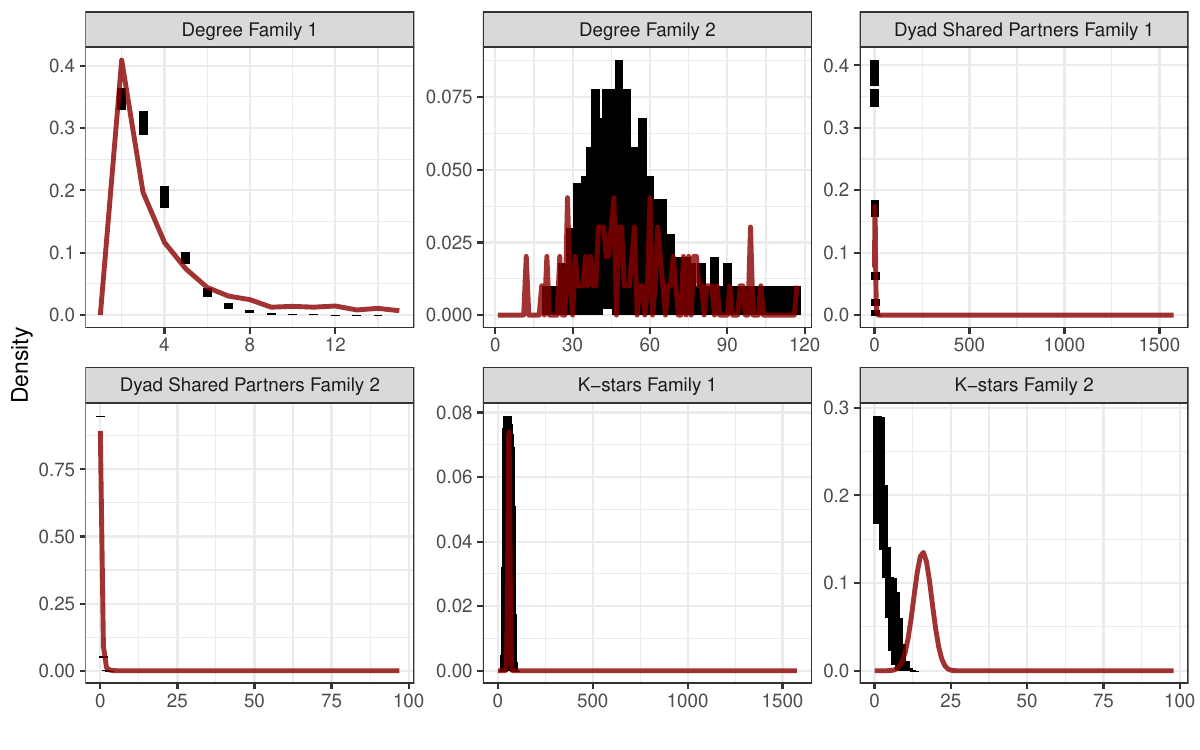}
  \vspace{-.35in}
  \caption{\textbf{Posterior predictive goodness-of-fit checks, out-of-sample}. Vertical black rectangles represent the interquantile range across 50 replicate networks. The red line in each panel denotes the observed value in a network formed by a random 25\% sample of cosponsorship decisions during the 107th Senate. The model generally replicates structural features well, shown by overlap between black bars and red lines. However, $k$-stars of bills are consistently underpredicted in the out-of-sample set.}
  \label{fig:gof-107}
\end{figure}

We assess model performance by generating 50 replicates of the
out-of-sample network from our model's posterior predictive
distribution. We compare network-level statistics (e.g., degree
distributions of senators and bills, family-specific distributions of
edge-shared partners and $k$-stars) between these replicates and our
test set. Figure~\ref{fig:gof-107} displays the results, with black
bars showing the interquartile range across replicates and red lines
indicating observed values in the test set. The model generally fits
the out-of-sample network well, with most red lines falling within
their corresponding black bars, capturing structural features without
explicit specification. However, it tends to underpredict the
distribution of $k$-stars among bills. The model also demonstrates
high predictive accuracy and calibration, predicting observed edges
with higher probabilities 70\% of the time, as shown in
Figure S.8 
 in the Online SI.

Having established that the model generally fits the data well even out of
sample, we obtain the estimates of all parameters and hyper-parameters in
Equation~\eqref{eq:joint} for this $K_1=K_2=3$ model fitted to the entire
bipartite cosponsorship network in the 107th Senate. More
specifically, we compute
various quantities of interest in the form of predicted probabilities
of block interactions and block memberships. As our discussion hinges
on these derived quantities, we present all estimated values in
Tables S.4 and S.5 
 in the Online SI.

\subsection{Pathways to legislative collaboration}
\label{sec:empirics}

\begin{figure}[t!]
  \centering \spacingset{1}
  \includegraphics[width=\textwidth, trim={0 3cm 0 2.5cm},clip]{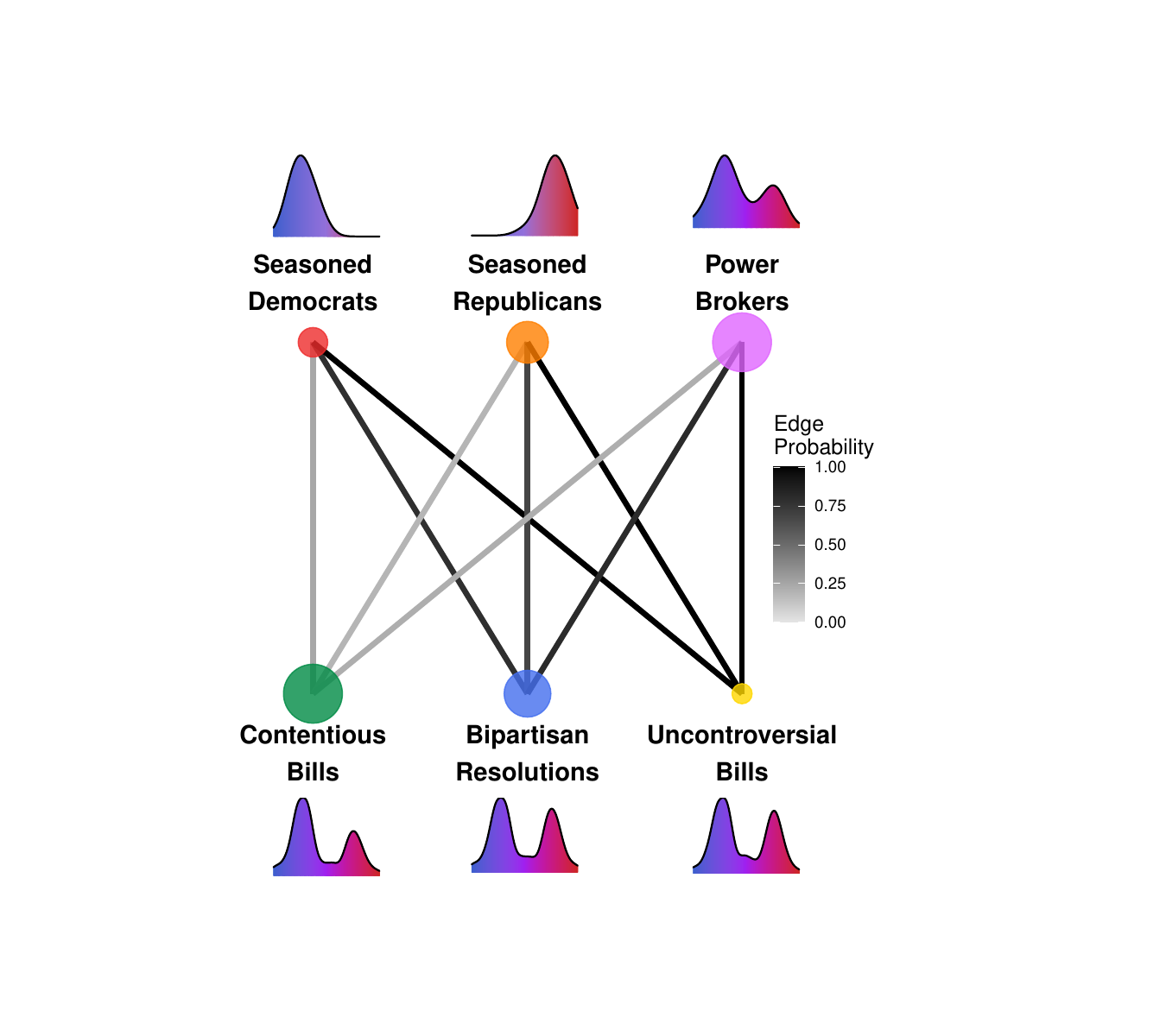}
  \vspace{-.25in}
  \caption{{\bf Blockmodel of senator and legislation latent group
      connection probabilities.} Block size is proportional to the
    number of the nodes expected to instantiate the corresponding
    latent group, and connections between them are shaded denoting cosponsorship probabilities between group members (darker shades indicate higher connection likelihoods). Senator groups tend to engage more with a 'Uncontroversial' legislation group but less with a larger 'Contentious' one.  Next to each block, we also present a density of ideological positions of
    member senators (top row) and bill sponsors (bottom row),
    revealing that while ideology can help distinguish across types of
    senator coalitions, it cannot discriminate across relevant types
    of legislation.}
  \label{fig:block}
\end{figure}

What kinds of coalitions are at play when it comes to making
cosponsorship decisions, and how do these coalitions interact when
considering different types of legislation?  Figure~\ref{fig:block}
presents the estimated blockmodel for the 107th Senate, illustrating the likelihood of cosponsorship ties between senator and legislation latent groups. Edge shading reflects cosponsorship probabilities, and node sizes indicate group instantiation frequencies. Ideological distributions of senator and bill groups are also presented. The density for
each senator group represents the distribution of ideal points of its
members while the density for a bill group is that of its members'
sponsors. 

\begin{figure}[t!]
  \centering \spacingset{1} 
  \includegraphics[scale=0.7, trim={0 2cm 0 3cm}, clip]{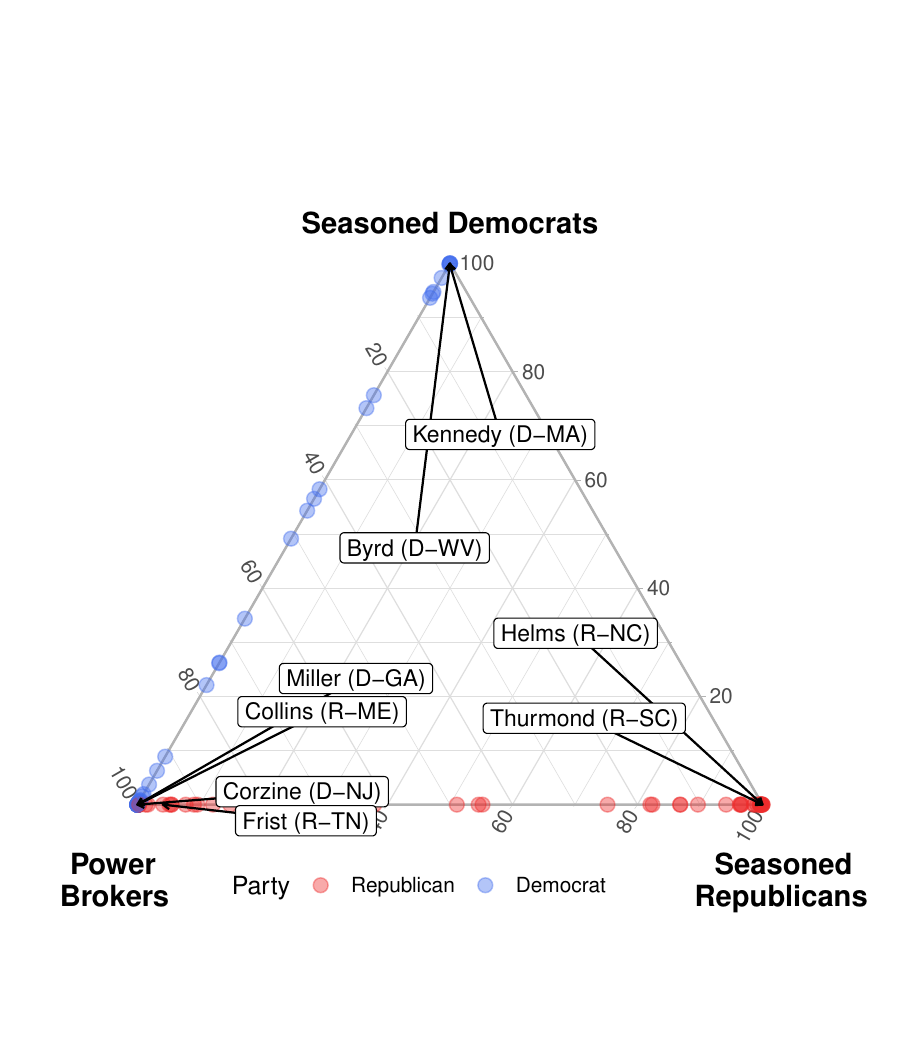}
  \caption{{\bf Ternary plot of senator latent group membership
      probabilities.} For clarity of presentation, example senators
    are colored by party. Senators in group~1 (top corner) are more
    likely to be Democrats, while senators in group~2 (right corner)
    are more likely to be Republicans; Group~3 (left corner) senators
    hail from both sides of the aisle and are likely to be junior and
    involved in cross-partisan bill sharing.  }\label{fig:senategroup}
\end{figure}

As expected, there are estimated senator groups (depicted as the top
row of circles in Figure~\ref{fig:block}) that align well with party
memberships. Members of these partisan coalitions include seasoned
Republicans (such as Strom Thurmond (R-SC) and Jesse Helms (R-NC)),
and seasoned Democrats (like Robert Byrd (D-WV) and Edward Kennedy
(D-MA)).

In addition, however, our model identifies a substantial group of
senators (depicted in purple) who stand out as having different
cosponsorship patterns than their more partisan
counterparts. Exemplars of this group, whom we call the \emph{junior
  power brokers}, include Jon Corzine (D-NJ), Tom Carper (D-DE), Susan
Collins (R-ME), Bill Frist (R-TN), Zell Miller (D-GA), and Hillary
Clinton (D-NY) --- all junior Senators at the
time. Figure~\ref{fig:senategroup} presents the estimated mixed
memberships of all Senators (i.e., their probability of acting as part
of any of the discovered latent groups), highlighting a few of the
most notable legislators of the session. The top 10 members of each
senator latent group by mixed membership probability are presented in
Online SI Table S.2. 

\begin{figure}[t!]
  \centering \spacingset{1}
\includegraphics[width=\textwidth]{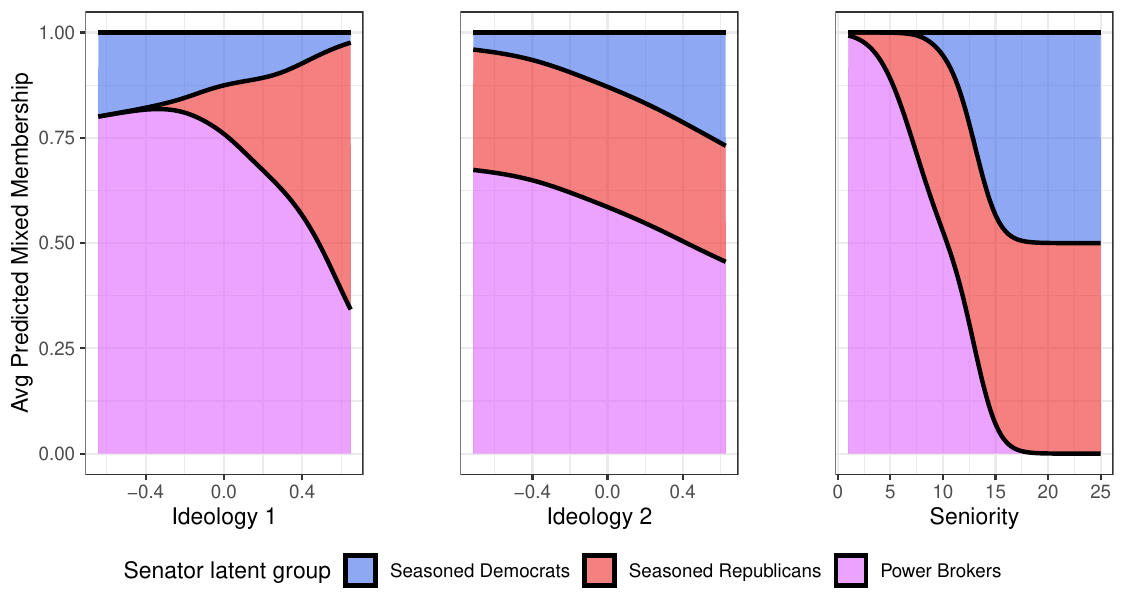}
\caption{{\bf Predicted Mixed Memberships of senator predictors.}  The
  $y$-axes plot average predicted mixed memberships across the three
  possible senator latent groups, given each shift in value of a
  senator predictor in the $x$-axes; for instance at low values of
  Ideology (dimension)~1, the average predicted memberships for being
  in group~1 (Seasoned Democrats) and group~3 (power-brokers) are
  highest; as Ideology~1 values increase (corresponding to increase in
  the conservative direction), average predicted group~2 membership
  (Seasoned Republicans) increases and supplants group 1 entirely.}
\label{fig:covMMsenator}
\end{figure}

Indeed, this third bipartisan group is likely to be formed by senators
who have little experience in the Senate coming from all over the
ideological spectrum, as evidenced by the distribution of ideological
positions depicted over the corresponding group in
Figure\ref{fig:block}.\footnote{Plotted quantities are obtained by
  computing
  $\expec[\text{SoftMax}(\bm{x}_p^\top\hat{\boldsymbol{\beta}}_{1g})]$,
  where the expectation is taken over the observed values of all but a
  focal variable (e.g., ideology), and the
  $\hat{\boldsymbol{\beta}}_{1g}$ are estimated monadic
  coefficients. Full table of estimates of monadic coefficients is in
  Online SI Table S.5.}
 We explore this more directly
in the left-most panel of Figure~\ref{fig:covMMsenator}, which shows
how the probability of group membership changes as a function of
ideology.  Although the group of junior power brokers (depicted in
purple) is primarily predicted to be comprised of left-leaners on the
first DW-Nominate dimension, positions along the second ideological
dimension (seen in the central panel of Figure~\ref{fig:covMMsenator})
--- often interpreted as capturing cross-cutting salient issues of the
day \citep{poole2017ideology} --- help distinguish this group of
senators from their staunch Democratic counterparts.

Many of these junior power brokers later ascended to leadership roles within their parties. For instance, during the latter part of the 107th Congress, the Senate Republican Conference underwent leadership changes. Trent Lott, facing criticism for racially insensitive remarks, resigned and was swiftly replaced by Bill Frist (R-TN), a prominent healthcare advocate and a top member of the junior power brokers identified by our model.

Similarly, many of them were pivotal ``last'' votes in large
contentious bills that required just an extra nudge or two for
passage. For example, a major bill in the 107th was the \emph{Farm
  Bill}, designed to repeal the \emph{Freedom to Farm Act} of
1996. While politics over agriculture had historically been regional
rather than ideological, the \emph{Freedom to Farm Act} was a
significant deviation from that norm. Veteran senators Tom Daschle
(D-SD) and Agriculture Committee Chairman Tom Harkin (D-IA)
collaborated to bring the bill together, and a series of negotiations
began to bring the necessary senatorial support on board --- including
support for small dairy farmers included in the bill. In the end, the
largely Democratic set of supporters was complemented by key support
from relatively junior Republicans Susan Collins (R-ME) and Jeff
Sessions (R-AL) --- again identified by our model as likely members of
the power brokers group. This role as brokers is further supported by
analyses of the betweeness centrality of Senators who are likely to
instantiate this group, which tends to be higher than that of Senators
whose membership in the other two groups is more likely (see
Table S.6 
 in the Online SI).

In addition to offering a nuanced view of the collaboration strategies
of senators, the model is able to identify the types of legislation
which these groups of senators are likely to cosponsor. In this case,
the model uncovers three broad classes of bills and resolutions
(depicted in the bottom row of circles in Figure~\ref{fig:block}), and
the corresponding probabilities that members of any of the three
senator groups will support them through cosponsorship
decisions. Critically, while ideology plays an important role in
defining the latent senator groups that structure cosponsorship (with
right-skewed, left-skewed, and bimodal distributions characterizing
membership into the three groups at the top of
Figure~\ref{fig:block}), no such differences in ideology of sponsors
can help distinguish across the groups of legislation uncovered by our
model (as indicated by the similarly bimodal densities of sponsor
ideology across all three groups in the bottom of
Figure~\ref{fig:block}). 

We next show that investigating this nuance in bill composition can
help us understand how cosponsorship collaborations
took place during this nominally partisan session of Congress.

\subsection{Legislation types that facilitate cosponsorship}

The largest type of legislation uncovered by our model is also the
least likely to be supported by members of any senator group,
suggesting that the bulk of legislation introduced in the Senate
received little support from Senators other than the original
sponsor. This latent class of bills, which we labeled ``Contentious
Bills'' in Figure~\ref{fig:block}, consists of high-stakes bills on
controversial economic issues and social programs, including those
 that handle the allocation of public funds for such programs.
For example, the \emph{Senior Self-Sufficiency Act} (SN 107 2842), \emph{Bioterrorism Awareness Act} (SN 107 1548), and the
\emph{Nationwide Health Tracking Act of 2002} (SN 107 2054) belong to
this group.  Online SI Table S.3 
 presents details of legislation with the top ten mixed membership probabilities
in each of the three latent groups.

\begin{figure}[t!]
  \spacingset{1}
  \centering \vspace{-0.5in}
\includegraphics[width=\textwidth]{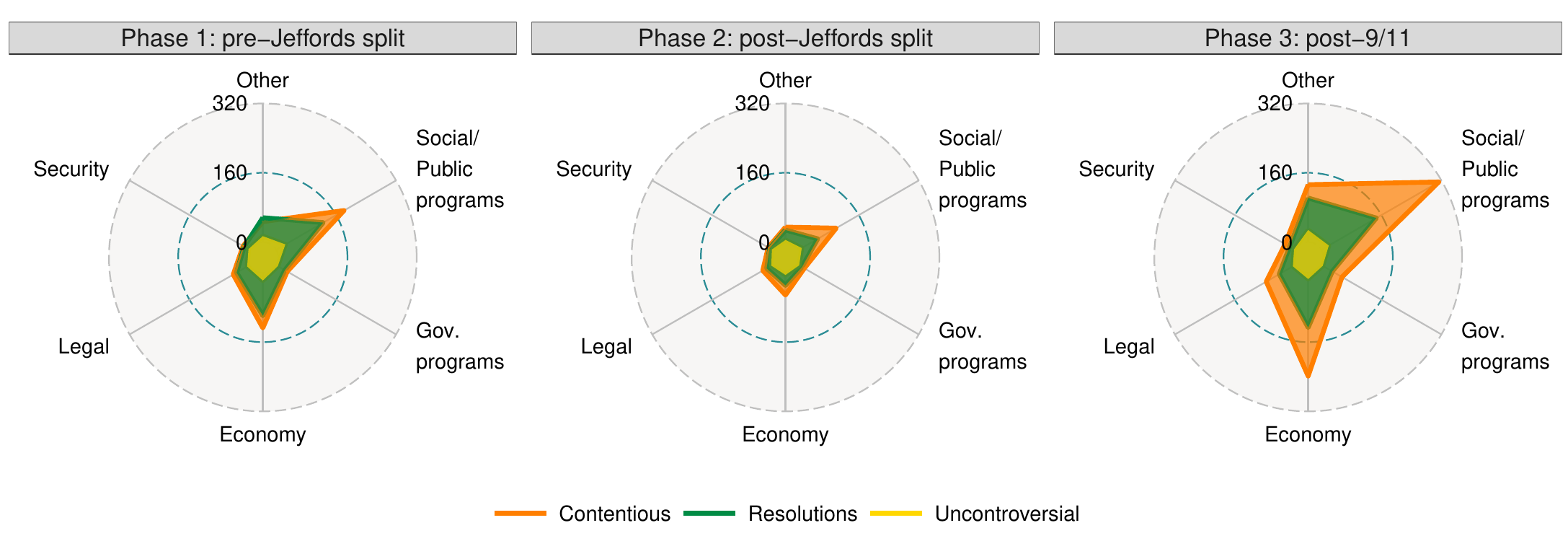}
\caption{{\bf Radar graphs of predicted legislation by topic within
    each phase of Congress, by bill latent group.} Panels are phases~1
  (pre-Jeffords split),~2 (post-Jeffords split), and~3 (post 9/11) in
  the Congress, from left to right. Each radar plot includes bill
  topics as poles, with estimated number of bills in the topic plotted
  against each pole, by latent group. Phase~2 produces the fewest
  pieces of legislation, while Phase~3 produces the most. Over time,
  the predicted number of bills in the ``Contentious Bills'' group
  (orange polygon) increases, especially in domains related to social
  public programs and the economy. The number of bills in the
  ``Bipartisan Resolutions'' group grew more slowly than that in the
  ``Contentious Bills'' block (green polygon), but has similarly
  favored social/public programs and the economy. Finally, the number
  of bills in the ``Popular \& Uncontroversial'' (yellow polygon)
  changed the least throughout the session. }
  \label{fig:billgroup}
\end{figure}

The size of the ``Contentious Bills'' group grew during the third and
last phase of the 107th Senate, after the 9/11 attacks. This is easily
seen in Figure~\ref{fig:billgroup}, which presents radar plots of
predicted legislation memberships by phase of the Congress (panels
from left to right present bills from the pre-Jeffords' split phase,
post-Jeffords' split second phase, and post 9/11 phase). Each radar
graph positions the six observed substantive topics along spokes of a
wheel, and plots the predicted number of bills on that topic as a
point along the corresponding spoke: the farther away from the wheel
center, the more bills are predicted to be on that topic. Doing this
for each of the three latent groups results in the three shaded
polygons presented in each panel of the figure.\footnote{The vertices
  of each polygon are obtained by summing each latent group's
  estimated mixed membership proportions for a given topic in a single
  phase --- a way to think of \textit{bills in each group allocated
    towards each topic} --- and plotting these against each topic
  pole's total number of bills.} The dominance of bills in the
``Contentious Bills'' group in the third phase, depicted in orange, is
readily apparent.

How, then, were the different senator groups able to find common
ground in order to avoid stalemate? Clues can be gleaned in the
definition of the other two latent bill groups uncovered by our model
--- the groups we have labeled ``Bipartisan Resolutions'' and
``Uncontroversial Bills'' in both Figures~\ref{fig:block}
and~\ref{fig:billgroup}. The former is particularly illuminating: its
topical composition almost mirrors that of the ``Contentious Bills''
(i.e., it is comprised of pieces of legislation that deal with
controversial public social programs and economic issues, as indicated
by the similarly-proportioned shapes of green and orange polygons in
Figure~\ref{fig:billgroup}), but it is likely to be comprised of
\emph{resolutions} (primarily concurrent and simple), rather than bills --- thus offering lower-stakes
opportunities for presenting bipartisan positions that do not result
in codified law.  Such resolutions, which exclude continuing
resolutions and are often sponsored by
more senior senators, offer opportunities to build bridges across
partisan divides without incurring the costs associated with creating
laws. Table S.3 
 in the Online SI contains the top pieces of legislation in the group.

In turn, legislation in the comparatively smaller ``Uncontroversial''
group (shown in yellow in Figures~\ref{fig:block}
and~\ref{fig:billgroup}) also draws consistent cosponsorship support
from all senator groups, as pieces in it tend to be either
uncontroversial resolutions or bills on popular social programs. As a
result, such legislation is widely cosponsored by Republicans and
Democrats alike. For instance, the Senate joint resolution over the
Sept. 11 attacks (SJ 107 22) has the highest mixed membership
probability in this group, followed closely by bills such as the
\emph{Railroad Retirement and Survivors' Improvement Act of 2001} (SN
107 697).  These legislations form a small but steady core of that
supplements low-stakes efforts (such as those in the ``Bipartisan
Resolutions'' block), that can nevertheless result in substantial
legislation, such as the \emph{Family Opportunity Act of 2002} (SN 107
321).
 
The importance of this meaningful cooperation mechanism revealed by
the blockmodel is particularly notable, as the model was able to
identify it net of two important drivers of cosponsorship likelihood:
\emph{quid pro quo} behaviors (represented by dyadic predictors),
measured as the coefficient on the (log) proportion of ``reciprocity''
(\textit{Log Reciprocity}), and the shared committee experience of a
given senator-bill dyad (\textit{Shared Committee}). For the former,
and given the log transformation, our model suggests that a 1\%
increase in the reciprocity (i.e., the proportion of times the sponsor
of a piece of legislation acted as a cosponsor for a given senator’s
bill in the previous Congress) is associated with a roughly 2\% increase in
the odds of cosponsorship.  In the case of the latter, we find that
sharing a committee is significantly and positively associated with
collaboration, making cosponsorship about 3 times more likely. These
results, which are fully explored in Online SI Table S.4,
 are consistent with previous research on the
determinants of legislative collaboration --- lending further credence
to our general findings.

In sum, Senators appear to have leveraged a mix of low-stakes
resolutions over potentially contentious issues and a small but
important set of bills for which there was bipartisan support.  This
enabled them to build cross-partisan bridges and keep the 107th term
from devolving into stalemate.  Our model identified these novel
patterns of cooperation after accounting for other, more traditional
forces that have been found to drive collaboration and
cosponsorship. Our application offers clues not only about to which
kinds of legislators are best positioned to do so, but about which
kinds of \emph{legislation} make such collaborations possible. These
and similar insights would be lost when analyses aggregate over bills
and their characteristics, as we now demonstrate.

\subsection{Comparison with the unipartite network model}

We conclude our empirical section by comparing the results of our
model against those of a unipartite network model.  For direct
comparison, we use a unipartite (and static) version of our model, known as
\texttt{dynMMSBM} \citep{oliv:prat:imai:22}.  As discussed in
Section~\ref{sec:agg}, we project the bipartite network data
onto a unipartite weighted network, in which the weight of edges
between senators is given by the number of bills they cosponsor
together.\footnote{\texttt{dynMMSBM} can accommodate such weighted networks by
using a binomial likelihood, where the weights are modeled as
``successes'' among a number of trials equal to the number of bills
two senators could have cosponsored together.}   

For fair comparison, we keep the model specification as similar to the
one used in our bipartite network analysis as possible: three latent groups of
senators, and the same set of senator-specific covariates. In
addition, we use the estimated senator mixed-membership vectors from
\texttt{biMMSBM} as the initial values of the corresponding vectors in
the unipartite model. This increases our confidence that any
difference in the learned grouping of legislators is a result of
the data aggregation process, rather than the differences of model
specification and estimation.

\begin{figure}[t]
  \centering \spacingset{1}
  \begin{subfigure}[b]{0.48\textwidth}
         \centering
         \includegraphics[width=0.98\textwidth, trim = 80 90 80 60, clip]{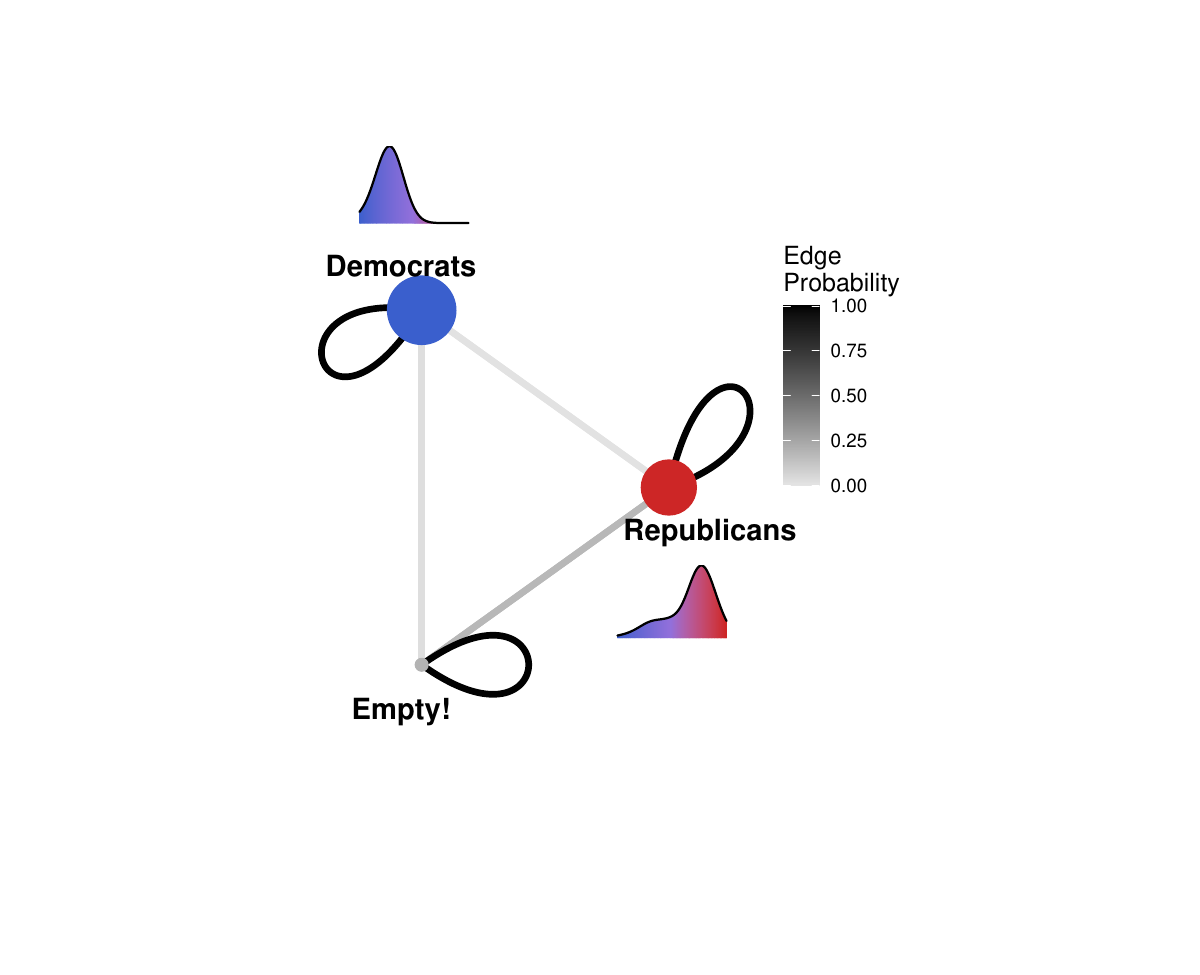}
         \caption{Unipartite Senator Groups Blockmodel}
         \label{fig:uni_block}
     \end{subfigure}
     \hfill
     \begin{subfigure}[b]{0.48\textwidth}
         \centering
         \includegraphics[width=0.98\textwidth]{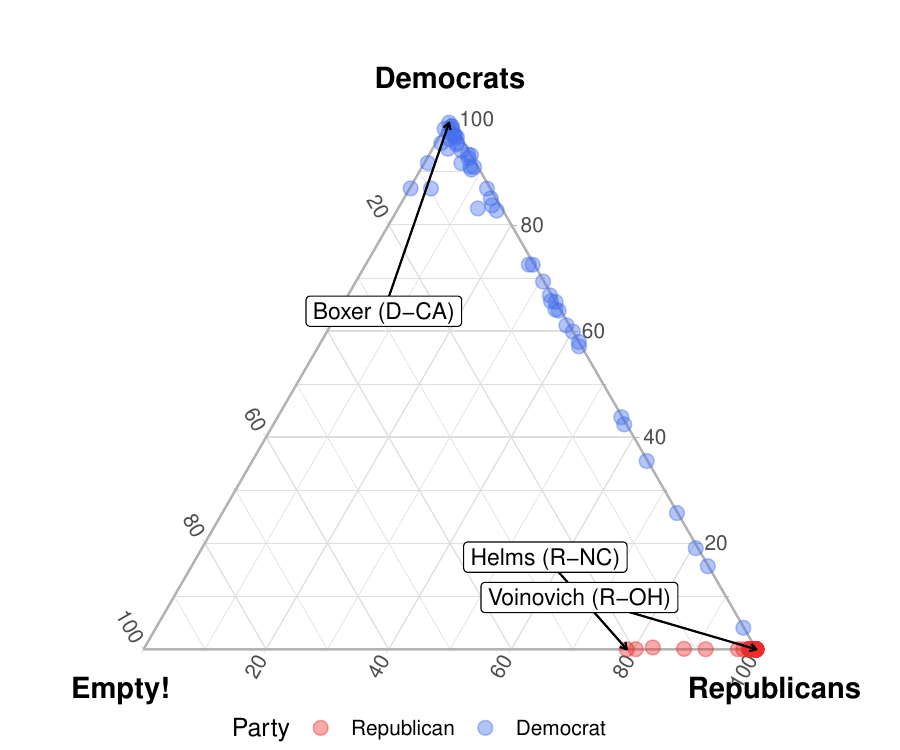}
         \caption{Unipartite Senator Mixed-Memberships}
         \label{fig:uni_tern}
     \end{subfigure}
     \caption{{\bf Results from fitting a \texttt{MMSBM} to the
         projected cosponsorship network.} After fitting a unipartite
       \texttt{MMSBM} to the projected cosponsorship network in the
       107th Senate, we recover a blockmodel of group interaction
       probabilities (left panel) and a set of group membership
       probabilities for Senators (right panel). As expected, the
       model produces groups that are mostly aligned with partisanship
       and ideology, as indicated by the ideological distributions of
       likely members depicted next to each estimated block. We also
       find evidence of high intra-group collaboration probabilities
       (especially among Democrats), and very low probabilities of
       connecting across partisan groups, painting a picture of a
       highly polarized session.}
  \label{fig:uni_res}
\end{figure}

Figure~\ref{fig:uni_res} presents the main quantities of
interest from this comparison: the estimated blockmodel
between the three discovered groups of senators in the left panel (i.e., the propensity
of members of any of these groups to collaborate with a member of another latent group) and the
composition of these groups in the right panel (i.e., the extent to which different
senators are likely to be in any of the three groups). 

In contrast to the results shown in Figures~\ref{fig:block}~and~\ref{fig:senategroup}, the left and right panels of
Figure~\ref{fig:uni_res} offer a picture of a polarized, relatively
non-cooperative session of Congress, divided clearly along partisan
lines --- exactly as we expected from our discussion of the issues
brought about by the aggregation process involved in projecting a
bipartite network onto a unipartite one (see Section~\ref{sec:agg}).

While the unipartite model could accommodate three different groups of
senators under its specification,  the data  support membership  primarily into  two of  those
latent blocks, leaving the third group essentially empty. Furthermore,
membership into these two blocks is strongly aligned with partisanship, as the
bulk of  Democratic senators  (depicted as blue  circles on  the right
panel of  Figure~\ref{fig:uni_res}) concentrate  on one vertex  of the
ternary plot (with  liberal Democrat Barbara Boxer  being estimated as
the senator most  likely to instantiate this latent  group), while the
majority  of Republican  senators (depicted  as red  circles) tend  to
concentrate  on another.  Those senators  who tend  to have  a minimal
likelihood  of   instantiating  the  ``Empty!''   block   tend  to  be
Republican ideologues, led by the notoriously conservative Jessee Helms
(R-NC).

Moreover, the unipartite model identifies only a moderate amount of
collaboration across the aisle, yielding the estimated probability of
about .18 that a member of the latent ``Democratic'' group is
connected in the projected unipartite network to a menber of the
``Republican'' group.  In contrast, the probability of a connection
between two members of the \emph{same} partisan latent group is
estimated to be about .99, as indicated by the darker-shaded loops in
the blockmodel on the left of Figure~\ref{fig:uni_res}.  While
Democrats enjoyed a slight majority after senator Jeffords decided to
leave the Republican party (suggesting that the cohesive majority
managed to move legislation along), this picture of polarization
painted by the unipartite model gives few clues as to why the 107th
Senate was able to remain as productive as it was.

In sum, relying on the projected network of cosponsorships not only
misses the rich and nuanced information about legislative
collaboration that individual bills have to offer, but it also
distorts what we can learn about the nature of legislative
coalitions. Using a unipartite model to study naturally bipartite data
results in an artificially inflated sense of clustering and group
cohesion --- a risk that becomes even more pressing when we are
interested in group dynamics and polarization that drive processes of
network formation.

\section{Conclusion}
\label{sec:conclude}

We have developed a new approach to modeling bipartite networks that
allows researchers to incorporate more complete information available
to them rather than projecting these networks onto unipartite
networks, leading to biased inference.  As bipartite networks are
quite common in the social sciences, we see natural applications in a
number of different domains. For example, our model could be used to
explore questions relating to country-trade product networks, state
memberships in organizations, posts on social media platforms and
hashtags, or product recommendation systems.

In the future, and given the prevalence of bipartite networks observed
over time, fruitful extensions of our proposed approach would allow
researchers to incorporate dynamics into the generative model of
bipartite network formation\if0\blind \citep[for an extension
incorporating dynamics in the unipartite case,
see][]{oliv:prat:imai:22}\fi.  In addition, we could explore larger
multi-mode networks, integrating entities like lobbying firms into
cosponsorship networks or examining relationships among NGOs, IGOs,
and countries internationally. These networks allow for co-clustering
of diverse actors sharing indirect connections, necessitating improved
tools for studying relational data beyond traditional single-mode
representations and avoiding aggregation bias.

\clearpage

\pdfbookmark[1]{References}{References}
\spacingset{1.6}
\printbibliography 

@article{oliv:prat:imai:22,
  author = 	 {Olivella, Santiago and Pratt, Tyler and Imai, Kosuke},
  title = 	 {Dynamic Stochastic Blockmodel Regression for Social Networks: Application to International Conflicts},
  journal =  {Journal of the American Statistical Association},
  year = 	 {2022},
  OPTkey = 	 {},
  pages = 	 {1068--1081},
  volume =       {117},
  number =       {539},
  OPTnumber = 	 {},
  OPTaddress = 	 {},
  OPTmonth = 	 {},
  OPTnote = 	 {},
  OPTannote = 	 {}
}

@Manual{handcock:etal:2023,
  author = {Mark S. Handcock and David R. Hunter and Carter T. Butts and Steven M. Goodreau and Pavel N. Krivitsky and Martina Morris},
  title = {ergm: Fit, Simulate and Diagnose Exponential-Family Models for Networks},
  organization = {The Statnet Project (\url{https://statnet.org})},
  year = {2023},
  note = {R package version 4.6.0},
  url = {https://CRAN.R-project.org/package=ergm},
}

@Article{hunter:etal:2008,
  title = {{ergm}: A Package to Fit, Simulate and Diagnose Exponential-Family Models for Networks},
  author = {David R. Hunter and Mark S. Handcock and Carter T. Butts and Steven M. Goodreau and Martina Morris},
  journal = {Journal of Statistical Software},
  year = {2008},
  volume = {24},
  number = {3},
  pages = {1--29},
  doi = {10.18637/jss.v024.i03},
}

@Article{krivitsky:etal:2023,
  title = {{ergm} 4: New Features for Analyzing Exponential-Family Random Graph Models},
  author = {Pavel N. Krivitsky and David R. Hunter and Martina Morris and Chad Klumb},
  journal = {Journal of Statistical Software},
  year = {2023},
  volume = {105},
  number = {6},
  pages = {1--44},
  doi = {10.18637/jss.v105.i06},
}

@article{blei:etal:2017,
  title={Variational inference: A review for statisticians},
  author={Blei, David M and Kucukelbir, Alp and McAuliffe, Jon D},
  journal={Journal of the American statistical Association},
  volume={112},
  number={518},
  pages={859--877},
  year={2017},
  publisher={Taylor \& Francis}
}

@article{kirkland:gross:2014,
  title={Measurement and theory in legislative networks: The evolving topology of Congressional collaboration},
  author={Kirkland, Justin H and Gross, Justin H},
  journal={Social networks},
  volume={36},
  pages={97--109},
  year={2014},
  publisher={Elsevier}
}

@article{newman:etal:2001,
  title={Random graphs with arbitrary degree distributions and their applications},
  author={Newman, Mark EJ and Strogatz, Steven H and Watts, Duncan J},
  journal={Physical review E},
  volume={64},
  number={2},
  pages={026118},
  year={2001},
  publisher={APS}
}

@article{guillaume:latapy:2004,
  title={Bipartite structure of all complex networks},
  author={Guillaume, Jean-Loup and Latapy, Matthieu},
  journal={Information processing letters},
  volume={90},
  number={5},
  pages={215--221},
  year={2004},
  publisher={Elsevier}
}

@article{hunter2008,
  title={Goodness of fit of social network models},
  author={Hunter, David R and Goodreau, Steven M and Handcock, Mark S},
  journal={Journal of the american statistical association},
  volume={103},
  number={481},
  pages={248--258},
  year={2008},
  publisher={Taylor \& Francis}
}

@article{hoffman_etal2013,
  title={Stochastic variational inference},
  author={Hoffman, Matthew D and Blei, David M and Wang, Chong and Paisley, John},
  journal={The Journal of Machine Learning Research},
  volume={14},
  number={1},
  pages={1303--1347},
  year={2013},
  publisher={JMLR. org}
}

@article{gross_kirkland2019,
  title={Rivals or Allies? A Multilevel Analysis of Cosponsorship within State Delegations in the US Senate},
  author={Gross, Justin H. and Justin Kirkland},
  journal={Congress \& the Presidency},
  volume={46},
  number={2},
  pages={183--213},
  year={2019},
  publisher={Taylor \& Francis}
}

@article{agneessens2004,
  title={Choices of theatre events: p{$\ast$} models for affiliation networks with attributes},
  author={Agneessens, Filip and Roose, Henk and Waege, Hans},
  journal={Metodoloski zvezki},
  volume={1},
  number={2},
  pages={419},
  year={2004},
  publisher={Anuska Ferligoj}
}

@article{karrer2011,
  title={Stochastic blockmodels and community structure in networks},
  author={Karrer, Brian and Newman, Mark EJ},
  journal={Physical review E},
  volume={83},
  number={1},
  pages={016107},
  year={2011},
  publisher={APS}
}

@article{minhas2019,
  title={Inferential approaches for network analysis: Amen for latent factor models},
  author={Minhas, Shahryar and Hoff, Peter D and Ward, Michael D},
  journal={Political Analysis},
  volume={27},
  number={2},
  pages={208--222},
  year={2019},
  publisher={Cambridge University Press}
}

@article{desmarais2012,
  title={Statistical inference for valued-edge networks: The generalized exponential random graph model},
  author={Desmarais, Bruce A and Cranmer, Skyler J},
  journal={PloS one},
  volume={7},
  number={1},
  pages={e30136},
  year={2012},
  publisher={Public Library of Science San Francisco, USA}
}

@article{platt:1999,
  title={Probabilistic outputs for support vector machines and comparisons to regularized likelihood methods},
  author={Platt, John and others},
  journal={Advances in large margin classifiers},
  volume={10},
  number={3},
  pages={61--74},
  year={1999},
  publisher={Cambridge, MA}
}

@article{wang2009,
  title={Exponential random graph (p$\ast$) models for affiliation networks},
  author={Wang, Peng and Sharpe, Ken and Robins, Garry L and Pattison, Philippa E},
  journal={Social Networks},
  volume={31},
  number={1},
  pages={12--25},
  year={2009},
  publisher={Elsevier}
}

@article{wang2013,
  title={Exponential random graph model specifications for bipartite networks—A dependence hierarchy},
  author={Wang, Peng and Pattison, Philippa and Robins, Garry},
  journal={Social networks},
  volume={35},
  number={2},
  pages={211--222},
  year={2013},
  publisher={Elsevier}
}

@article{skvoretz1999,
  title={Logit models for affiliation networks},
  author={Skvoretz, John and Faust, Katherine},
  journal={Sociological Methodology},
  volume={29},
  number={1},
  pages={253--280},
  year={1999},
  publisher={Wiley Online Library}
}

@article{airoldi2008,
  title={Mixed membership stochastic blockmodels},
  author={Airoldi, Edoardo Maria and Blei, David M and Fienberg, Stephen E and Xing, Eric P},
  journal={Journal of machine learning research},
  year={2008}
}

@inproceedings{foulds_etal2013,
  title={Stochastic collapsed variational Bayesian inference for latent Dirichlet allocation},
  author={Foulds, James and Boyles, Levi and DuBois, Christopher and Smyth, Padhraic and Welling, Max},
  booktitle={Proceedings of the 19th ACM SIGKDD international conference on Knowledge discovery and data mining},
  pages={446--454},
  year={2013}
}

@inproceedings{dulac_etal2020,
  title={Mixed-Membership Stochastic Block Models for Weighted Networks},
  author={Dulac, Adrien and Gaussier, Eric and Largeron, Christine},
  booktitle={Conference on Uncertainty in Artificial Intelligence},
  pages={679--688},
  year={2020},
  organization={PMLR}
}

@Manual{olivella_etal_2021,
  title = {NetMix: Dynamic Mixed-Membership Network Regression Model},
  author = {Olivella, Santiago and Lo, Adeline and Pratt, Tyler and Imai, Kosuke},
  year = {2021},
  url = {https://CRAN.R-project.org/package=NetMix},
  note = {R package version 0.2.0.9013}
}

@article{gopalan_blei2013,
  title={Efficient discovery of overlapping communities in massive networks},
  author={Gopalan, Prem K and Blei, David M},
  journal={Proceedings of the National Academy of Sciences},
  volume={110},
  number={36},
  pages={14534--14539},
  year={2013},
  publisher={National Acad Sciences}
}

@article{frank_strauss1986,
  title={Markov graphs},
  author={Frank, Ove and Strauss, David},
  journal={Journal of the american Statistical association},
  volume={81},
  number={395},
  pages={832--842},
  year={1986},
  publisher={Taylor \& Francis}
}

@article{tam_cho_legislative_2010,
	title = {Legislative {Success} in a {Small} {World}: {Social} {Network} {Analysis} and the {Dynamics} of {Congressional} {Legislation}},
	volume = {72},
	issn = {0022-3816},
	shorttitle = {Legislative {Success} in a {Small} {World}},
	doi = {10.1017/S002238160999051X},
	number = {1},
	urldate = {2021-02-13},
	journal = {The Journal of Politics},
	author = {Tam Cho, Wendy K. and Fowler, James H.},
	year = {2010},
	pages = {124--135},
}

@article{kirkland_relational_2011,
	title = {The {Relational} {Determinants} of {Legislative} {Outcomes}: {Strong} and {Weak} {Ties} {Between} {Legislators}},
	volume = {73},
	issn = {0022-3816, 1468-2508},
	shorttitle = {The {Relational} {Determinants} of {Legislative} {Outcomes}},
	doi = {10.1017/S0022381611000533},
	language = {en},
	number = {3},
	urldate = {2021-02-13},
	journal = {The Journal of Politics},
	author = {Kirkland, Justin H.},
	year = {2011},
	pages = {887--898},
}

@article{desmarais_etal_2015,
  title={Persistent policy pathways: Inferring diffusion networks in the American states},
  author={Desmarais, Bruce A and Harden, Jeffrey J and Boehmke, Frederick J},
  journal={American Political Science Review},
  volume={109},
  number={2},
  pages={392--406},
  year={2015},
  publisher={Cambridge University Press}
}

@article{larson2017,
  title={Networks and interethnic cooperation},
  author={Larson, Jennifer M},
  journal={The Journal of Politics},
  volume={79},
  number={2},
  pages={546--559},
  year={2017},
  publisher={University of Chicago Press Chicago, IL}
}

@article{kim:liau:imai:20,
  title={Measuring Trade Profile with Granular Product-Level Data},
  author={Kim, In Song and Liao, Steven and Imai, Kosuke},
  journal={American Journal of Political Science},
  volume={64},
  number={1},
  pages={102--117},
  year={2020},
  publisher={Wiley Online Library}
}

@article{arnold_friendship_2000,
	title = {Friendship and {Votes}: {The} {Impact} of {Interpersonal} {Ties} on {Legislative} {Decision} {Making}},
	volume = {32},
	issn = {0160-323X},
	shorttitle = {Friendship and {Votes}},
	doi = {10.1177/0160323X0003200206},
	language = {en},
	number = {2},
	urldate = {2021-02-13},
	journal = {State and Local Government Review},
	author = {Arnold, Laura W. and Deen, Rebecca E. and Patterson, Samuel C.},
	year = {2000},
	pages = {142--147},
}

@article{battaglini_etal_2020,
  title={Effectiveness of connected legislators},
  author={Battaglini, Marco and Sciabolazza, Valerio Leone and Patacchini, Eleonora},
  journal={American Journal of Political Science},
  volume={64},
  number={4},
  pages={739--756},
  year={2020},
  publisher={Wiley Online Library}
}

@article{fowler_legislative_2006,
	title = {Legislative cosponsorship networks in the {US} {House} and {Senate}},
	volume = {28},
	issn = {0378-8733},
	language = {en},
	number = {4},
	urldate = {2021-03-30},
	journal = {Social Networks},
	author = {Fowler, James H.},
	year = {2006},
	pages = {454--465}
}

@article{krutz_issues_2005,
	title = {Issues and {Institutions}: "{Winnowing}" in the {U}.{S}. {Congress}},
	volume = {49},
	issn = {0092-5853, 1540-5907},
	shorttitle = {Issues and {Institutions}},
	doi = {10.1111/j.0092-5853.2005.00125.x},
	language = {en},
	number = {2},
	urldate = {2021-04-13},
	journal = {American Journal of Political Science},
	author = {Krutz, Glen S.},
	year = {2005},
	pages = {313--326},
}

@article{brandenberger_trading_2018,
	title = {Trading favors---{Examining} the temporal dynamics of reciprocity in congressional collaborations using relational event models},
	volume = {54},
	issn = {03788733},
	doi = {10.1016/j.socnet.2018.02.001},
	urldate = {2021-04-21},
	journal = {Social Networks},
	author = {Brandenberger, Laurence},
	year = {2018},
	pages = {238--253},
}

@article{govaert_clustering_2003,
	title = {Clustering with block mixture models},
	volume = {36},
	issn = {00313203},
	doi = {10.1016/S0031-3203(02)00074-2},
	language = {en},
	number = {2},
	urldate = {2021-03-03},
	journal = {Pattern Recognition},
	author = {Govaert, Gérard and Nadif, Mohamed},
	year = {2003},
	pages = {463--473},
}

@article{neal:2020,
  title={A sign of the times? Weak and strong polarization in the US Congress, 1973--2016},
  author={Neal, Zachary P},
  journal={Social networks},
  volume={60},
  pages={103--112},
  year={2020},
  publisher={Elsevier}
}

@article{neal_backbone_2014,
	title = {The backbone of bipartite projections: {Inferring} relationships from co-authorship, co-sponsorship, co-attendance and other co-behaviors},
	volume = {39},
	issn = {0378-8733},
	shorttitle = {The backbone of bipartite projections},
	doi = {10.1016/j.socnet.2014.06.001},
	language = {en},
	urldate = {2021-03-06},
	journal = {Social Networks},
	author = {Neal, Zachary},
	year = {2014},
	pages = {84--97},
}

@article{maoz_etal_2006,
  title={Structural equivalence and international conflict: A social networks analysis},
  author={Maoz, Zeev and Kuperman, Ranan D and Terris, Lesley and Talmud, Ilan},
  journal={Journal of Conflict Resolution},
  volume={50},
  number={5},
  pages={664--689},
  year={2006},
  publisher={Sage Publications Sage CA: Thousand Oaks, CA}
}

@article{zhou_analysis_2019,
	title = {Analysis of spectral clustering algorithms for community detection: the general bipartite setting},
	volume = {20},
	issn = {1533-7928},
	shorttitle = {Analysis of spectral clustering algorithms for community detection},
	url = {http://jmlr.org/papers/v20/18-170.html},
	number = {47},
	urldate = {2021-03-08},
	journal = {Journal of Machine Learning Research},
	author = {Zhou, Zhixin and Amini, Arash A.},
	year = {2019},
	pages = {1--47},
}

@article{chatterjee2013,
  title={Estimating and understanding exponential random graph models},
  author={Chatterjee, Sourav and Diaconis, Persi},
  journal={The Annals of Statistics},
  volume={41},
  number={5},
  pages={2428--2461},
  year={2013},
  publisher={Institute of Mathematical Statistics}
}

@article{schweinberger2011,
  title={Instability, sensitivity, and degeneracy of discrete exponential families},
  author={Schweinberger, Michael},
  journal={Journal of the American Statistical Association},
  volume={106},
  number={496},
  pages={1361--1370},
  year={2011},
  publisher={Taylor \& Francis}
}

@article{larremore_efficiently_2014,
	title = {Efficiently inferring community structure in bipartite networks},
	volume = {90},
	doi = {10.1103/PhysRevE.90.012805},
	number = {1},
	urldate = {2021-03-08},
	journal = {Physical Review E},
	author = {Larremore, Daniel B. and Clauset, Aaron and Jacobs, Abigail Z.},
	year = {2014},
	pages = {012805},
}

@article{razaee_matched_2019,
	title = {Matched {Bipartite} {Block} {Model} with {Covariates}},
	volume = {20},
	issn = {1533-7928},
	url = {http://jmlr.org/papers/v20/17-153.html},
	number = {34},
	urldate = {2021-03-08},
	journal = {Journal of Machine Learning Research},
	author = {Razaee, Zahra S. and Amini, Arash A. and Li, Jingyi Jessica},
	year = {2019},
	pages = {1--44},
}

@article{kim2020,
    title={Mapping Political Communities: A Statistical Analysis of Lobbying Networks in Legislative Politics},
    DOI={10.1017/pan.2020.29}, 
    journal={Political Analysis},
    publisher={Cambridge University Press},
    author={Kim, In Song and Kunisky, Dmitriy}, 
    year={2020},
    pages={1–20}
}

@techreport{teh_collapsed_2007,
	title = {A {Collapsed} {Variational} {Bayesian} {Inference} {Algorithm} for {Latent} {Dirichlet} {Allocation}},
	language = {en},
	urldate = {2021-04-13},
	institution = {UC Irvine School of Information and Computer Science},
	author = {Teh, Yee W. and Newman, David and Welling, Max},
	year = {2007},
}

@article{wasserman_pattison_1996,
  title={Logit models and logistic regressions for social networks: I. An introduction to Markov graphs andp},
  author={Wasserman, Stanley and Pattison, Philippa},
  journal={Psychometrika},
  volume={61},
  number={3},
  pages={401--425},
  year={1996},
  publisher={Springer}
}

@article{harward_calculus_2010,
	title = {The {Calculus} of {Cosponsorship} in the {U}.{S}. {Senate}},
	volume = {35},
	issn = {1939-9162},
	url = {https://onlinelibrary.wiley.com/doi/abs/10.3162/036298010790821950},
	doi = {https://doi.org/10.3162/036298010790821950},
	language = {en},
	number = {1},
	urldate = {2021-04-12},
	journal = {Legislative Studies Quarterly},
	author = {Harward, Brian M. and Moffett, Kenneth W.},
	year = {2010},
	pages = {117--143},
}

@article{white_murphy_2016,
  title={Mixed-membership of experts stochastic blockmodel},
  author={White, Arthur and Murphy, Thomas Brendan},
  journal={Network Science},
  volume={4},
  number={1},
  pages={48--80},
  year={2016},
  publisher={Cambridge University Press}
}

@article{handcock2007,
  title={Model-based clustering for social networks},
  author={Handcock, Mark S and Raftery, Adrian E and Tantrum, Jeremy M},
  journal={Journal of the Royal Statistical Society: Series A (Statistics in Society)},
  volume={170},
  number={2},
  pages={301--354},
  year={2007},
  publisher={Wiley Online Library}
}

@article{cao2009,
  title={Networks of intergovernmental organizations and convergence in domestic economic policies},
  author={Cao, Xun},
  journal={International Studies Quarterly},
  volume={53},
  number={4},
  pages={1095--1130},
  year={2009},
  publisher={Blackwell Publishing Ltd Oxford, UK}
}

@article{muraoka2020,
  title={The Cosponsorship Patterns of Reserved Seat Legislators},
  author={Muraoka, Taishi},
  journal={Legislative Studies Quarterly},
  volume={45},
  number={4},
  pages={555--580},
  year={2020},
  publisher={Wiley Online Library}
}

@article{zhang_community_2008,
  author={Zhang, Yan and Friend, Andrew J and Traud, Amanda L and Porter, Mason A and Fowler, James H and Mucha, Peter J},
  title = {Community structure in Congressional cosponsorship networks},
  volume = {387},
  number = {7},
  journal = {Physica A: Statistical Mechanics and its Applications},
  year = {2008},
  pages = {1705--1712}
}

@article{rippere_polarization_2016,
	title = {Polarization {Reconsidered}: {Bipartisan} {Cooperation} through {Bill} {Cosponsorship}},
	volume = {48},
	issn = {0032-3497, 1744-1684},
	shorttitle = {Polarization {Reconsidered}},
	doi = {10.1057/pol.2016.4},
	language = {en},
	number = {2},
	urldate = {2021-07-07},
	journal = {Polity},
	author = {Rippere, Paulina S.},
	year = {2016},
	pages = {243--278},
}

@article{lawless_nice_2018,
	title = {Nice {Girls}? {Sex}, {Collegiality}, and {Bipartisan} {Cooperation} in the {US} {Congress}},
	volume = {80},
	issn = {0022-3816, 1468-2508},
	shorttitle = {Nice {Girls}?},
	doi = {10.1086/698884},
	language = {en},
	number = {4},
	urldate = {2021-07-07},
	journal = {The Journal of Politics},
	author = {Lawless, Jennifer L. and Theriault, Sean M. and Guthrie, Samantha},
	year = {2018},
	pages = {1268--1282},
}

@article{bratton_networks_2011,
	title = {Networks in the {Legislative} {Arena}: {How} {Group} {Dynamics} {Affect} {Cosponsorship}},
	volume = {36},
	issn = {1939-9162},
	shorttitle = {Networks in the {Legislative} {Arena}},
	url = {https://onlinelibrary.wiley.com/doi/abs/10.1111/j.1939-9162.2011.00021.x},
	doi = {10.1111/j.1939-9162.2011.00021.x},
	language = {en},
	number = {3},
	urldate = {2021-07-07},
	journal = {Legislative Studies Quarterly},
	author = {Bratton, Kathleen A. and Rouse, Stella M.},
	year = {2011},
	pages = {423--460},
}

@article{cirone_cabinets_2018,
	title = {Cabinets, {Committees}, and {Careers}: {The} {Causal} {Effect} of {Committee} {Service}},
	volume = {80},
	issn = {0022-3816, 1468-2508},
	shorttitle = {Cabinets, {Committees}, and {Careers}},
	doi = {10.1086/697252},
	language = {en},
	number = {3},
	urldate = {2021-07-07},
	journal = {The Journal of Politics},
	author = {Cirone, Alexandra and Van Coppenolle, Brenda},
	year = {2018},
	pages = {948--963},
}

@article{lancichinetti_etal_2015,
  title={High-reproducibility and high-accuracy method for automated topic classification},
  author={Lancichinetti, Andrea and Sirer, M Irmak and Wang, Jane X and Acuna, Daniel and K{\"o}rding, Konrad and Amaral, Lu{\'\i}s A Nunes},
  journal={Physical Review X},
  volume={5},
  number={1},
  pages={011007},
  year={2015},
  publisher={APS}
}

@article{peixoto2014,
  title={Hierarchical block structures and high-resolution model selection in large networks},
  author={Peixoto, Tiago P},
  journal={Physical Review X},
  volume={4},
  number={1},
  pages={011047},
  year={2014},
  publisher={APS}
}

@inproceedings{huang_etal_2005,
  title={Link prediction approach to collaborative filtering},
  author={Huang, Zan and Li, Xin and Chen, Hsinchun},
  booktitle={Proceedings of the 5th ACM/IEEE-CS Joint Conference on Digital Libraries (JCDL'05)},
  pages={141--142},
  year={2005},
  organization={IEEE}
}

@article{hoff2002,
  title={Latent space approaches to social network analysis},
  author={Hoff, Peter D and Raftery, Adrian E and Handcock, Mark S},
  journal={Journal of the american Statistical association},
  volume={97},
  number={460},
  pages={1090--1098},
  year={2002},
  publisher={Taylor \& Francis}
}

@article{porter_network_2005,
	title = {A network analysis of committees in the {U}.{S}. {House} of {Representatives}},
	volume = {102},
	number = {20},
	journal = {Proceedings of the National Academy of Sciences},
	author = {Porter, Mason A. and Mucha, Peter J. and Newman, M. E. J. and Warmbrand, Casey M.},
	year = {2005},
	pages = {7057--7062},
}

@book{poole2017ideology,
  title={Ideology \& congress: A political economic history of roll call voting},
  author={Poole, Keith T and Rosenthal, Howard},
  year={2017},
  publisher={Routledge}
}

@book{sunstein:2009,
  title={Republic.com 2.0},
  author={Sunstein, Cass R.},
  year={2009},
  publisher={Princeton University Press}
}

@book{sunstein:2018,
  title={\# Republic: Divided democracy in the age of social media},
  author={Sunstein, Cass},
  year={2018},
  publisher={Princeton University Press}
}

@book{riker:1962,
  author    = {William H. Riker},
  title     = {The Theory of Political Coalitions},
  publisher = {Yale University Press},
  year      = {1962},
  address   = {New Haven},
}

@article{bernhard:sulkin:2013,
  title={Commitment and consequences: Reneging on cosponsorship pledges in the US House},
  author={Bernhard, William and Sulkin, Tracy},
  journal={Legislative Studies Quarterly},
  volume={38},
  number={4},
  pages={461--487},
  year={2013},
  publisher={Wiley Online Library}
}

@article{koger:2003,
  title={Position taking and cosponsorship in the US House},
  author={Koger, Gregory},
  journal={Legislative studies quarterly},
  volume={28},
  number={2},
  pages={225--246},
  year={2003},
  publisher={Wiley Online Library}
}

@article{gonzalez:wang:2016,
  title={Networked discontent: The anatomy of protest campaigns in social media},
  author={Gonz{\'a}lez-Bail{\'o}n, Sandra and Wang, Ning},
  journal={Social networks},
  volume={44},
  pages={95--104},
  year={2016},
  publisher={Elsevier}
}

@article{larson:etal:2019,
  title={Social networks and protest participation: Evidence from 130 million Twitter users},
  author={Larson, Jennifer M and Nagler, Jonathan and Ronen, Jonathan and Tucker, Joshua A},
  journal={American Journal of Political Science},
  volume={63},
  number={3},
  pages={690--705},
  year={2019},
  publisher={Wiley Online Library}
}

@article{rosenman:etal:2023,
  title={Recalibration of Predicted Probabilities Using the “Logit Shift”: Why Does It Work, and When Can It Be Expected to Work Well?},
  author={Rosenman, Evan TR and McCartan, Cory and Olivella, Santiago},
  journal={Political Analysis},
  volume={31},
  number={4},
  pages={651--661},
  year={2023}
}

@article{campbell1982cosponsoring,
  title={Cosponsoring legislation in the US Congress},
  author={Campbell, James E},
  journal={Legislative Studies Quarterly},
  pages={415--422},
  year={1982},
  publisher={JSTOR}
}

@article{fong2020expertise,
  title={Expertise, networks, and interpersonal influence in congress},
  author={Fong, Christian},
  journal={The Journal of Politics},
  volume={82},
  number={1},
  pages={269--284},
  year={2020},
  publisher={The University of Chicago Press Chicago, IL}
}

@article{grossmann2013lobbying,
  title={Lobbying and congressional bill advancement},
  author={Grossmann, Matt and Pyle, Kurt},
  journal={Interest Groups \& Advocacy},
  volume={2},
  number={1},
  pages={91--111},
  year={2013},
  publisher={Springer}
}

@article{wilson1997cosponsorship,
  title={Cosponsorship in the US Congress},
  author={Wilson, Rick K and Young, Cheryl D},
  journal={Legislative Studies Quarterly},
  pages={25--43},
  year={1997},
  publisher={JSTOR}
}

@article{anderson2003keys,
  title={The keys to legislative success in the US House of Representatives},
  author={Anderson, William D and Box-Steffensmeier, Janet M and Sinclair-Chapman, Valeria},
  journal={Legislative Studies Quarterly},
  volume={28},
  number={3},
  pages={357--386},
  year={2003},
  publisher={Wiley Online Library}
}

@article{bernhard2013commitment,
  title={Commitment and consequences: Reneging on cosponsorship pledges in the US House},
  author={Bernhard, William and Sulkin, Tracy},
  journal={Legislative Studies Quarterly},
  volume={38},
  number={4},
  pages={461--487},
  year={2013},
  publisher={Wiley Online Library}
}

@article{holman2022let,
  title={Let’s Work Together: Bill Success via Women’s Cosponsorship in US State Legislatures},
  author={Holman, Mirya R and Mahoney, Anna and Hurler, Emma},
  journal={Political Research Quarterly},
  volume={75},
  number={3},
  pages={676--690},
  year={2022},
  publisher={SAGE Publications Sage CA: Los Angeles, CA}
}

@article{baltakys2023inference,
  title={Inference of monopartite networks from bipartite systems with different link types},
  author={Baltakys, Kestutis},
  journal={Scientific Reports},
  volume={13},
  number={1},
  pages={1072},
  year={2023},
  publisher={Nature Publishing Group UK London}
}

@article{marrs2020inferring,
  title={Inferring influence networks from longitudinal bipartite relational data},
  author={Marrs, Frank W and Campbell, Benjamin W and Fosdick, Bailey K and Cranmer, Skyler J and B{\"o}hmelt, Tobias},
  journal={Journal of Computational and Graphical Statistics},
  volume={29},
  number={3},
  pages={419--431},
  year={2020},
  publisher={Taylor \& Francis}
}

@article{neal2014backbone,
  title={The backbone of bipartite projections: Inferring relationships from co-authorship, co-sponsorship, co-attendance and other co-behaviors},
  author={Neal, Zachary},
  journal={Social Networks},
  volume={39},
  pages={84--97},
  year={2014},
  publisher={Elsevier}
}

@article{rippere2016polarization,
  title={Polarization reconsidered: Bipartisan cooperation through bill cosponsorship},
  author={Rippere, Paulina S},
  journal={Polity},
  volume={48},
  number={2},
  pages={243--278},
  year={2016},
  publisher={The University of Chicago Press Chicago}
}

@book{harbridge_is_2015,
	address = {New York, NY},
	title = {Is bipartisanship dead? policy agreement and agenda-setting in the {House} of {Representatives}},
	isbn = {9781107079953},
	shorttitle = {Is bipartisanship dead?},
	publisher = {Cambridge University Press},
	author = {Harbridge, Laurel},
	year = {2015},
	note = {OCLC: ocn907082845},
}

@article{harbridge:etal:2023,
  title={The bipartisan path to effective lawmaking},
  author={Harbridge-Yong, Laurel and Volden, Craig and Wiseman, Alan E},
  journal={The Journal of Politics},
  volume={85},
  number={3},
  pages={1048--1063},
  year={2023},
  publisher={The University of Chicago Press Chicago, IL}
}

@article{desposato:etal:2011,
  title={Using cosponsorship to estimate ideal points},
  author={Desposato, Scott W and Kearney, Matthew C and Crisp, Brian F},
  journal={Legislative Studies Quarterly},
  volume={36},
  number={4},
  pages={531--565},
  year={2011},
  publisher={Wiley Online Library}
}

@article{latapy2008basic,
  title={Basic notions for the analysis of large two-mode networks},
  author={Latapy, Matthieu and Magnien, Cl{\'e}mence and Del Vecchio, Nathalie},
  journal={Social networks},
  volume={30},
  number={1},
  pages={31--48},
  year={2008},
  publisher={Elsevier}
}

@article{hoff2021additive,
  title={Additive and Multiplicative Effects Network Models},
  author={Hoff, Peter D},
  journal={Statistical science},
  volume={36},
  number={1},
  pages={34--50},
  year={2021},
  publisher={Institute of Mathematical Statistics}
}

@article{promel2002note,
  title={A note on triangle-free and bipartite graphs},
  author={Pr{\"o}mel, Hans J{\"u}rgen and Schickinger, Thomas and Steger, Angelika},
  journal={Discrete mathematics},
  volume={257},
  number={2-3},
  pages={531--540},
  year={2002},
  publisher={Elsevier}
}

@article{liu2018cycle,
  title={Cycle lengths and minimum degree of graphs},
  author={Liu, Chun-Hung and Ma, Jie},
  journal={Journal of Combinatorial Theory, Series B},
  volume={128},
  pages={66--95},
  year={2018},
  publisher={Elsevier}
}

\clearpage
\appendix

\section{Appendix}
\renewcommand{\thefigure}{A.\arabic{figure}}
\renewcommand{\thetable}{A.\arabic{table}}
\setcounter{figure}{0}
\setcounter{table}{0}

\subsection{Projecting Bipartite Networks onto Unipartite Networks is a Common Practice}
\label{app:projection}

\begin{table}[h!] \spacingset{1}\footnotesize
\caption{Applications with naturally bipartite applications in top field journals in 2000s. ``Projected'' indicates unipartite network considered for empirical application.}\label{tab:netapp}
\centering
\begin{tabular}[t]{>{\raggedright\arraybackslash}p{5.4cm}>{\raggedright\arraybackslash}p{2cm}>{\raggedright\arraybackslash}p{6.5cm}>{\raggedright\arraybackslash}p{1cm}}
\toprule
\textcolor{black}{\textbf{Author}} & \textcolor{black}{\textbf{Journal}} & \textcolor{black}{\textbf{Network; Nodes}} & \textcolor{black}{\textbf{Projected}}\\
\midrule
\addlinespace[0.3em]
\multicolumn{4}{l}{\textbf{Alliances}}\\
\hspace{1em}\textcolor{black}{Franzese et al. (2012)} & \textcolor{black}{PA} & \textcolor{black}{country alliances: countries $\leftrightarrow$ alliance treaties} & \textcolor{black}{Yes}\\
\hspace{1em}\textcolor{black}{\cellcolor{gray!6}{Kinne \& Bunte (2018)}} & \textcolor{black}{\cellcolor{gray!6}{BJPS}} & \textcolor{black}{\cellcolor{gray!6}{defense cooperation agreements (DCA) network; countries $\leftrightarrow$ DCAs}} & \textcolor{black}{\cellcolor{gray!6}{Yes}}\\
\addlinespace[0.3em]
\multicolumn{4}{l}{\textbf{Communication}}\\
\hspace{1em}\textcolor{black}{Aar\o{}e \& Peterson (2018)} & \textcolor{black}{BJPS} & \textcolor{black}{media flows; individuals $\leftrightarrow$ stories} & \textcolor{black}{Yes}\\
\hspace{1em}\textcolor{black}{\cellcolor{gray!6}{Boucher \& Thies (2019)}} & \textcolor{black}{\cellcolor{gray!6}{JOP}} & \textcolor{black}{\cellcolor{gray!6}{Twitter; Twitter users $\leftrightarrow$ tweets}} & \textcolor{black}{\cellcolor{gray!6}{Yes}}\\
\hspace{1em}\textcolor{black}{Siegel \& Badaan (2020)} & \textcolor{black}{APSR} & \textcolor{black}{Twitter; Twitter users $\leftrightarrow$ tweets} & \textcolor{black}{Yes}\\
\addlinespace[0.3em]
\multicolumn{4}{l}{\textbf{Conflict}}\\
\hspace{1em}\textcolor{black}{\cellcolor{gray!6}{Rozenas et al. (2019)}} & \textcolor{black}{\cellcolor{gray!6}{PA}} & \textcolor{black}{\cellcolor{gray!6}{conflict \& treaty network; actors $\leftrightarrow$ treaties}} & \textcolor{black}{\cellcolor{gray!6}{Yes}}\\
\hspace{1em}\textcolor{black}{Nieman et al. (2021)} & \textcolor{black}{JOP} & \textcolor{black}{troop placements; major $\leftrightarrow$ minor powers} & \textcolor{black}{Yes}\\
\addlinespace[0.3em]
\multicolumn{4}{l}{\textbf{Congress \& Parliament}}\\
\hspace{1em}\textcolor{black}{\cellcolor{gray!6}{Cho \& Fowler (2010)}} & \textcolor{black}{{\cellcolor{gray!6}JOP}} & \textcolor{black}{{\cellcolor{gray!6}legislative cosponsorship; legislators $\leftrightarrow$ bills}} & \textcolor{black}{{\cellcolor{gray!6}Yes}}\\
\hspace{1em}\textcolor{black}{Cranmer \& Desmarais (2011)} & \textcolor{black}{{PA}} & \textcolor{black}{{legislative cosponsorship; legislators $\leftrightarrow$ bills}} & \textcolor{black}{{Yes}}\\
\hspace{1em}\textcolor{black}{\cellcolor{gray!6}{Box-Steffensmeier et al. (2018)}} & \textcolor{black}{\cellcolor{gray!6}{AJPS}} & \textcolor{black}{\cellcolor{gray!6}{Dear Colleague letters; legislators $\leftrightarrow$ interest groups}} & \textcolor{black}{\cellcolor{gray!6}{Yes}}\\
\hspace{1em}\textcolor{black}{Zelizer (2019)} & \textcolor{black}{APSR} & \textcolor{black}{cue taking network; legislators $\leftrightarrow$ bills} & \textcolor{black}{Yes}\\
\hspace{1em}\textcolor{black}{\cellcolor{gray!6}{Battaglini et al. (2020)}} & \textcolor{black}{\cellcolor{gray!6}{AJPS}} & \textcolor{black}{\cellcolor{gray!6}{legislative cosponsorship; legislators $\leftrightarrow$ bills}} & \textcolor{black}{\cellcolor{gray!6}{Yes}}\\
\hspace{1em}\textcolor{black}{Kim \& Kunisky (2020)} & \textcolor{black}{PA} & \textcolor{black}{Congressional lobbying; special interest groups $\leftrightarrow$ politicians} & \textcolor{black}{No}\\
\addlinespace[0.3em]
\multicolumn{4}{l}{\textbf{International organizations}}\\
\hspace{1em}\textcolor{black}{\cellcolor{gray!6}{Martinsen et al. (2020)}} & \textcolor{black}{\cellcolor{gray!6}{BJPS}} & \textcolor{black}{\cellcolor{gray!6}{welfare governance network; bureaucrats $\leftrightarrow$ states}} & \textcolor{black}{\cellcolor{gray!6}{Yes}}\\
\addlinespace[0.3em]
\multicolumn{4}{l}{\textbf{International political economy}}\\
\hspace{1em}\textcolor{black}{Bodea \& Hicks (2015)} & \textcolor{black}{JOP} & \textcolor{black}{central bank independence for countries/firms; countries $\leftrightarrow$ investors} & \textcolor{black}{Yes}\\
\hspace{1em}\textcolor{black}{\cellcolor{gray!6}{Kim et al. (2019)}} & \textcolor{black}{\cellcolor{gray!6}{AJPS}} & \textcolor{black}{\cellcolor{gray!6}{trade network; countries $\leftrightarrow$ products}} & \textcolor{black}{\cellcolor{gray!6}{Yes}}\\
\addlinespace[0.3em]
\multicolumn{4}{l}{\textbf{Policies}}\\
\hspace{1em}\textcolor{black}{\cellcolor{gray!6}{Fischer \& Sciarini (2016)}} & \textcolor{black}{\cellcolor{gray!6}{JOP}} & \textcolor{black}{\cellcolor{gray!6}{state policy collaboration; political actors $\leftrightarrow$ policies}} & \textcolor{black}{\cellcolor{gray!6}{Yes}}\\
\hspace{1em}\textcolor{black}{Gilardi et al. (2020)} & \textcolor{black}{AJPS} & \textcolor{black}{policy adoption/issue definition; states $\leftrightarrow$ policies} & \textcolor{black}{Yes}\\
\addlinespace[0.3em]
\multicolumn{4}{l}{\textbf{Political elites}}\\
\hspace{1em}\textcolor{black}{\cellcolor{gray!6}{Nyhan \& Montgomery (2015)}} & \textcolor{black}{\cellcolor{gray!6}{AJPS}} & \textcolor{black}{\cellcolor{gray!6}{campaign consultants; consultants $\leftrightarrow$ candidates}} & \textcolor{black}{\cellcolor{gray!6}{Yes}}\\
\hspace{1em}\textcolor{black}{Pietryka \& Debats (2017)} & \textcolor{black}{APSR} & \textcolor{black}{voters-elite network; voters $\leftrightarrow$ elites} & \textcolor{black}{Yes}\\
\hspace{1em}\textcolor{black}{\cellcolor{gray!6}{Weschle (2017)}} & \textcolor{black}{\cellcolor{gray!6}{BJPS}} & \textcolor{black}{\cellcolor{gray!6}{party-societal group network; parties $\leftrightarrow$ societal groups}} & \textcolor{black}{\cellcolor{gray!6}{Yes}}\\
\hspace{1em}\textcolor{black}{Weschle (2018)} & \textcolor{black}{APSR} & \textcolor{black}{political and social actors; political $\leftrightarrow$ social actors} & \textcolor{black}{Yes}\\
\hspace{1em}\textcolor{black}{\cellcolor{gray!6}{Jiang \& Zeng (2019)}} & \textcolor{black}{\cellcolor{gray!6}{JOP}} & \textcolor{black}{\cellcolor{gray!6}{elite network; elite (lower) $\leftrightarrow$ elite (upper) politicians}} & \textcolor{black}{\cellcolor{gray!6}{Yes}}\\
\hspace{1em}\textcolor{black}{Box-Steffensmeier et al. (2020)} & \textcolor{black}{AJPS} & \textcolor{black}{campaign donor list sharing; legislators $\leftrightarrow$ lists} & \textcolor{black}{Yes}\\
\addlinespace[0.3em]
\multicolumn{4}{l}{\textbf{Village networks}}\\
\hspace{1em}\textcolor{black}{\cellcolor{gray!6}{Larson (2017)}} & \textcolor{black}{\cellcolor{gray!6}{JOP}} & \textcolor{black}{\cellcolor{gray!6}{ethnic cooperation; individuals $\leftrightarrow$ ethnic groups}} & \textcolor{black}{\cellcolor{gray!6}{Yes}}\\
\hspace{1em}\textcolor{black}{Haim, Nanes \& Davidson (2021)} & \textcolor{black}{JOP} & \textcolor{black}{police community connections; police $\leftrightarrow$ citizens} & \textcolor{black}{Yes}\\
\bottomrule
\end{tabular}
\end{table}

\end{document}


\newcommand\const{\mathrm{const.}}
\newcommand*\diff{\mathop{}\!\mathrm{d}}
\newcommand*\expec{\mathop{}\mathbb{E}}
\newcommand\numberthis{\addtocounter{equation}{1}\tag{\theequation}}
\newcommand{\appropto}{\mathrel{\vcenter{
  \offinterlineskip\halign{\hfil$##$\cr
    \propto\cr\noalign{\kern2pt}\sim\cr\noalign{\kern-2pt}}}}}

\newcommand\dynMMSBM{\textsf{dynMMSBM}}
\newcommand\dist{\buildrel\rm d\over\sim}
\newcommand\ind{\stackrel{\rm indep.}{\sim}}
\newcommand\iid{\stackrel{\rm i.i.d.}{\sim}}
\newcommand\logit{{\rm logit}}
\renewcommand\r{\right}
\renewcommand\l{\left}
\newcommand\E{\mathbb{E}}
\newcommand\PP{\mathbb{P}}
\newcommand{\argmax}{\operatornamewithlimits{argmax}}

\newcommand\spacingset[1]{\renewcommand{\baselinestretch}%
  {#1}\small\normalsize}

\spacingset{1.25}

\newcommand{\tit}{Online Supplementary Information\\ A Statistical Model of Bipartite Networks:
  Application to Cosponsorship in the United States Senate}


\if0\blind

{\title{\tit\thanks{The methods described in this paper can be
     implemented via the open-source statistical software, {\sf
       NetMix}, available at
     \url{https://CRAN.R-project.org/package=NetMix}. The authors are grateful for comments from Alison Craig, Skyler Cranmer, Sarah Shugars and the participants of the Harvard IQSS Applied Statistics seminar.
     }}

 \author{Adeline Lo\thanks{Assistant Professor of Political Science, UW-Madison. Email: \href{mailto:aylo@wisc.edu}{\texttt{aylo@wisc.edu}}, URL:\href{https://www.loadeline.com/}{\tt https://www.loadeline.com/}
     }  \quad \quad Santiago Olivella\thanks{Associate Professor of Political Science, UNC-Chapel Hill. Email: \href{mailto:olivella@unc.edu}{\texttt{olivella@unc.edu}}} \quad \quad Kosuke
   Imai\thanks{Professor, Department of Government and Department of Statistics, Harvard
     University.  1737 Cambridge Street, Institute for Quantitative
     Social Science, Cambridge 02138. Email:
     \href{mailto:imai@harvard.edu}{\texttt{imai@harvard.edu}}, URL:\href{https://imai.fas.harvard.edu/}{\tt https://imai.fas.harvard.edu/}
     }
     }

  \date{October 2024}

\maketitle
}\fi

\if1\blind \title{\bf \tit} \maketitle
\fi

\pdfbookmark[1]{Title Page}{Title Page}

\thispagestyle{empty}
\setcounter{page}{0}

\renewcommand{\thesection}{S.\arabic{section}}
\renewcommand{\thefigure}{S.\arabic{figure}}
\renewcommand{\thetable}{S.\arabic{table}}
\renewcommand{\theequation}{S.\arabic{equation}}

\newpage
\spacingset{1.83}

\tableofcontents
\clearpage

\section{Additional methodological details}
\label{app:model}

\subsection{Plate Diagram of the Proposed Model}
\label{app:dag}


\begin{figure}[h]
  \centering \spacingset{1}
\begin{tikzpicture}
\node[obs] (y) {$\mathbf{y}_{pq}$}; 
\node[latent, above=of y, xshift=-1.0cm] (B) {$\mathbf{B}$};
\node[latent, above=of y, xshift=1.0cm] (gamma) {$\boldsymbol{\gamma}$};
\node[latent, left=of y] (z_p) {$\mathbf{z}_{pq}$};
\node[latent, right=of y] (u_q) {$\mathbf{u}_{qp}$};
\node[latent, left=of z_p] (pi_p) {$\boldsymbol{\pi}_p$};
\node[latent, right=of u_q] (psi_q) {$\boldsymbol{\psi}_q$};
\node[obs, below=of pi_p] (x) {$\mathbf{x}_p$};
\node[obs, below=of psi_q] (w) {$\mathbf{w}_q$};
\node[obs, below=of y] (d) {$\mathbf{d}_{pq}$};
\node[latent, above=of pi_p] (beta1) {$\boldsymbol{\beta}_{1}$};
\node[latent, above=of psi_q] (beta2) {$\boldsymbol{\beta}_{2}$};
\plate[inner sep=0.25cm, yshift=.15cm] {plate1} {(y) (z_p) (u_q) (d)} {$\{p,q\} \in E$};
\plate[inner sep=0.25cm, yshift=.2cm] {plate2} {(pi_p)(x)} {$p \in V_1$};
\plate[inner sep=0.25cm, yshift=.2cm] {plate3} {(psi_q)(w)} {$q \in V_2$};
\edge {z_p, u_q, B, d, gamma} {y};
\edge {pi_p} {z_p};
\edge {psi_q} {u_q};
\edge {w, beta2} {psi_q};
\edge {x, beta1} {pi_p};
\end{tikzpicture}\caption{{\bf Plate diagram of the proposed model}. Observed data represented as shaded nodes; hyperparameters presented as nodes outside plates.}\label{fig:dag}
\end{figure}

\subsection{Motivation for the proposed model components}
\label{app:motivation}

The model components in Figure~\ref{fig:dag} are most easily justified
when considering the case of cosponsorship, legislative productivity,
and collaboration . Any modeling tool that hopes to produce plausible
answers to the puzzle of productivity and collaboration in times of
nominal partisan division should allow researchers to explore the
heterogeneity of both legislator and bill attributes, as allowed by the $\bf{x}_p$ and $\bf{w}_q$ terms in our model. Much as senators
might differentially decide to cosponsor based on their personal
characteristics, so too might legislation attract cosponsorship based
on its different content and the specific context in which it is
introduced. A useful model of cosponsorships thus allows the
politics of coalition formation around certain types of legislation to
account for this heterogeneity.

Furthermore, it is important for such a model to capture how the
interaction of different bill and senator types can lead to
cosponsorship collaborations beyond what would be expected from
understanding the coalitional dynamics that drive much of legislative
politics, as allowed by the $\bf{d}_{pq}$ term in
Figure~\ref{fig:dag}. The extant literature has identified two
substantial predictors of cosponsorship decision that are defined at
the senator-bill dyad level, and that result in the kind of
non-coalitional homophily and heterophily that is common in social
networks of different kinds.

First, individual senators often trade favors
with one another, such that cosponsorship may result from quid pro
quo behavior and norms of individual reciprocity --- \emph{you cosponsored
my bill before, I'll cosponsor yours now}
\citep{brandenberger_trading_2018, harbridge:etal:2023}. Similarly, scholars have
increasingly stressed the role of Senate committees in forming support
around legislation, finding evidence that sitting in a committee
involved in the life cycle of a bill can affect a legislator's support
for it \citep{porter_network_2005,cirone_cabinets_2018}. Accordingly,
a model that aims to capture the full set of forces behind a bipartite
network such as that formed by cosponsorships should account for these
kinds of naturally dyadic features that complement group-based
drivers of edge formation.

\subsection{Details of the Estimation Algorithm}
\label{app:svi}

To approximate the collapsed posterior proportional to
Equation (6)
, we first define a factorized distribution of the
joint dyad-specific latent group membership variables
(i.e. $\mathbf{Z}$ and $\mathbf{U}$) as follows:
\begin{equation}
m(\mathbf{Z},\mathbf{U}\mid \boldsymbol{\Phi}) = 
\prod_{p,q \in V_{1} \times V_{2}} m(z_{pq}, u_{pq} \mid \boldsymbol{\phi}_{pq}) 
\label{eq:variational}
\end{equation}
where $\boldsymbol{\Phi}=\{\boldsymbol{\phi}_{pq}\}_{p, q\in V_1 \times V_2}$ are sum-to-one, ($K_1\times K_2$)-dimensional variational parameters. 

The goal of variational inference is to find, in the space of
functions of the form given by Equation~\eqref{eq:variational}, one
that closely approximates (in KL divergence terms) the target
posterior. This is equivalent to maximizing the evidence lower bound
$\mathcal{L}(\boldsymbol{\Phi})$ to Equation (6)
 with respect to
vectors $\boldsymbol{\phi}_{pq}$:
\begin{align}
\begin{split}
    \hat{\boldsymbol{\phi}}_{pq} &= \argmax_{\boldsymbol{\phi}_{pq}}\underbrace{\expec_{m}\left[\log f(\mathbf{Y},\mathbf{Z},\mathbf{U}, \mid \mathbf{B}, \boldsymbol{\beta}, \boldsymbol{\gamma})\right] - \expec_m\left[
    \log m(\mathbf{Z},\mathbf{U}\mid \boldsymbol{\Phi})\right]}_{\mathcal{L}(\boldsymbol{\Phi})}\\
    & \appropto \{(\alpha_{pg} + C'_{pg})(\alpha_{qg} + C'_{qh})(\theta_{pq,g,h}^{y_{pq}}(1-\theta_{pq,g,h})^{1-y_{pq}})\}_{g,h \in K_1 \times K_2}
    \label{eq:phi}
    \end{split}
\end{align}
where
$C_{pg}^\prime= \sum_{q' \in
  V_2^{-q}}\sum_{h'=1}^{K_2}\phi_{pq',g,h'}$ is the expected value of
the marginal count $C_{pg}$ under the variational distribution (and
similarly for $C_{qh}^\prime$). The approximation in the last line
results from using a zeroth-order Taylor series expansion of the
expectation in place of calculating the computationally expensive
integral over the Poisson-Binomial distribution of the count
statistics.

In addition to finding the posterior over mixed-membership vectors,
we take an empirical Bayes approach and maximize the lower
bound $\mathcal{L}(\boldsymbol{\Phi})$ to obtain values of relevant
hyper-parameters $\mathbf{B}$, $\bm{\beta}$ and $\bm{\gamma}$. After
appropriate initialization (see below), then, the full
lower bound optimization proceeds iteratively by first updating
the variational parameters according to Equation~\eqref{eq:phi} (the
E-step), and maximizing $\mathcal{L}(\boldsymbol{\Phi})$ w.r.t. to the
hyper-parameters, holding $\boldsymbol{\phi}_{pq}$ (and derived global
counts $C_{pg}$ and $C_{qh}$) constant at their most recent value (the
M-step), until the change in the lower bound is below a user-specified
tolerance.  As there are no closed-form solution for these optimal
values, we rely on a numerical optimization routine (required gradients are available below). In what follows, we provide details on those steps. 

\subsubsection{E-step}
\subsubsection*{E-step: $\mathbf{Z}$ and $\mathbf{U}$}

Variational parameters $\boldsymbol{\phi}_{pq}$ are updated by
restricting Equation~\eqref{eq:phi} to terms that depend only on
$\mathbf{z}_{pq}$ and $\mathbf{u}_{pq}$ and taking the logarithm of
the resulting expression,
\begin{align*}
& \log P(\mathbf{Y},\mathbf{Z},
\mathbf{U}\mid\mathbf{B}, \boldsymbol{\beta}_1, \boldsymbol{\beta}_2, \boldsymbol{\gamma},  \mathbf{X}_1, \mathbf{X}_2, \mathbf{D})  \\
= & \  z_{pq,g}\sum_{h=1}^{K_2} u_{qq,h}
\l\{Y_{pq}\log(\theta_{pqgh})+(1-Y_{pq})\log(1-\theta_{pqgh})\r\} + \log\Gamma(\alpha_{pg}+ C_{pg}) +\const
\end{align*}
Note that $C_{pg}=C^\prime_{pg}+\mathds{1}(z_{pq,g}=g)$ and
that, for $x\in\{0,1\}$, $\Gamma(y + x)=y^x\Gamma(y)$. As
$\mathds{1}(z_{pq,g}=g)\in \{0,1\}$, we can re-express $\log\Gamma(\alpha_{pg}+ C_{pg})=z_{pq,g}\log(\alpha_{pg}+C^\prime_{pg}) + \log\Gamma(\alpha_{pg} + C^\prime_{pg})$ and simplify the expression:
\begin{align*}
& \ z_{pq,g}\sum_{h=1}^{K_2} u_{qp,h} \l\{Y_{pq}\log (\theta_{pqgh})+(1-Y_{pq})\log(1-\theta_{pqgh})\r\}  + z_{pq,g}\log \left(\alpha_{pg}+C^\prime_{pg}\right) + \const
\end{align*}
Then take the expectation under the variational distribution $\widetilde{Q}$:
\begin{align*}
&  \expec_{\widetilde{Q}}\{\log P(\mathbf{Y},\mathbf{Z},
\mathbf{U}\mid \mathbf{B},
\boldsymbol{\beta}_1, \boldsymbol{\beta}_2, \boldsymbol{\gamma}, \mathbf{D},
\mathbf{X}_1, \mathbf{X}_2)\} \\
= & \ z_{pq,g}\sum_{h=1}^{K_2}\expec_{\widetilde{Q}}(u_{q p,g})\bigl(Y_{pqt}\log(\theta_{pqgh})+(1-Y_{pq})\log(1-\theta_{pqgh})\bigr) + z_{pq,g}
\expec_{\widetilde{Q}}\left\{\log\left(\alpha_{pg}+C_{pg}^\prime\right)\right\}
+ \const
\end{align*}
The exponential of this corresponds to the (unnormalized) parameter vector of a multinomial distribution
$\widetilde{Q}(\mathbf{z}_{pq}\mid \boldsymbol{\phi}_{pq})$.

The update for $\mathbf{u}_{qp}$ is similarly derived.  Restrict
Equation~\eqref{eq:phi} to terms that depend only on
$\mathbf{u}_{qp}$ (for specific $p$, $q$ nodes in $V$) and taking
the logarithm of the resulting expression,
\begin{align*}
& \log P(\mathbf{Y},\mathbf{Z},
\mathbf{U}\mid \mathbf{B},
\boldsymbol{\beta}_1, \boldsymbol{\beta}_2, \boldsymbol{\gamma},  \mathbf{X}_1, \mathbf{X}_2, \mathbf{D})  \\
= & \  u_{q  p,h}\sum_{g=1}^{K_1} z_{pq,g}
\l\{Y_{pq}\log(\theta_{pqgh})+(1-Y_{pq})\log(1-\theta_{pqgh})\r\} + \log\Gamma(\alpha_{qh}+ C_{qh}) +\const
\end{align*}
Re-express
$\log\Gamma(\alpha_{qh}+ C_{qh})=u_{q
  p,h}\log(\alpha_{qh}+C^\prime_{qh}) + \log\Gamma(\alpha_{qh} +
C^\prime_{qh})$ and simplify the expression:
\begin{align*}
& \ u_{q  p,h}\sum_{g=1}^{K_1} z_{p  q,g} \l\{Y_{pq}\log (\theta_{pqgh})+(1-Y_{pq})\log(1-\theta_{pqgh})\r\}  + u_{q  p,h}\log \left(\alpha_{2qh}+C^\prime_{qh}\right) + \const
\end{align*}
Take the expectation under the variational distribution $\widetilde{Q}$:
\begin{align*}
&  \expec_{\widetilde{Q}}\{\log P(\mathbf{Y},\mathbf{Z},
\mathbf{U}\mid \mathbf{B},
\boldsymbol{\beta}_1, \boldsymbol{\beta}_2, \boldsymbol{\gamma}, \mathbf{D},
\mathbf{X}_1, \mathbf{X}_2)\} \\
= & \ u_{q  p,h}\sum_{g=1}^{K_1}\expec_{\widetilde{Q}}(z_{p  q,g})\bigl(Y_{pq}\log(\theta_{pqgh})+(1-Y_{pq})\log(1-\theta_{pqgh})\bigr) + u_{q  p,h}\
\expec_{\widetilde{Q}}\left\{\log\left(\alpha_{2qh}+C_{qh}^\prime\right)\right\}
+ \const
\end{align*}

\subsubsection{M-step}
\label{app:mstep}

\subsubsection*{Lower Bound}
Expression for the lower bound,
\begin{align*}
\mathcal{L}(\tilde{Q}) & \ = \ 
\expec_{\widetilde{Q}}[\log P(\mathbf{Y},\mathbf{Z},\mathbf{U}\mid \mathbf{B}, \boldsymbol{\gamma},\boldsymbol{\beta}_1, \boldsymbol{\beta}_2,
\mathbf{X}_1, \mathbf{X}_2, \mathbf{D})] -\expec_{\widetilde{Q}}[\log \widetilde{Q}(\mathbf{Z},\mathbf{U}\mid
\boldsymbol{\Phi})]\\
& \ = \   \sum_{p\in V_{1}} \biggl[\log\Gamma\left(\xi_{p}\right) - \log\Gamma\left(\xi_{p} + N_{2}\right) \biggr]
+  \sum_{q\in V_{2}} \biggl[\log\Gamma\left(\xi_{q}\right) - \log\Gamma\left(\xi_{q} + N_{1}\right) \biggr] \\
&\quad + \sum_{p\in V_{1}}\sum^{K_1}_{g=1}\biggl[\expec[\log\Gamma(\alpha_{pg}+ C_{pg})] - \log\Gamma(\alpha_{pg})\biggr]
+ \sum_{q\in V_{2}}\sum^{K_2}_{h=1}\biggl[\expec[\log\Gamma(\alpha_{qh}+ C_{qh})] - \log\Gamma(\alpha_{qh})\biggr]\\
&\quad +\sum_{(p,q)\in V_1\times V_2}
\sum^{K_1}_{g=1}\sum^{K_2}_{h=1}
\phi_{pq,g}\phi_{pq,h}\l\{Y_{pq}\log \theta_{pqgh} +(1-Y_{pq})\log(1-\theta_{pqgh})\r\}\\
&\quad -
\sum^K_{g,h=1}  \frac{(B_{gh}-\mu_{gh})^2}{2\sigma_{gh}^2}
-\sum^{J_d}_{j=1}\frac{(\gamma_{j}-\mu_\gamma)^2}{2\sigma^2_\gamma}- \sum^{K_1}_{g=1}\sum^{J_{1x}}_{j=1}
\frac{(\beta_{1gj}-\mu_{\beta_1})^2}{2\sigma^2_{\beta_1}}
- \sum^{K_2}_{h=1}\sum^{J_{2x}}_{j=1}
\frac{(\beta_{2hj}-\mu_{\beta_2})^2}{2\sigma^2_{\beta_2}}
\\
&\quad -
\sum_{(p,q)\in
	V_{t}} \sum_{g=1}^{K_1}\sum_{h=1}^{K_2} \{\phi_{pq,g}\log \phi_{pq,g}-
\phi_{qp,h}\log(\phi_{qp,h}) \}
\end{align*}

\subsubsection*{M-step 1: update for $\mathbf{B}$}

Restricting the lower bound to terms that contain $B_{gh}$ (blockmodel), we obtain
\begin{align*}
\mathcal{L}(\widetilde{Q})
& \ = \ \sum_{p,q\in
	E_{t}}  \sum^{K_1}_{g=1}\sum^{K_2}_{h=1}\phi_{pq,g}\phi_{qp,h}\{Y_{pq}\log \theta_{pqgh} +(1-Y_{pq})\log(1-\theta_{pqgh})\}\\
&\qquad- \sum^K_{g,h=1}\frac{(B_{gh}-\mu_{gh})^2}{2\sigma_{gh}^2} + \const
\end{align*}
Optimize this lower bound with respect to $\mathbf{B}_{gh}$ using a
gradient-based numerical optimization method. The corresponding
gradient is:
\begin{align*}
\frac{\partial\mathcal{L}_{B_{gh}}}{\partial B_{gh}}
& \ = \ \sum _{p,q\in V_1\times V_2}  \phi_{p  q,g} \phi_{qp,h}\left(Y_{pq}-\theta_{pqgh}\right)- \frac{B_{gh}-\mu_{B_{gh}}}{\sigma_{B_{gh}}^2}
\end{align*}

\subsubsection*{M-step 2: update for $\boldsymbol{\gamma}$}

Restricting the lower bound to terms containing
$\boldsymbol{\gamma}$ (dyadic coefficients), and recalling that
$\theta_{pqtgh}=[1+\exp(-B_{gh}-\mathbf{d}_{pqt}\boldsymbol{\gamma})]^{-1}$, then:
\begin{align*}
\mathcal{L}(\tilde{Q})
& \ = \ \sum_{p,q\in V_1\times V_2} 
\sum^{K_1}_{g=1}\sum^{K_2}_{h=1} \phi_{p  q,g}\phi_{qp,h}\l\{Y_{pq}\log \theta_{pqgh}+(1-Y_{pq})\log(1-\theta_{pqgh})\r\}\\
&\quad -\sum^{J_d}_{j}\frac{(\gamma_{j}-\mu_\gamma)^2}{2\sigma^2_\gamma} + \const
\end{align*}
To optimize this expression w.r.t. $\gamma_{j}$ (the $j$th
element of the $\boldsymbol{\gamma}$ vector), we again use a numerical
optimization algorithm based on the following gradient,
\begin{align*}
\frac{\partial \mathcal{L}(\widetilde{Q})}{\gamma_{j}}
& \ = \  \sum_{p,q\in V_1\times V_2}  \sum^{K_1}_{g=1}\sum^{K_2}_{h=1}
\phi_{p  q,g} \phi_{qp,h} d_{pqj}\left(Y_{pq}-\theta_{pqgh}\right)- \frac{\gamma_{j}-\mu_{\gamma}}{\sigma_{\gamma}^2}
\end{align*}

\subsubsection*{M-step 3: update for $\boldsymbol{\beta}_{1}$, $\boldsymbol{\beta}_{2}$}
Let $\alpha_{pg}=\exp\l(\mathbf{x}_{1}^\top\boldsymbol{\beta}_{1g}\r)$, $\xi_{p}=\sum_{g=1}^{K_1}\alpha_{pg}$, $\alpha_{qh}=\exp\l(\mathbf{x}_{2q}^\top\boldsymbol{\beta}_{2h}\r)$, and $\xi_{q}=\sum_{h=1}^{K_2}\alpha_{qh}$. To find the optimal value of $\boldsymbol{\beta}_{1g}$, roll all terms not involving the coefficient vector into a constant:
\begin{align*}
\mathcal{L}(\widetilde{Q})
& \ = \ \sum_{p\in V_{1}}\l[\log\Gamma(\xi_{1p})-\log\Gamma(\xi_{p}+N_{2})\r] \\
& \ \quad + \sum_{p\in V_{1}}\sum^{K_1}_{g=1}\left[\expec_{\widetilde{Q}_2}[\log\Gamma(\alpha_{pg}+C_{pg})]-\log\Gamma(\alpha_{pg})\right]\\
&\qquad - \sum_{g=1}^{K_1}\sum^{J_{1x}}_{j=1} \frac{(\beta_{1gj}-\mu_{\beta_1})^2}{2\sigma^2_{\beta_1}} + \const
\end{align*}
No closed form solution exists for an optimum w.r.t. $\beta_{1gj}$, but a gradient-based
algorithm can be implemented to maximize the above. The corresponding
gradient w.r.t. each element of $\boldsymbol{\beta}_{1g}$ is:
\begin{align*}
\frac{\partial\mathcal{L}(\tilde{Q})}{\partial \beta_{1gj}}
& \ = \ \sum_{p\in V_{1}}\alpha_{pg}x_{1pj}\Bigl(\expec_{\widetilde{Q}_2}[\breve{\psi}(\alpha_{pg}+C_{pg})-\breve{\psi}(\alpha_{pg})]\\
& \  \qquad + \l[\breve{\psi}(\xi_{1p})-\breve{\psi}(\xi_{1p}+N_{1})\r]\Bigr)\\
& \ \qquad - \frac{\beta_{1gj} - \mu_{\beta_1}}{\sigma_{\beta_1}^2}
\end{align*}
where $\breve{\psi}(\cdot)$ is the digamma function. Again,
we can approximate expectations of non-linear functions of random
variables using a zeroth-order Taylor series expansion. The M-step for the regression coefficients of the second family is similarly defined.

\subsection{Stochastic variational inference}

On the $t$th iteration, our algorithm completes the following steps:
\begin{enumerate}
    \item Sample a subset of dyads $E^t \subset E$, with corresponding sets of vertices $V^t_1=\{p: p,q \in E^t\}$ and $V^t_2=\{q: p,q \in E^t\}$.
    \item Update all $\boldsymbol{\phi}_{pq: p,q\in E_t}$ according to Equation~\eqref{eq:phi}, and compute a set of intermediate global count statistics (after normalization),
    \[
    \widehat{C}_{pg} = \frac{N_2}{|V^t_2|}\sum_{q'\in V_2^t}\sum_{h'=1}^{K_2}\phi_{pq',g,h'}; \quad \widehat{C}_{qh} = \frac{N_1}{|V^t_1|}\sum_{p'\in V_1^t}\sum_{g'=1}^{K_1}\phi_{p'q,g',h}
    \]
    weighted to match the amount of information contained in the original network.   
    \item Update global count statistics matrices using an online average that follows an appropriately decreasing step-size schedule:
    \[
     \mathbf{C}^{(t)}_{p} = (1-\rho_{p,t}) \mathbf{C}^{(t-1)}_{p} + \rho_{p,t}\widehat{\mathbf{C}}_{p}; \quad \mathbf{C}^{(t)}_{q} = (1-\rho_{q,t}) \mathbf{C}^{(t-1)}_{q} + \rho_{q,t}\widehat{\mathbf{C}}_{q}
    \]
    where step-size $\rho_{p,t}=(\tau + t)^\kappa$ such that $\tau>0$ and $\kappa \in (0.5, 1]$.
    \item Update values of hyper-parameters $\Lambda = \{\boldsymbol{\beta}, \boldsymbol{\gamma}, \mathbf{B}\}$ by taking an ``online'' step in the direction of the (noisy) Euclidean gradient of $\mathcal{L}(\boldsymbol{\Phi})$ w.r.t $\Lambda$:
    \[
    \lambda^{(t)} = \lambda^{(t-1)} + \rho_{\lambda, t}\nabla_\Lambda \mathcal{L}^{(t)}(\boldsymbol{\Phi})
    \]
    with appropriate gradients given in Appendix. 
\end{enumerate}

Although different dyad sampling heuristics used for Step 1 can result in unbiased gradient estimates \citep[see][for a few examples]{gopalan_blei2013}, we follow the scheme proposed by \cite{dulac_etal2020}, which is both simple to implement and has been found to work well in sparse settings. The procedure is: 
    \begin{enumerate}
        \item Sample node $i$ in $V$ uniformly at random.
        \item Form a set $s_1 = \{i,j : y_{ij}=1, \forall j\in V\}$ (i.e., set of all connected dyads involving $i$). Form $M$ sets $s^m_0=\{i,j : y_{ij}=0, \exists j \in V\}$ (i.e., a set of some disconnected dyads involving $i$), where each set is of equal cardinality, and the disconnected dyads are sampled uniformly at random and with replacement. 
        \item Sample, with equal probability, either $s_1$ or any of the $s_0^m$ sets. This set of dyads constitutes a subnetwork. 
        \end{enumerate}

        In our application, we set $M=10$ and set $|s_0^m|$ be $1/M$ times the number of non-links between $i$ and every other vertex in the network.

        After the algorithm converges, we can recover the mixed-membership vectors by computing their posterior predictive expectations:
\[
\widehat{\pi}_{pg} = \frac{C_{pg} + \alpha_{pg}}{N_2 + \sum_{g'=1}^{K_1}\alpha_{pg'}}, \quad \widehat{\psi}_{qh} = \frac{C_{qh} + \alpha_{qh}}{N_1 + \sum_{h'=1}^{K_2}\alpha_{qh'}}
\]

\subsection{Initial values for $\phi$ and $\psi$}

Implementation of the model requires defining good starting
values for the mixed-membership vectors. While spectral clustering methods offer good
starting values for $\pi$ and $\psi$ in the unipartite setting,
applying it to non-square affiliation matrices poses interesting challenges. To produce high-quality initial values in a viable amount of time we rely on the co-clustering approach of \citep{govaert_clustering_2003}, which estimates a simpler, single-membership SBM using a fast EM algorithm. 

\subsection{Standard error computation}

We obtain measures of uncertainty around regression coefficients
$\boldsymbol{\beta}$ and $\boldsymbol{\gamma}$ by evaluating the
curvature of the lower bound at the estimated optimal values for these
hyper-parameters. When considering terms that involve these
hyper-parameters, the lower bound reduces to the expected value of the
log-posterior taken with respect to the variational distribution
$\widetilde{Q}$. Thus, evaluating the Hessian of the lower
bound (and the corresponding covariance matrix of the
hyper-parameters) requires evaluating that expectation. Details of the
required Hessian are below.

\subsubsection{Hessian for $\gamma$}
Restricted to terms that involve $\boldsymbol{\gamma}$, the typical element of the required Hessian is given by 
\begin{align*}
    \frac{\partial^2\mathcal{L}(\widetilde{Q})}{\partial\gamma_j\partial\gamma_{j\prime}}&=\sum_{p,q\in V_1\times V_2}-d_{pqj}d_{pqj\prime}\left[\bar{\theta}_{pq}(1-\bar{\theta}_{pq})\right] - \sigma_{\gamma}^{-2}\delta_{jj\prime}
\end{align*}
where $\delta_{jj\prime}$ is the Kronecker delta function, and the term
\[
\bar{\theta}_{pq} = \expec_{\widetilde{Q}}[\theta_{pq}]=\boldsymbol{\hat{\phi}}_{pq}^\top\mathbf{\hat{B}}\boldsymbol{\hat{\phi}}_{qp} + \mathbf{d}_{pq}^\top\boldsymbol{\hat{\gamma}}
\]
is a closed-form solution to the expectation over the variational distribution of the model's parameters. 
\subsubsection{Hessian for $\beta_1$ and $\beta_2$}
In turn, and focusing on Family 1 coefficients, we can characterize the Hessian of the lower bound w.r.t. $\boldsymbol{\beta}_1$ with
\begin{align*}
\frac{\partial^2\mathcal{L}(\widetilde{Q})}{\partial\beta_{gj}\partial\beta_{gj\prime}} &= \sum_{p\in V_1}x_{pj}x_{pj\prime}\alpha_{pg}\biggl(\breve{\psi}(\xi_{p})-\breve{\psi}(\alpha_{pg})+\expec_{\widetilde{Q}}[\breve{\psi}(\alpha_{pg}+C_{pg})]-\breve{\psi}(\xi_{p}+N_2)\\
&\qquad + \alpha_{pg} \left(\breve{\psi}_1(\xi_{pg})-\breve{\psi}_1(\alpha_{pg})+\expec_{\widetilde{Q}}[\breve{\psi}_1(\alpha_{pq}+C_{pq})]-\breve{\psi}_1(\xi_{p}+N_2)\right)\biggr)\\
&\qquad - \sigma^{-2}_{\beta_1}\delta_{jj\prime}
\end{align*}
for coefficients in the same group $g$, and 
\[
\frac{\partial^2\mathcal{L}(\widetilde{Q})}{\partial\beta_{gj}\partial\beta_{hj\prime}} = \sum_{p\in V_1}x_{pj}x_{pj\prime}\alpha_{pg}\alpha_{ph}\left(\breve{\psi}_1(\xi_{p})-\breve{\psi}_1(\xi_{p}+N_2)\right)
\]
for coefficients associated with different latent groups $g$ and $h$. As before, we use $\breve{\psi}(\cdot)$ to denote the digamma function, and $\breve{\psi}_1(\cdot)$ to denote the trigamma function.

Unlike the Hessian for $\boldsymbol{\gamma}$, there are no closed-form solutions for the expectations involved in these expressions. To approximate them, we take $S$ samples from the Poisson-Binomial distribution of $C_{pg}$, $C^{(s)}_{pg}, s\in 1,\ldots,S$, and let
\[
\expec_{\widetilde{Q}}[\breve{\psi}(\alpha_{pg}+C_{pg})]\approx \frac{1}{S}\sum_s\breve{\psi}(\alpha_{pq}+C^{(s)}_{pq});\quad \expec_{\widetilde{Q}}[\breve{\psi}_1(\alpha_{pq}+C_{pq})] \approx \frac{1}{S}\sum_s\breve{\psi}_1(\alpha_{pq}+C^{(s)}_{pq})
\]
The Hessian for the coefficients associated with Family 2, $\boldsymbol{\beta}_2$, is similarly approximated.

\section{Simulation results}
\label{app:sims}

\paragraph{Setup.}
We simulate bipartite networks with unbalanced numbers of
Senator and Bills nodes under 6 different scenarios, defined by overall
network size and difficulty of the mixed-membership learning
problem. More specifically, we define small (i.e. 300 total nodes) and
large (i.e. 3000 total nodes) networks, each with twice as many Bill
nodes as there are Senator nodes. In all instances, we define the
edge-generating process according to our model, using
$K_1=K_2=2$ latent groups for each of the node types, a single monadic
predictor drawn independently from $N(0, 1.5)$, and a
single irrelevant dyadic predictor drawn from a standard Normal distribution.

\begin{table}[h]
  \centering
  \begin{tabular}{lcc}
    \toprule
    \textbf{Scenario Difficulty} & \textbf{Blockmodel} &
                                                         \textbf{Monadic
                                                         Coefficients}\\
    \midrule
    Easy & 
    $\begin{bmatrix} 
    0.85 & 0.01\\ 
    0.01 & 0.99
    \end{bmatrix}$
    & $\begin{bmatrix}-4.50 & -4.50\\ 0.00 & 0.00\end{bmatrix}$
    \\
    Medium & $\begin{bmatrix}0.65 & 0.35\\ 0.20 & 0.75\end{bmatrix}$ &$\begin{bmatrix}0.05 & 0.75\\ -0.75 & -1.00\end{bmatrix}$\\
    Hard & $\begin{bmatrix}0.65 & 0.40\\ 0.50 & 0.45\end{bmatrix}$ &$\begin{bmatrix}0.00 & 0.00\\ -0.75 & -1.00\end{bmatrix}$\\
    \bottomrule
  \end{tabular}
  \caption{Simulation scenarios, defining various levels of estimation
    difficulty}   \label{tab:simvals}
\end{table}

To simulate different levels of estimation difficulty, we vary both the blockmodel and the coefficients associated
with the mixed-membership vectors, which are set to be equal across
the two node types. In the
``easy'' scenario, memberships are barely mixed, and there is a clear
difference in edge probabilities between different groups of the
different node types. In contrast, the ``hard'' scenario is such
that all nodes have a roughly equal probability of instantiating each
block, and there is little difference in the probabilities of forming
edges between blocks, as given by the blockmodel. The ``medium''
scenario offers a more realistic, in-between estimation problem. The
specific values for scenarios are given in
Table~\ref{tab:simvals}.

\begin{figure}[t]
    \centering\spacingset{1}
    \includegraphics[width=\textwidth]{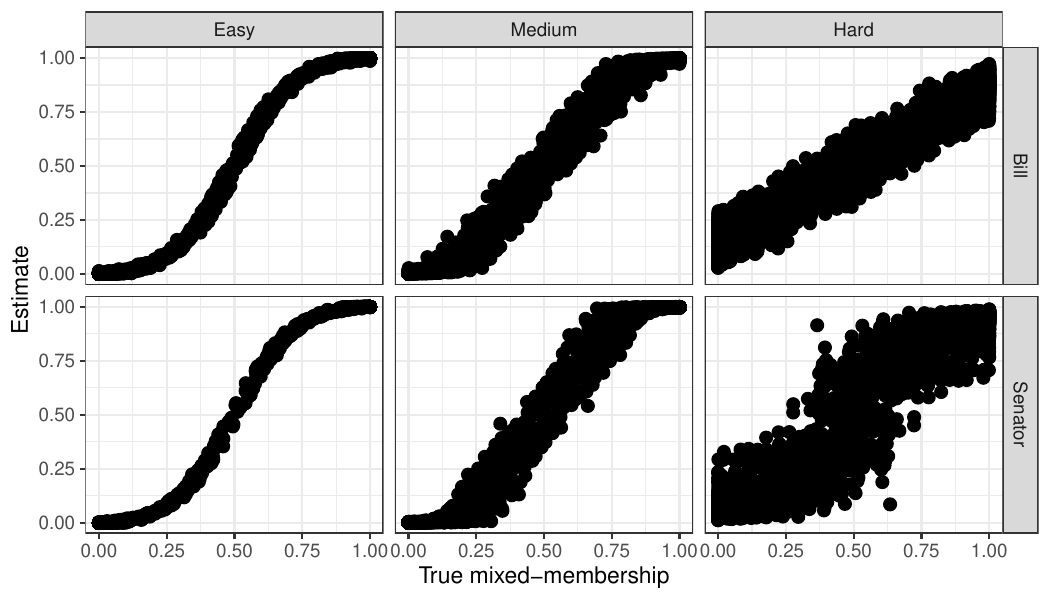}
    \caption{\textbf{Mixed-Membership Recovery}: Estimated and true mixed-membership vectors under the easy, medium, and
      hard estimation scenarios. In all instances, recovery is
      excellent.} 
    \label{fig:SimCors}
\end{figure}

\paragraph{Results.}
We begin by evaluating the accuracy of mixed-membership estimation by
comparing true and estimated mixed-membership vectors (after
re-labeling the latter to match the known, simulated group labels
using the Hungarian algorithm). Correlations across node types and
difficulty scenarios are demonstrated in Figure~\ref{fig:SimCors}.
Overall, our model retrieves these mixed-membership vectors with a
high degree of accuracy --- even in regimes in which block memberships
play a small role in the generation of edges, and regardless of
whether there is an asymmetry in the number of nodes in each family.

\begin{figure}[t]
    \centering \spacingset{1}
    \includegraphics[width=\textwidth]{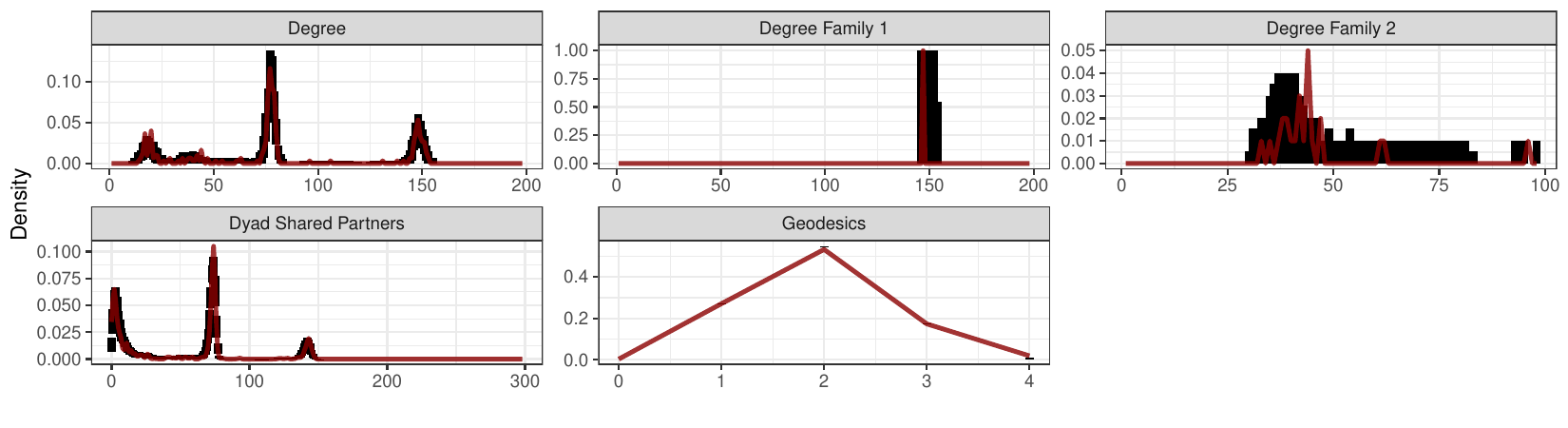}
    \caption{\textbf{Posterior Predictive Goodness-Of-Fit}: The figure shows, for a randomly chosen simulated scenario, the extent to which the model can recover structural features of the observed network. The solid red line traces the observed distribution of the different network motifs; solid black rectangles show the central 90\% distribution of values observed across 100 network replicates, obtained from the estimated model posterior. A good fit is indicated by lines that always fall within black regions.}
    \label{fig:SimGOF}
\end{figure}

Next, we evaluate the accuracy of estimated node-level and dyadic coefficients by simulating derived quantities of interest based on them, and compare
these simulated quantities to their
true counterparts, as one would when conducting a goodness-of-fit
analysis based on posterior predictive distributions. As is typical in
network modeling, these derived quantities are structural features of
the network \citep[e.g.][]{hunter2008}. Figure~\ref{fig:SimGOF}
depicts three such features --- node degree (by node family), the number of partners shared by each dyad, and the minimum geodesic distance between nodes in the network. In each panel, the red line traces the true
distribution of these network statistics, while the black vertical
bars track their distribution across 100 network replicates, each
generated using the estimated coefficients. If the latter are
correctly estimated, network replicates should have characteristics
that reflect that on which the estimation is based, and the red line
should fall squarely within each vertical black rectangle. Overall, network characteristics are well recovered by our model, although recovery of the degree distribution for the bill nodes (i.e., the largest family) is less accurate than that of the legislator nodes (i.e., family with fewer vertices). 

\begin{figure}[h]
  \centering \spacingset{1}
  \includegraphics[width=0.65\textwidth]{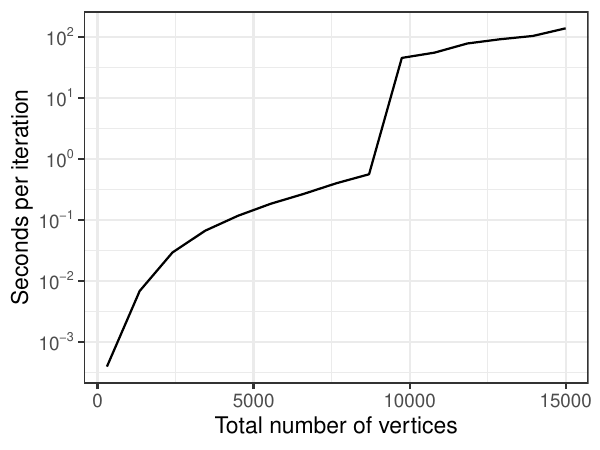}
  \caption{\textbf{Time per iteration for networks of different sizes}:
    For medium difficulty scenario networks, the
    plot shows time per iteration (in seconds) taken to fit our model
    to networks of different sizes.} 
  \label{fig:SimTime}
\end{figure}

To evaluate scalability of our approach, we conduct simulations under
the ``medium'' difficulty scenario, as described above. We hold all
conditions constant, and increase the total size of the vertex set
from 300 to 15,000, keeping twice as many vertices in the largest
family as in the smallest family. In all instances, we let the models
run until convergence, using the stochastic variational inference
procedure described above (in each iteration, we sample 40\% of
nodes). Models took between ~5 seconds and ~9 hours to fully estimate,
taking anywhere between 100 and 450 iterations to converge. As the
time to convergence is affected by the stochastic nature of the
estimation, Figure~\ref{fig:SimTime} presents the time per iteration
(in seconds) taken to fit our model to networks of different
sizes. Overall, although time per iteration increases as the network
size grows, even the largest network in our simulation can be reliably
fit in under 10 hours on a desktop computer.
  
\begin{figure}[h]
  \centering \spacingset{1}
  \includegraphics[width=\textwidth]{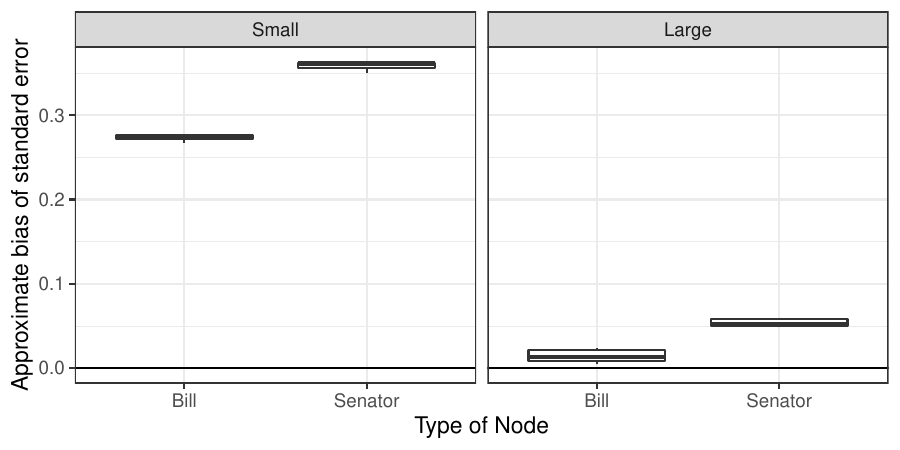}
  \caption{\textbf{Approximate Bias in Standard Errors}: For each simulated
    network, the figure shows the extent to which our approximate
    standard errors differ from the standard deviation of
    coefficients estimated on simulated networks.} 
  \label{fig:covSim}
\end{figure}  
  
Finally, we evaluate the frequentist properties of our estimates of
uncertainty in regression parameters by evaluating the extent to which
they reflect the variability we can expect from repeated network
sampling. To do so, we sample 100
networks from each of our 6 scenarios, for a total of 600 simulated networks. 
 Figure~\ref{fig:covSim} shows, for each simulation scenario,
the difference between our standard errors and the standard deviation
across coefficients estimated on each of the network replicates. For
small networks, our standard errors can be conservative ---
particularly for the set of coefficients associated with the smaller
group of Senator nodes. As the number
of nodes increases, however, our estimated uncertainty more
accurately reflects the variability we would expect to see under
repeated sampling.

\FloatBarrier

\section{Additional empirical results}

\subsection{Cosponsorship degree distributions 107th Senate}
\label{app:degrees}

\begin{figure}[t!]
  \centering \spacingset{1}
  \includegraphics[width=\textwidth]{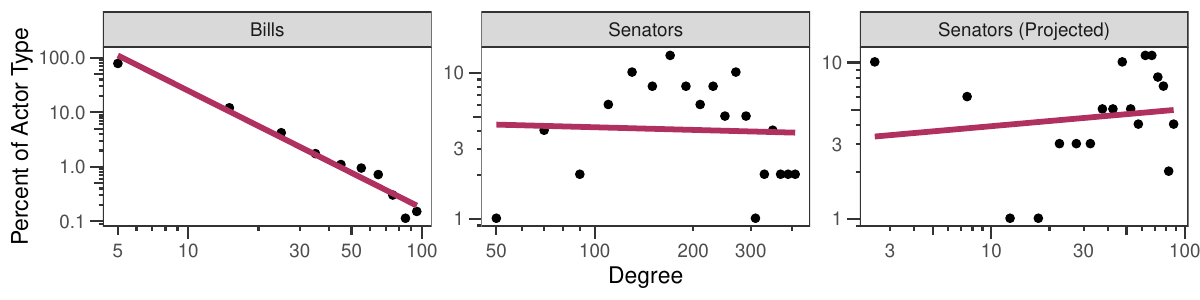}
  \caption{\textbf{Bill and Senator degree distributions.} Degree
    distributions presented for Bills (left) and Senators (center, in
    the bipartite network, and right, in the projected network) with a
    power law distribution overlaid as a red curve (approximated by a
    linear log-log model). Whereas the power law fits the bill degree
    distribution quite well, the degree distributions among senators
    (both in the bipartite graph and in the projected network) differ
    dramatically from it, illustrating the kind of heterogeneity that
    can be lost when aggregating over bills in the projection from
    bipartite to unipartite.}
  \label{fig:senbill-degree-distr}
\end{figure}

Figure~\ref{fig:senbill-degree-distr} presents Senator and Bill degree
distributions from the 107th Congress. Bipartite degree distribution
calculations differ from unipartite ones in that they are separately
conducted for each family, so that senators can display different
degree distributions compared to bills. Previous studies of
cosponsorship patterns in the U.S. congress have found this to be the
case \citep[e.g.][]{fowler_legislative_2006}, and our data reveal
similar differences.

Figure~\ref{fig:senbill-degree-distr} displays a summary of these
distributions, plotting the midpoints of the degree histograms for
each vertex type. We plot both degrees and their observed relative
frequencies (expressed as percentages) in the log scale. When degree
distributions follow the common power-law distribution (whereby
$p(x)\propto x^{-\lambda}$ for a given degree $x$ and $\lambda>0$)
that many other networks exhibit, such log-log plots tend to align
with negatively-sloped linear predictions.

This is clearly the case for the degree distribution of bills,
depicted on the left panel of Figure~\ref{fig:senbill-degree-distr}.
The plot also includes a red line with the predictions of linear
log-log model, which shows a good approximation to the pure power-law
model if the estimated slope is negative. For bills, the fit of a
power-law is almost perfect, indicating that while most bills tend to
attract few cosponsors, there is a long and heavy tail of bills
attracting a large number of them. It is precisely this heterogeneity
that can result in substantial aggregation bias when projecting the
originally bipartite network. It also means that the cosponsorship
network is likely to exhibit \emph{scale invariance} for the
distribution of bill cosponsors (i.e., we can expect to see a
similarly shaped degree distribution if we consider a subset of bills)
--- justifying the analysis of subnetworks formed by sampling the set
of bills (as we do below).\footnote{As pointed out by
  \citet[fowler_legislative_2006], the distribution also exhibits an
  interesting deviation from a typical power-law right around the 50
  senator mark, indicating the strategic value of having a majority of
  senators cosponsoring a piece of legislation.}

In contrast, the degree distributions of Senators (i.e., the
distributions over the number of bills Senators cosponsor) are far
from being accurately described by a power-law.  This is indicated by
the fact that the red line does not fit to the points well on the
central and right panels of
Figure~\ref{fig:senbill-degree-distr}. Indeed, these distributions are
quite different from that of bills, suggesting that there is
substantial heterogeneity in the number and strength of connections
between senators.  This highlights the importance of considering the
entire set of senators when studying the structural characteristics of
the cosponsorship network, and considering any subset of legislators
could result in a misrepresentation of the collaboration network. This
difference with respect to the degree distribution of bills also
highlights the kind of information that is lost when bills are
aggregated over in the process of projecting from bipartite to
unipartite networks, as is also evident on the right-most panel of
Figure~\ref{fig:senbill-degree-distr}, which appears now to be a
mixture of two distributions.

The bipartite network also can accomodate certain types of statistics
for descriptive analysis that is not applicable for unipartite
networks.  They include within-family edge-shared partners and
family-specific k-stars. Conversely, several common network statistics
for unipartite graphs do not apply in bipartite network settings, such
as triangles (bipartite graphs cannot have triangles
\citep{promel2002note}), or must be adjusted, such as path lengths
(which must be even) or minimum degrees \citep{liu2018cycle}. These
considerations will play a role when conducting posterior predictive
checks of model fit, as they typically rely on evaluating how well a
model captures these and other structural features of the modeled
network.

\subsection{Model performance comparison}
\label{app:ergm_comp}

We compare our proposed approach to the most popular, readily
available alternative model for bipartite networks: the bipartite
ERGM, implemented in the R package \texttt{ergm}
\citep{handcock:etal:2023, hunter:etal:2008, krivitsky:etal:2023}. The
bipartite ERGM uses a set of constraints and bipartite-specific
network statistics to adapt the canonical, one-model model to
bipartite networks. Our goal is to evaluate how well the bipartite
ERGM can predict the cosponsorship network of the 107th Senate, using
network statistics only. We then compare it to how biMMSBM can fit the
same data using only the blockmodel and latent mixed membership
vectors.

While we tried fitting the ERGM to the full dataset, we found all of
the specifications we tried resulted in failed convergence; the lack
of scalability appears to be a major limitation of ERGM. As a result,
we focus on a subgraph formed by a random sample of 10\% of observed
edges. To this subgraph, we fit a model that includes a term for the
edge density, the census of 3-stars among senators (i.e., stars
involving exactly 1 bill and three senators), a geometrically weighted
census of dyad-shared partners among bills (i.e., the distribution
over numbers of shared senators for any pair of bills), and
geometrically weighted degree distributions for both senators and
bills.\footnote{We arrived at this particular specification through
  \emph{much} trial and error, iterating over specifications that
  invariably hung or failed to converge. This failure-prone process,
  often elided from descriptions of empirical exercises that rely on
  ERGM-type models, is sometimes touted as a feature. Arriving at a
  specification that works, however, can only offer a weak proof of
  existence, and even a perfectly specified model can result in an
  ill-defined probabilistic models \citep[see, for
  instance,][]{chatterjee2013}.} The latter terms are so-called
dyad-dependent terms, and are defined specifically for bipartite
networks \citep{wang2009, wang2013}. After fitting the ERGM, all
measures of MCMC performance indicated convergence.

For biMMSBM, we use 5 latent communities in each family (a number
arrived at by evaluating the AUROC generated by alternative models, as
we do in the main analysis. We compare models with $2-2$, $3-3$,
$3-5$, and $5-5$ latent communities).

\begin{figure}[ht]
  \centering \spacingset{1}
  \includegraphics[scale=0.9]{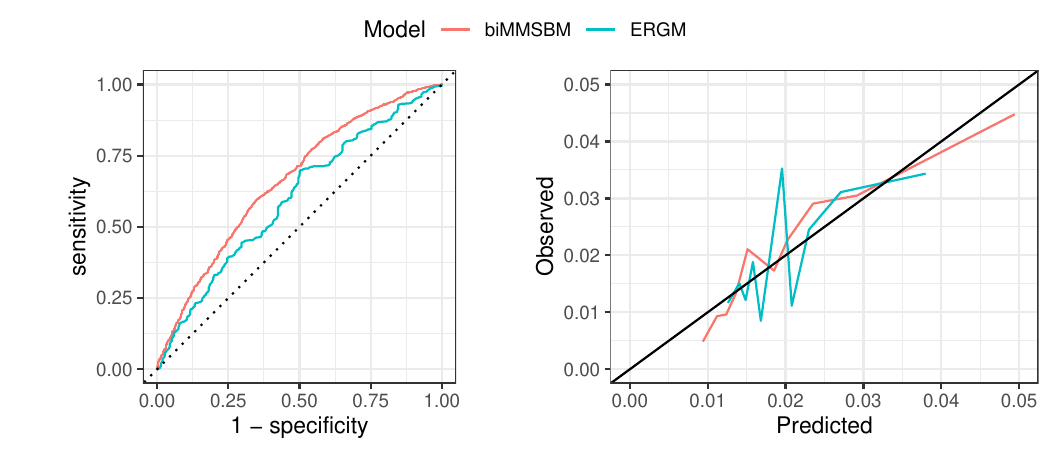}
  \caption{Measures of predictive accuracy of bipartite ERGM and
    biMMSBM on subset of cosponsorship network in the 107th
    Senate. The left panel shows the ROC curves for the ERGM (blue)
    and biMMSBM (salmon); curves further from the 45-degree reference
    line indicate better model classification accuracy. The right
    panel shows the calibration of the same models, with lines closer
    to the 45-degree line indicating a better match between predicted
    edge probabilities and observed edge proportions (that is, better
    calibration). Using both criteria, biMMSBM offers a better
    predictive fit to the cosponsorship data.} 
  \label{fig:ergm_comp}
\end{figure}

Figure~\ref{fig:ergm_comp} offers evidence of better predictive
accuracy obtained by biMMSBM on this network, as indicated by a both a
higher overall AUROC (0.66 vs. 0.59 obtained by the ERGM; left panel),
and far better calibration of predicted probabilities \citep[after
running a Platt correction for both sets of
predictions;[]{platt:1999}, as indicated by the alignment of the
biMMSBM set of predicted probabilities with the corresponding
empirical proportions of edges (right panel). These are both in-sample
measures of fit, as the fragilityt of the ERGM estimation prevented us
from performing an out-of-sample evaluation.

\subsection{Goodness of Fit}
\label{sec:gof}

Additionally, besides overall network statistics, we assess predictive quality using two metrics: accuracy, quantified by the area under the receiver operating characteristic curve (ROC), and calibration, comparing observed frequencies in a test set with model-predicted probabilities.

\begin{figure}[t]
  \centering \spacingset{1}
  \includegraphics[width=\textwidth]{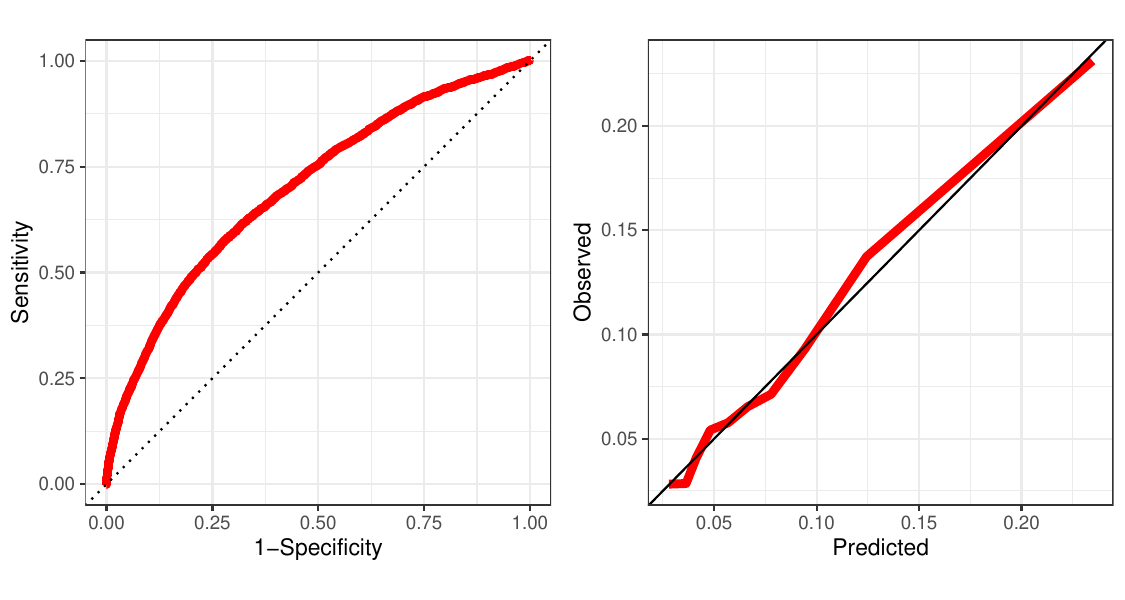}
  \caption{\textbf{Out-of-Sample goodness-of-fit based on edge
      prediction quality:} The figure demonstrates that our biMMSBM
    model with $K_1=K_2=3$ latent groups per family can
    adequately predict observed edges in the bipartite cosponsorship
    network. The left panel shows the Receiver-Operating
    Characteristic Curve in red, with curves closer to the upper-left
    corner indicating better predictive accuracy (here, the area under
    the curve is 0.70, out of a maximum of 1). The right panel shows
    the calibration of predicted probabilities (after standard Platt
    correction) in red, aligning with the 45-degree line for better predictive calibration.}
  \label{fig:gof_outcome}
\end{figure}

In Figure~\ref{fig:gof_outcome}, we show the out-of-sample ROC (left
panel) as well as the prediction probability calibration plot (right
panel).\footnote{We evaluate calibration after applying the standard
  Platt correction, which fits a logistic regression of observed
  outcomes on the uncorrected predicted probabilities, and use the
  transformed scores \citep{platt:1999, rosenman:etal:2023}.} The area
under the ROC curve is .7 (with the red curve on the left
bending away from the dashed diagonal), and the model's predictions
are well calibrated (with the red curve on the right almost aligning
with the solid black diagonal).

\subsection{Model outputs}
\subsubsection{Group memberships}
\begin{table}[h!tb] \spacingset{1}
\centering
\begin{tabular}[t]{lll}
\toprule
\cmidrule(l{3pt}r{3pt}){1-3}
\textbf{Senior Democrats} & \textbf{Senior Republicans} & \textbf{Junior Power Brokers}\\
\midrule
\cellcolor{gray!6}{Byrd, Robert C. [WV]} & \cellcolor{gray!6}{Helms, Jesse [NC]} & \cellcolor{gray!6}{Corzine, Jon [NJ]}\\
Inouye, Daniel K. [HI] & Thurmond, Strom [SC] & Carnahan, Jean [MO]\\
\cellcolor{gray!6}{Hollings, Ernest F. [SC]} & \cellcolor{gray!6}{Lott, Trent [MS]} & \cellcolor{gray!6}{Carper, Thomas R. [DE]}\\
Kennedy, Edward M. [MA] & Stevens, Ted [AK] & Dayton, Mark [MN]\\
\cellcolor{gray!6}{Breaux, John B. [LA]} & \cellcolor{gray!6}{Cochran, Thad [MS]} & \cellcolor{gray!6}{Miller, Zell [GA]}\\
\addlinespace
Sarbanes, Paul S. [MD] & Hatch, Orrin G. [UT] & Clinton, Hillary Rodham [NY]\\
\cellcolor{gray!6}{Biden Jr., Joseph R. [DE]} & \cellcolor{gray!6}{Domenici, Pete V. [NM]} & \cellcolor{gray!6}{Bayh, Evan [IN]}\\
Baucus, Max [MT] & Grassley, Charles E. [IA] & Nelson, E. Benjamin [NE]\\
\cellcolor{gray!6}{Akaka, Daniel K. [HI]} & \cellcolor{gray!6}{Smith, Bob [NH]} & \cellcolor{gray!6}{Stabenow, Debbie [MI]}\\
Leahy, Patrick J. [VT] & Nickles, Don [OK] & Feingold, Russell D. [WI]\\
\bottomrule
\end{tabular}
\caption{\textbf{Senators with largest mixed-membership probabilities in each latent group.}}
\label{tab:topsenators}
\end{table}

\begin{table}[h!tb] \spacingset{1}
 \rowcolors{2}{gray!6}{white}
\begin{tabular}{@{}p{2in}p{2in}p{2in}@{}}
\toprule
\multicolumn{1}{c}{\textbf{Contentious Bills}} &
  \multicolumn{1}{c}{\textbf{Bipartisan Resolutions}} &
  \multicolumn{1}{c}{\textbf{Uncontroversial}} \\ \midrule
SN\_107\_2842 Senior Self-Sufficiency Act; allocation of \$1M grants to provide supportive services to elderly in noninstitutional residences. &
  SJ\_107\_1 Joint resolution proposing amendment to Constitution of US relating to voluntary school prayer. &
  SJ\_107\_22 Joint resolution expressing sense of Senate/House regarding terrorist attacks on September 11, 2001. \\
SN\_107\_1548 Bioterrorism Awareness Act. &
  SE\_107\_82 Resolution authorize production of records by Permanent Subcommittee on Investigations of Committee on Governmental Affairs and representation by Senate Legal Counsel. &
  SE\_107\_169 Resolution relative to death of Honorable Mike Mansfield, formerly Senator from Montana. \\
SN\_107\_2899 Atchafalaya National Heritage Area Act. &
  SE\_107\_9 Resolution notifying Pres. of election of President pro tempore. &
  SE\_107\_292 Resolution expressing support for Pledge of Allegiance. \\
SN\_107\_3176 Renewal Community Tax Benefit Improvement Act. &
  SE\_107\_10 Resolution notifying House of election of President pro tempore. &
  SE\_107\_354 Resolution relative to death of Paul Wellstone, Senator from Minnesota. \\
SN\_107\_3045 Finger Lakes Initiative Act of 2002. &
  SE\_107\_54 Resolution authorizing expenditures by the committees of the Senate. &
  SE\_107\_160 Resolution designating Oct 2001, as ``Family History Month''. \\
SN\_107\_1637 Bill to waive limitations in paying costs of projects in response to 9/11. &
  SE\_107\_77 Resolution to authorize production of records by Permanent Subcommittee on Investigations of Committee on Governmental affairs. &
  SE\_107\_66 Resolution regarding release of 24 US military personnel currently detained by China. \\
SN\_107\_2634 225th Anniversary of American Revolution Commemoration Act (establishes educational program). &
  SE\_107\_28 Resolution to authorize testimony/legal representation in State of Idaho V. Fredrick Leroy Leas, Sr. &
  SN\_107\_321 Family Opportunity Act of 2002. \\
SN\_107\_2054 Nationwide Health Tracking Act of 2002. &
  SE\_107\_84 Resolution to authorize representation by Senate Legal Counsel in Timothy A. Holt V. Phil Gramm. &
  SN\_107\_677 Housing Bond and Credit Modernization and Fairness Act of 2001. \\
SN\_107\_1649 Vancouver National Historic Reserve Preservation Act of 2002 (assigns funding). &
  SC\_107\_10 Concurrent resolution regarding the Republic of Korea's unlawful bailout of Hyundai Electronics. &
  SN\_107\_697 Railroad Retirement and Survivors' Improvement Act of 2001. \\
SN\_107\_3092 Children's Health Protection and Eligibility Act of 2002 (authorizes expenditures). &
  SJ\_107\_4 Joint resolution proposing amendment to Constitution relating to contributions and expenditures intended to affect elections. &
                                                                                                                                             SN\_107\_1707 Medicare Physician Payment Fairness Act of 2001. \\ \bottomrule
\end{tabular}
 \caption{{\bf Bills with largest mixed-membership probabilities in each latent group.}}
 \label{tab:topbills2}
\end{table}

   \subsubsection{Model estimated coefficients}
 

\begin{table}[H] \spacingset{1}
\centering
\begin{tabular}[t]{lccc}
\toprule
  \textbf{Blockmodel estimates} & 1 Contentious Bills & 2 Bipartisan Resolutions & 3 Uncontroversial Bills\\
\midrule
1 Democrat & 0.2474 & 0.7844 & 1.0000\\
2 Republican & 0.1918 & 0.6765 & 0.9999\\
3 Junior Power Brokers & 0.2181 & 0.8095 & 1.0000\\
\bottomrule
\end{tabular}
\begin{tabular}[t]{cclcc}
\toprule
 &  & \multicolumn{1}{c}{Coefficient Name} & Estimate & SE\\
\midrule
\addlinespace[0.3em]
\multicolumn{5}{l}{\textbf{Dyadic predictors}}\\
\hspace{1em} &  & No reciprocity history & -6.6168 & 0.1040\\
\cmidrule{3-5}
\hspace{1em} &  & Log proportional reciprocity & 2.0015 & 0.0146\\
\cmidrule{3-5}
\hspace{1em} &  & Shared committee & 1.5037 & 0.0504\\
\midrule
\addlinespace[0.3em]
\textbf{Model Summary Statistics} & & & & \\
Lower bound & -591.1986 & & & \\
Number of dyads & 260667 & & & \\
\% Obs. in Each Family 1 Block & 0.138 & 0.285 & 0.576\\
\% Obs. in Each Family 2 Block & 0.578 & 0.360 & 0.062\\
\bottomrule
\end{tabular}
\caption{\textbf{biMMSBM Estimated Coefficients: Blockmodel, Dyadic.} Point estimates and approximate standard errors of coefficients in the dyadic regression equation show that reciprocity norms and shared committee duties between bill sponsors and potential co-sponsors enhance co-sponsorship likelihood.
}
\label{tab:estimcoef1}
\end{table}

\spacingset{1.2}
\begin{table}[h]
\fontsize{6}{8}\selectfont
\begin{minipage}[t]{0.5\textwidth}
\begin{tabular}{ccccc}
\toprule
 & Group & Coefficient Name & Estimate & SE\\
\midrule
\addlinespace[0.3em]
\multicolumn{5}{l}{\textbf{Senator predictors}}\\
\hspace{1em} & 1 Democrat & Intercept & 4.7867 & 1.3322\\
\cmidrule{3-5}\nopagebreak
\hspace{1em} &  & Seniority & 0.5993 & 0.6318\\
\cmidrule{3-5}\nopagebreak
\hspace{1em} &  & Ideology 1 & -5.5500 & 1.2965\\
\cmidrule{3-5}\nopagebreak
\hspace{1em} &  & Ideology 2 & 5.0876 & 1.2912\\
\cmidrule{3-5}\nopagebreak
\hspace{1em} &  & Party-Republican & -3.7291 & 1.3772\\
\cmidrule{3-5}\nopagebreak
\hspace{1em} &  & Sex-Male & 0.4388 & 1.2850\\
\cmidrule{2-5}\nopagebreak
\hspace{1em} & 2 Republican & Intercept & 1.5207 & 1.4515\\
\cmidrule{3-5}\nopagebreak
\hspace{1em} &  & Seniority & 0.3924 & 0.6318\\
\cmidrule{3-5}\nopagebreak
\hspace{1em} &  & Ideology 1 & 10.3447 & 1.2935\\
\cmidrule{3-5}\nopagebreak
\hspace{1em} &  & Ideology 2 & -2.2911 & 1.2906\\
\cmidrule{3-5}\nopagebreak
\hspace{1em} &  & Party-Republican & 6.1726 & 1.4587\\
\cmidrule{3-5}\nopagebreak
\hspace{1em} &  & Sex-Male & 0.8331 & 1.2850\\
\cmidrule{2-5}\nopagebreak
\hspace{1em} & 3 Junior Power Brokers & Intercept & 20.3757 & 1.3299\\
\cmidrule{3-5}\nopagebreak
\hspace{1em} &  & Seniority & -0.5417 & 0.6318\\
\cmidrule{3-5}\nopagebreak
\hspace{1em} &  & Ideology 1 & -3.3304 & 1.2916\\
\cmidrule{3-5}\nopagebreak
\hspace{1em} &  & Ideology 2 & -2.2946 & 1.2901\\
\cmidrule{3-5}\nopagebreak
\hspace{1em} &  & Party-Republican & -0.5346 & 1.3436\\
\cmidrule{3-5}\nopagebreak
\hspace{1em} &  & Sex-Male & -0.0602 & 1.2847\\
\bottomrule 
\multicolumn{5}{l}{\textbf{Bill predictors}}\\
\hspace{1em} & 1 Contentious Bills & Intercept & 4.2305 & 0.1224\\
\cmidrule{3-5}\nopagebreak
\hspace{1em} &  & Topic:Legal & 0.6390 & 0.1033\\
\cmidrule{3-5}\nopagebreak
\hspace{1em} &  & Topic:Social programs Public goods & -0.0517 & 0.0650\\
\cmidrule{3-5}\nopagebreak
\hspace{1em} &  & Topic:Security & 0.8542 & 0.1557\\
\cmidrule{3-5}\nopagebreak
\hspace{1em} &  & Topic:Gov operations & 0.6023 & 0.1179\\
\cmidrule{3-5}\nopagebreak
\hspace{1em} &  & Topic:Other & -1.5853 & 0.0729\\
\cmidrule{3-5}\nopagebreak
\hspace{1em} &  & Sponsor Seniority & -0.0739 & 0.0057\\
\cmidrule{3-5}\nopagebreak
\hspace{1em} &  & Sponsor Ideology 1 & 0.6046 & 0.2433\\
\cmidrule{3-5}\nopagebreak
\hspace{1em} &  & Sponsor Ideology 2 & -0.2777 & 0.1449\\
\cmidrule{3-5}\nopagebreak
\hspace{1em} &  & Sponsor Party-Republican & -0.6889 & 0.1783\\
\cmidrule{3-5}\nopagebreak
\hspace{1em} &  & Sponsor Sex-Male & -0.4856 & 0.0842\\
\cmidrule{3-5}\nopagebreak
\end{tabular}

\end{minipage} \hspace{1cm} 
\begin{minipage}[t]{0.5\textwidth}
\begin{tabular}{ccccc}
\toprule
 & Group & Coefficient Name & Estimate & SE\\
\midrule
\addlinespace[0.3em]
\multicolumn{5}{l}{\textbf{Bill predictors (continued)}}\\
\hspace{1em} &  & Second Phase & 0.8622 & 0.0789\\
\cmidrule{3-5}\nopagebreak
\hspace{1em} &  & Third Phase & 0.7629 & 0.0553\\
\cmidrule{2-5}\nopagebreak
\hspace{1em} & 2 Bipartisan Resolutions & Intercept & 3.3078 & 0.1226\\
\cmidrule{3-5}\nopagebreak
\hspace{1em} &  & Topic:Legal & 0.6907 & 0.1036\\
\cmidrule{3-5}\nopagebreak
\hspace{1em} &  & Topic:Social programs Public goods & -0.1533 & 0.0652\\
\cmidrule{3-5}\nopagebreak
\hspace{1em} &  & Topic:Security & 0.9016 & 0.1552\\
\cmidrule{3-5}\nopagebreak
\hspace{1em} &  & Topic:Gov operations & 0.6219 & 0.1178\\
\cmidrule{3-5}\nopagebreak
\hspace{1em} &  & Topic:Other & -1.1869 & 0.0731\\
\cmidrule{3-5}\nopagebreak
\hspace{1em} &  & Sponsor Seniority & -0.0469 & 0.0057\\
\cmidrule{3-5}\nopagebreak
\hspace{1em} &  & Sponsor Ideology 1 & 0.3052 & 0.2438\\
\cmidrule{3-5}\nopagebreak
\hspace{1em} &  & Sponsor Ideology 2 & -0.0983 & 0.1457\\
\cmidrule{3-5}\nopagebreak
\hspace{1em} &  & Sponsor Party-Republican & -0.1289 & 0.1786\\
\cmidrule{3-5}\nopagebreak
\hspace{1em} &  & Sponsor Sex-Male & -0.2849 & 0.0842\\
\cmidrule{3-5}\nopagebreak
\hspace{1em} &  & Second Phase & 0.5371 & 0.0791\\
\cmidrule{3-5}\nopagebreak
\hspace{1em} &  & Third Phase & 0.4116 & 0.0554\\
\cmidrule{2-5}\nopagebreak
\hspace{1em} & 3 Uncontroversial Bills & Intercept & 1.6375 & 0.1383\\
\cmidrule{3-5}\nopagebreak
\hspace{1em} &  & Topic:Legal & 0.7306 & 0.1210\\
\cmidrule{3-5}\nopagebreak
\hspace{1em} &  & Topic:Social programs Public goods & -0.0564 & 0.0728\\
\cmidrule{3-5}\nopagebreak
\hspace{1em} &  & Topic:Security & 0.8924 & 0.2003\\
\cmidrule{3-5}\nopagebreak
\hspace{1em} &  & Topic:Gov operations & 0.4667 & 0.1461\\
\cmidrule{3-5}\nopagebreak
\hspace{1em} &  & Topic:Other & -0.8497 & 0.0814\\
\cmidrule{3-5}\nopagebreak
\hspace{1em} &  & Sponsor Seniority & -0.0624 & 0.0065\\
\cmidrule{3-5}\nopagebreak
\hspace{1em} &  & Sponsor Ideology 1 & 0.3024 & 0.2697\\
\cmidrule{3-5}\nopagebreak
\hspace{1em} &  & Sponsor Ideology 2 & 0.0398 & 0.1569\\
\cmidrule{3-5}\nopagebreak
\hspace{1em} &  & Sponsor Party-Republican & -0.2731 & 0.1990\\
\cmidrule{3-5}\nopagebreak
\hspace{1em} &  & Sponsor Sex-Male & -0.4408 & 0.0939\\
\cmidrule{3-5}\nopagebreak
\hspace{1em} &  & Second Phase & 0.3514 & 0.0918\\
\cmidrule{3-5}\nopagebreak
\hspace{1em} &  & Third Phase & 0.0778 & 0.0623\\
\cmidrule{3-5}\nopagebreak
\hspace{3em} &  &  &  &  \\
\addlinespace[3.75em]
\end{tabular}
\end{minipage}
\caption{\small \textbf{biMMSBM Estimated Coefficients: monadic.} Point estimates of coefficients in the scale of linear predictor, along with their corresponding approximate standard errors.
    }\label{tab:estimcoef2}
\end{table}

\subsection{Degree centrality and senator memberships}\label{sec:centrality}

\begin{table}[h]
\centering \spacingset{1}
\fontsize{9}{11}\selectfont
\begin{tabular}[t]{lccc}
\toprule
  & Estimate & Std. Error & t value\\
\midrule
Baseline: Power Brokers & 6401.593 & 861.9584 & 7.4268\\
Senior Democrats & -3009.3133 & 2112.9178 & -1.4242 \\
Senior Republicans & -1447.2995 & 1568.6299 & -0.9227\\
\midrule 
Multiple $R^2$: 0.02296 & Adjusted $R^2$: 0.00281 & F stat: 1.139 & \\
\bottomrule
\end{tabular}
\caption{\label{tab:between}\textbf{Regression of senator between centrality on group assignment probabilities.} Baseline is Group 3, which is positively correlated with between centrality.}
\end{table}

\subsection{Alternative model specifications}\label{sec:alternativemodels}

\paragraph{Separating block membership and covariate estimation}
Our proposed model allows for joint inference of block membership and
predictor effects. It also allows for inclusion of different predictors. To illustrate these possibilities,  we
restrict the \texttt{biMMSBM} model to infer only block membership or
only predictor effects. Findings are briefly summarized below:
\begin{enumerate}
\item \textbf{Block membership only}. Here no predictors contribute to
  variation in cosponsorship; instead information is highly
  concentrated on the block memberships, which are the same for every
  senator/bill for each respective family of blocks. We find that
  senators who fall into any of the senator latent groups are likely
  to have a tie with a bill that instantiates the second bill latent
  group (ranging 0.613-0.765); all other block-to-block interactions
  are limited (ranging 0.004-0.042). This model suggests low levels of
  tie formation for bills that fall into any latent group outside of
  group~2 --- a departure from the original joint estimation of
  results in our main model. The full table of model estimates is shown in
  Table~\ref{tab:estimcoefnocovsfull}.

\item \textbf{Same-state effects}. In this setting, we incorporate a
  dyadic predictor for same-state membership between a senator a sponsor
  of a bill.  In this specification, the same-state indicator appears as
  a strong predictor of a cosponsorship link. Despite this, it is
  still possible to discern similar results with respect to the
  discovered groups of legislators and bills. The full table of model
  estimates is shown in Table~\ref{tab:estimcoefcostatefull}.
\end{enumerate}

\begin{table}[t] \spacingset{1}
\caption{\textbf{Only Block Membership Full Model Coefficients.} Coefficient point estimates in linear predictor scale, with corresponding approximate standard errors.}
\centering
\begin{tabular}[t]{ccc}
\toprule
 & Coefficient & Standard.Error\\
\midrule
\addlinespace[0.3em]
\multicolumn{3}{l}{\textbf{Monadic: Senator intercepts}}\\
\hspace{1em} & 10.6352 & 0.4101\\
\cmidrule{2-3}
\hspace{1em} & 6.5183 & 0.4100\\
\cmidrule{2-3}
\hspace{1em} & 6.0022 & 0.4097\\
\cmidrule{1-3}
\addlinespace[0.3em]
\multicolumn{3}{l}{\textbf{Monadic: Bill intercepts}}\\
\hspace{1em} & 2.2122 & 0.0233\\
\cmidrule{2-3}
\hspace{1em} & 0.0932 & 0.0235\\
\cmidrule{2-3}
\hspace{1em} & 1.5949 & 0.0239\\
\bottomrule
\end{tabular}
\begin{tabular}[t]{lccc}
\toprule
  \textbf{Blockmodel} & Bill 1 & Bill 2 & Bill 3\\
  Senator 1 0.0103 &   0.7649 &   0.0020\\
Senator 2 & 0.0122 & 0.7555 & 0.0180\\
Senator 3 & 0.0041 & 0.6130 & 0.0416\\
\bottomrule
\end{tabular}
\label{tab:estimcoefnocovsfull}
\end{table}

\begin{landscape}
\spacingset{1.0}
\begin{table} 
\caption{\small \textbf{Same-State Full Model Coefficients.} Coefficient point estimates in linear predictor scale, with corresponding approximate standard errors.
    }
\fontsize{6}{8}\selectfont
\begin{minipage}[t]{0.5\textwidth}
\begin{tabular}{ccccc}
\toprule
 & Group & Coefficient.Name & Estimate & SE\\
\midrule
\addlinespace[0.3em]
\multicolumn{5}{l}{\textbf{Dyadic predictors}}\\
\hspace{1em} &  & No reciprocity history & -1.5580 & 0.0196\\
\cmidrule{3-5}
\hspace{1em} &  & Log proportional reciprocity & 0.4354 & 0.0044\\
\cmidrule{3-5}
\hspace{1em} &  & Shared committee & 0.6560 & 0.0208\\
\cmidrule{3-5}
\hspace{1em} &  & Same-state & 1.8788 & 0.0430\\
\cmidrule{1-5}
\addlinespace[0.3em]
\multicolumn{5}{l}{\textbf{Senator predictors}}\\
\hspace{1em} & 1 Democrat & Intercept & 9.6722 & 1.2719\\
\cmidrule{3-5}
\hspace{1em} &  & Seniority & 0.0229 & 0.3889\\
\cmidrule{3-5}
\hspace{1em} &  & Ideology 1 & -6.7897 & 1.2818\\
\cmidrule{3-5}
\hspace{1em} &  & Ideology 2 & 2.6573 & 1.2866\\
\cmidrule{3-5}
\hspace{1em} &  & Party-Republican & -4.8794 & 1.4720\\
\cmidrule{3-5}
\hspace{1em} &  & Sex-Male & -0.6568 & 1.2605\\
\cmidrule{2-5}
\hspace{1em} & 2 Republican & Intercept & 4.2601 & 1.2716\\
\cmidrule{3-5}
\hspace{1em} &  & Seniority & 0.8223 & 0.3889\\
\cmidrule{3-5}
\hspace{1em} &  & Ideology 1 & 1.7553 & 1.2792\\
\cmidrule{3-5}
\hspace{1em} &  & Ideology 2 & 4.7461 & 1.2862\\
\cmidrule{3-5}
\hspace{1em} &  & Party-Republican & 1.6111 & 1.3155\\
\cmidrule{3-5}
\hspace{1em} &  & Sex-Male & 1.0517 & 1.2606\\
\cmidrule{2-5}
\hspace{1em} & 3 Junior Power Brokers & Intercept & 20.1595 & 1.2715\\
  \cmidrule{3-5}
  \hspace{1em} &  & Seniority & -0.9228 & 0.3889\\
\cmidrule{3-5}
\hspace{1em} &  & Ideology 1 & -1.0970 & 1.2781\\
\cmidrule{3-5}
\hspace{1em} &  & Ideology 2 & -7.2593 & 1.2872\\
\cmidrule{3-5}
\hspace{1em} &  & Party-Republican & -1.3287 & 1.3154\\
\cmidrule{3-5}
  \hspace{1em} &  & Sex-Male & -1.1748 & 1.2605\\
\cmidrule{1-5}
\multicolumn{5}{l}{\textbf{Bill predictors}}\\
\hspace{1em} & 1 Contentious Bills & Intercept & 11.2669 & 0.1137\\
\cmidrule{3-5}
\hspace{1em} &  & Topic:Legal & 0.8587 & 0.0303\\
\cmidrule{3-5}
\hspace{1em} &  & Topic:Social programs Public goods & 0.8151 & 0.0555\\
\cmidrule{3-5}
\hspace{1em} &  & Topic:Security & 0.8096 & 0.0709\\
\cmidrule{3-5}
\hspace{1em} &  & Topic:Gov operations & 0.5672 & 0.0675\\
\cmidrule{3-5}
\hspace{1em} &  & Topic:Other & -5.5515 & 0.0981\\
\cmidrule{3-5}
\hspace{1em} &  & Sponsor Seniority & -0.1431 & 0.0147\\
\cmidrule{3-5}
\hspace{1em} &  & Sponsor Ideology 1 & 0.3879 & 0.5242\\
\cmidrule{3-5}
\hspace{1em} &  & Sponsor Ideology 2 & -0.5291 & 0.3436\\
\cmidrule{3-5}
\hspace{1em} &  & Sponsor Party-Republican & -1.3865 & 0.3804\\
\cmidrule{3-5}
\hspace{1em} &  & Sponsor Sex-Male & -2.6192 & 0.1364\\
\cmidrule{3-5}
\end{tabular}
\end{minipage} \hspace{1cm} 
\begin{minipage}[t]{0.5\textwidth}
\begin{tabular}{ccccc}
\toprule
 & Group & Coefficient Name & Estimate & SE\\
\midrule
\addlinespace[0.3em]
\multicolumn{5}{l}{\textbf{Bill predictors (continued)}}\\
\hspace{1em} &  & Second Phase & 3.6591 & 0.6810\\
\cmidrule{3-5}
\hspace{1em} &  & Third Phase & 1.9508 & 0.1370\\
\cmidrule{2-5}
\hspace{1em} & 2 Bipartisan Resolutions & Intercept & 10.5539 & 0.1137\\
\cmidrule{3-5}
\hspace{1em} &  & Topic:Legal & 0.7711 & 0.0308\\
\cmidrule{3-5}
\hspace{1em} &  & Topic:Social programs Public goods & 0.7775 & 0.0556\\
\cmidrule{3-5}
\hspace{1em} &  & Topic:Security & 1.2102 & 0.0711\\
\cmidrule{3-5}
\hspace{1em} &  & Topic:Gov operations & 0.9030 & 0.0678\\
\cmidrule{3-5}
\hspace{1em} &  & Topic:Other & -4.9057 & 0.0979\\
\cmidrule{3-5}
\hspace{1em} &  & Sponsor Seniority & -0.0878 & 0.0147\\
\cmidrule{3-5}
\hspace{1em} &  & Sponsor Ideology 1 & 0.5021 & 0.5239\\
\cmidrule{3-5}
\hspace{1em} &  & Sponsor Ideology 2 & -0.0069 & 0.3433\\
\cmidrule{3-5}
\hspace{1em} &  & Sponsor Party-Republican & -1.6197 & 0.3802\\
\cmidrule{3-5}
\hspace{1em} &  & Sponsor Sex-Male & -2.6768 & 0.1365\\
\cmidrule{3-5}
\hspace{1em} &  & Second Phase & 3.8117 & 0.6807\\
\cmidrule{3-5}
\hspace{1em} &  & Third Phase & 0.9561 & 0.1368\\
\cmidrule{2-5}
  \hspace{1em} & 3 Uncontroversial Bills & Intercept & 5.5493 & 0.1362\\
\hspace{1em} &  & Topic:Legal & 1.1332 & 0.0564\\
\cmidrule{3-5}
\hspace{1em} &  & Topic:Social programs Public goods & 0.7925 & 0.0676\\
\cmidrule{3-5}
\hspace{1em} &  & Topic:Security & 0.0947 & 0.1242\\
\cmidrule{3-5}
\hspace{1em} &  & Topic:Gov operations & 0.0754 & 0.0963\\
\cmidrule{3-5}
\hspace{1em} &  & Topic:Other & -3.3246 & 0.0793\\
\cmidrule{3-5}
\hspace{1em} &  & Sponsor Seniority & -0.0827 & 0.0166\\
\cmidrule{3-5}
\hspace{1em} &  & Sponsor Ideology 1 & -1.0612 & 0.5530\\
\cmidrule{3-5}
\hspace{1em} &  & Sponsor Ideology 2 & 0.0791 & 0.3645\\
\cmidrule{3-5}
\hspace{1em} &  & Sponsor Party-Republican & 0.2602 & 0.4025\\
\cmidrule{3-5}
\hspace{1em} &  & Sponsor Sex-Male & -1.7541 & 0.1588\\
\cmidrule{3-5}
\hspace{1em} &  & Second Phase & 1.5149 & 0.8123\\
\cmidrule{3-5}
\hspace{1em} &  & Third Phase & -0.0066 & 0.1424\\  
\cmidrule{3-5}
\midrule 
 \textbf{Blockmodel} & & & \\
 & Bill 1 & Bill 2 & Bill 3\\
\midrule
Senator 1 & 0.3652 & 0.9819 & 0.5175\\
Senator 2 & 0.1088 & 0.1960 & 0.9918\\
Senator 3 & 0.1399 & 0.2326 & 0.9939 \\
\bottomrule
\end{tabular}
\end{minipage}
\label{tab:estimcoefcostatefull}
\end{table}
\end{landscape}